\documentclass[aps,pra,twocolumn,superscriptaddress]{revtex4-1}

\usepackage{amsmath}
\usepackage{graphicx}
\usepackage{afterpage}
\usepackage{enumitem}
\usepackage{color}

\def\L{{\mathcal L}}

\begin{document}

\title{Entanglement renormalization for disordered systems}

\author{Andrew M.\ Goldsborough}
\email[]{andrew.goldsborough@mpq.mpg.de}
\homepage[]{www.goldsborough.org.uk}
\affiliation{Max-Planck-Institut f\"{u}r Quantenoptik, Hans-Kopfermann-Str. 1, D-85748 Garching, Germany}

\author{Glen Evenbly}
\email[]{glen.evenbly@usherbrooke.ca}
\homepage[]{www.glenevenbly.com}
\affiliation{D\'{e}partement de Physique and Institut Quantique, Universit\'{e} de Sherbrooke, Qu\'{e}bec, Canada}

\date{\today}

\begin{abstract}
We propose a tensor network method for investigating strongly disordered systems that is based on an adaptation of entanglement renormalization [G. Vidal, {Phys. Rev. Lett.} {\bf 99}, 220405 (2007)].
This method makes use of the strong disorder renormalization group to determine the order in which lattice sites are coarse-grained, which sets the overall structure of the corresponding tensor network ansatz, before optimization using variational energy minimization. Benchmark results from the disordered XXZ model demonstrates that this approach accurately captures ground state entanglement in disordered systems, even at long distances. This approach leads to a new class of efficiently contractible tensor network ansatz for $1D$ systems, which may be understood as a generalization of the multi-scale entanglement renormalization ansatz (MERA) for disordered systems.
\end{abstract}

\pacs{75.10.Jm, 05.30.-d, 02.70.-c}

\maketitle

\section{Introduction}
Disordered systems are naturally of interest in quantum many-body physics as real world materials are rarely perfect. Furthermore it is known that the presence of disorder can dramatically alter the properties of a system, as evidenced in the ground-breaking work of Anderson \cite{And58} which that showed that the amount of disorder in a non-interacting three dimensional electron system dictates whether the states are localized or extended. More recently there has been great excitement around the idea of many-body localization, which extends Anderson localization to interacting many-body systems \cite{BasAA06,NanH15,AltV15,AbaP17}.

Unfortunately, disordered systems can also be challenging to study using standard approaches. For instance the \emph{density matrix renormalization group} (DMRG) \cite{Whi92,OstR95,Sch05,Sch11}, which is otherwise regarded as an extremely reliable and robust method for $1D$ quantum systems, may have difficulty even for a relatively simple example like a $1D$ disordered Heisenberg model due to the presence of long range singlets in the ground state \cite{JuoCR97,JuoUCR99,GolR14,RugAC16}. On the other hand, specialist methods developed for disordered systems such as the \emph{strong disorder renormalization group} (SDRG) method of Ma, Dasgupta and Hu \cite{MaDH79,DasM80} may capture the qualitative features of the ground state entanglement, but generally only produce quantitatively precise results when the disorder is infinitely strong.

In this manuscript we propose and benchmark a new tensor network method designed for disordered systems. In most previous tensor network approaches the structure of the network, i.e.\ the pattern of how the tensors in the network are connected, is fixed independent of the problem under consideration whilst the parameters within the tensors are optimized variationally. Conversely, the method we introduce here is based on allowing both the structure of the tensor network as well as the parameters within the tensors to be adapted to the problem under consideration. This idea is similar to previous work \cite{GolR14,LinKCL17}, which adapted a coarse-graining transformation, based on blocking together sites using isometries, to specific instances of a disordered system, and thus produced an approximation to the quantum ground state as a disordered \emph{tree tensor network} (TTN). 
However, in the present work the coarse-graining transformation is based on entanglement renormalization \cite{Vid07}.
This differs by including unitary \emph{disentanglers}, such that the ground state is approximated as a generalized form of the multi-scale entanglement renormalization ansatz (MERA) \cite{Vid08_2,EveV09,PfeEV09,EvePPI10,EveV14}.
Here we use SDRG to adapt the overall structure of the network to the problem under consideration, which helps ensure the qualitative features of the ground state entanglement are captured, before then applying variational sweeping to optimize the tensors and improve the quantitative accuracy significantly.

The manuscript is organized as follows: Section \ref{sec:SDRG} provides an overview of some of the previous approaches to numerical strong disorder renormalization, and Sect.\ \ref{sec:TN} discusses the problems of trying to combine SDRG with a variational tensor network algorithm. Section \ref{sec:algorithm} describes the newly proposed tensor network method for disordered systems, also outlining the variational optimization algorithm that is employed. Finally, Sect.\ \ref{sec:results} presents benchmark results for the disordered XX and Heisenberg (XXX) models, which are compared to various other numerical methods.

\section{\label{sec:SDRG}Strong disorder renormalization}
\begin{figure}
    (a)\includegraphics[scale=0.25]{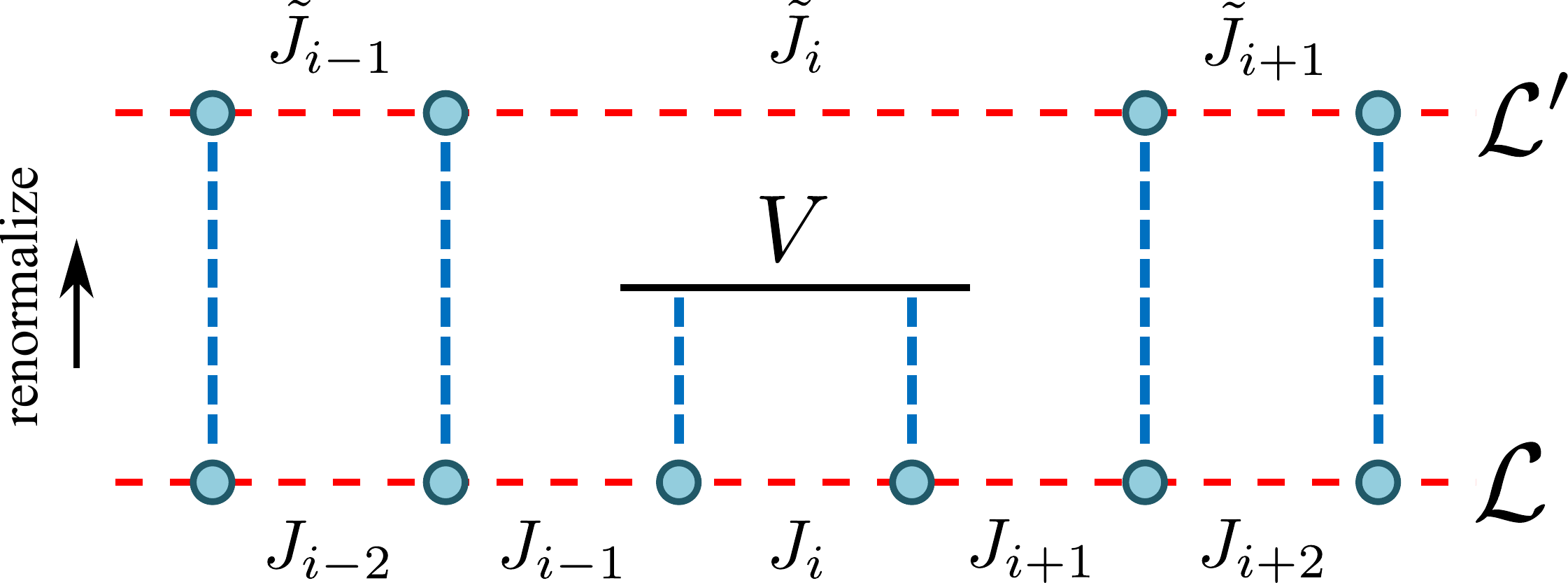} \vspace{0.4cm}\\
    (b)\includegraphics[scale=0.25]{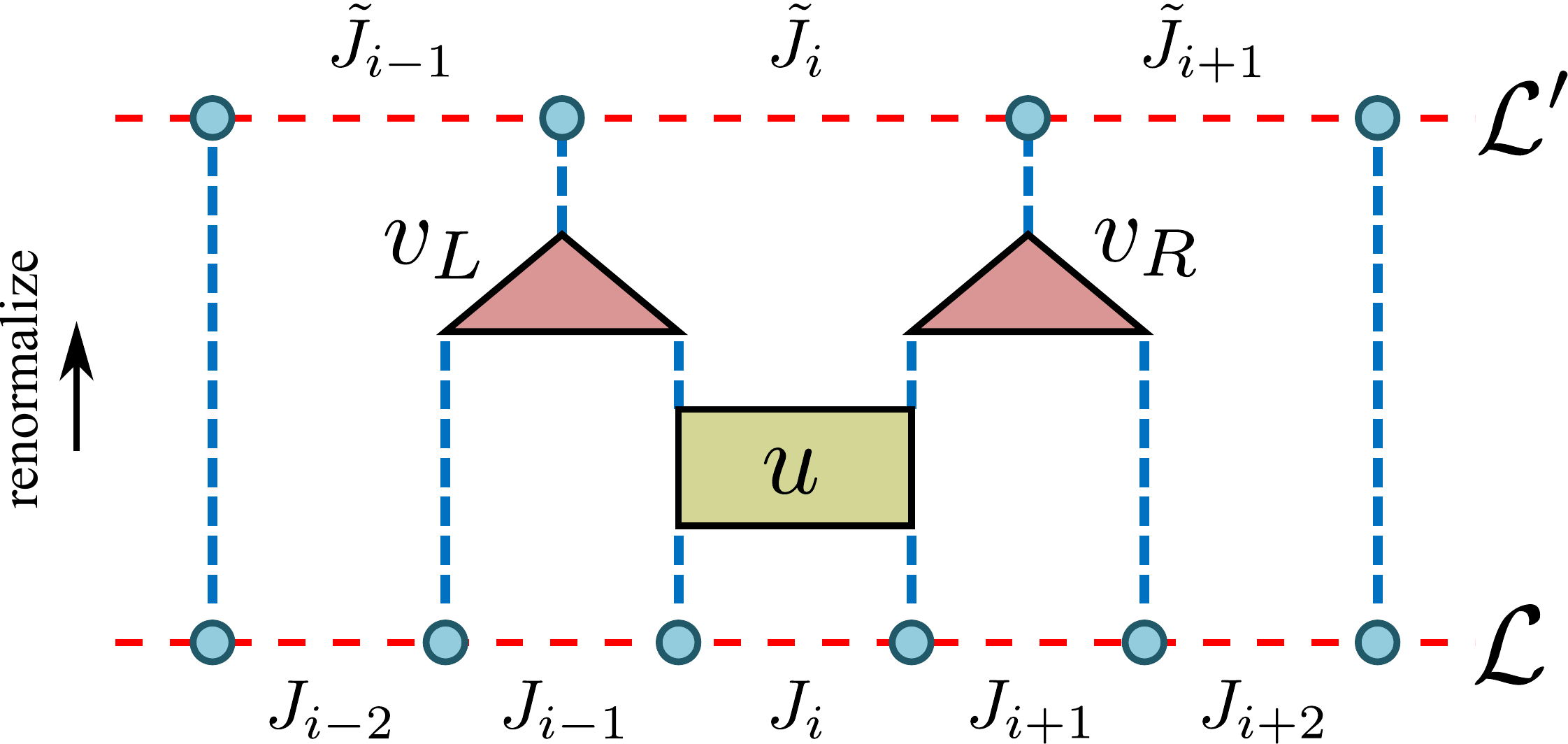} \vspace{0.4cm}\\
    (c)\includegraphics[scale=0.25]{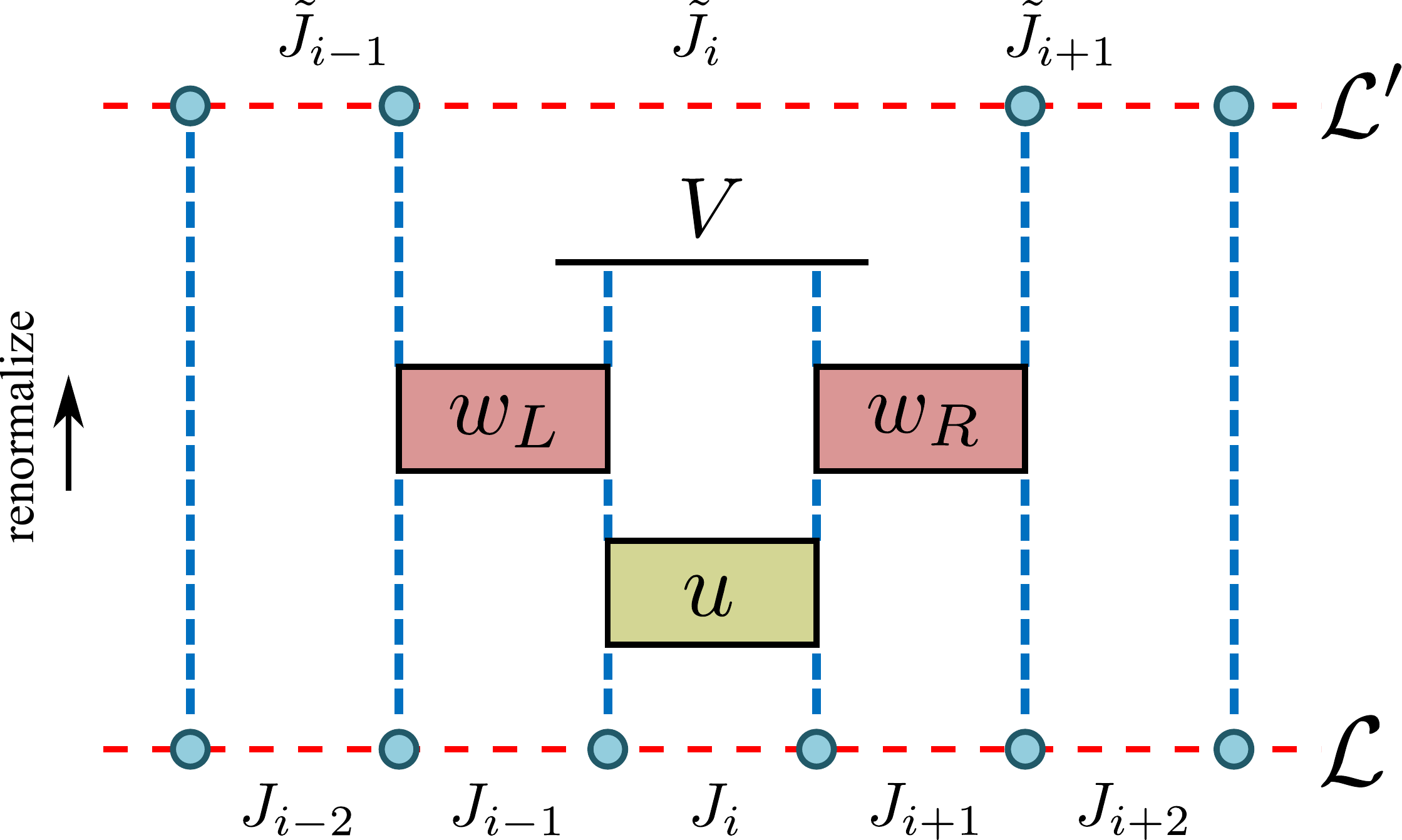}
    \caption{
        (a) Schematic of a step of the SDRG, which maps a local Hamiltonian of the form in Eq. \ref{eqn:Heisenberg} from an initial lattice $\mathcal{L}$ to a coarser lattice $\mathcal{L}'$. Here it was assumed that $J_{i}$ was the strongest coupling, such that the corresponding spins are fixed to form a singlet $V$, denoted by the horizontal line. 
        (b) A coarse-graining transformation from $\mathcal{L}$ to $\mathcal{L}'$ implemented via a unitary $u$ and a pair of isometries $v_L$ and $v_R$.
        (c) A coarse-graining transformation from $\mathcal{L}$ to $\mathcal{L}'$ implemented via unitaries $u$, $w_L$ and $w_R$. Similar to the transformation in (a) the spins associated to the strong coupling $J_{i}$ are fixed in a singlet state $V$.}
    \label{fig:MDHRG}
    \label{fig:dMERA2}
    \label{fig:dMERA3}
\end{figure}

In this section we review some of previous approaches proposed for strongly disordered systems. Let us consider the disordered XXZ model on a $1D$ lattice of spin-$\tfrac{1}{2}$ sites, with periodic boundary conditions (PBCs). The Hamiltonian is defined as
\begin{equation}
    H = \sum_{i} J_{i} \left\{ \frac{1}{2} \left[ s^{+}_{i} s^{-}_{i+1} + s^{-}_{i} s^{+}_{i+1} \right] + \Delta_{z} s^{z}_{i} s^{z}_{i+1} \right\}
\label{eqn:Heisenberg}
\end{equation}
where $s^x$, $s^y$ and $s^z$ are spin-$\tfrac{1}{2}$ matrices, with $s^\pm = s^x \pm i s^y$. Notice that the Heisenberg model when is recovered when the anisotropy is fixed at $\Delta_{z} = 1$, while the XX model is when $\Delta_{z} = 0$. Disorder is introduced in the couplings $J_{i}$ which are allowed to vary in strength with position; here we choose these couplings randomly between $0<J_{i}<J_{\text{max}}$ according to some probability distribution $P(J)$. 

\begin{figure}
    (a)\includegraphics[width=0.93\columnwidth]{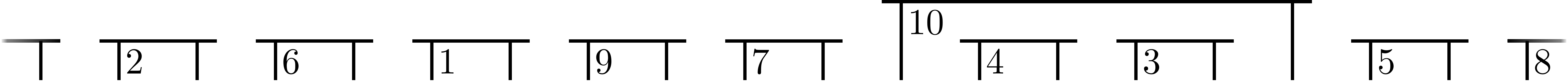} \vspace{0.4cm}\\ 
    (b)\includegraphics[width=0.93\columnwidth]{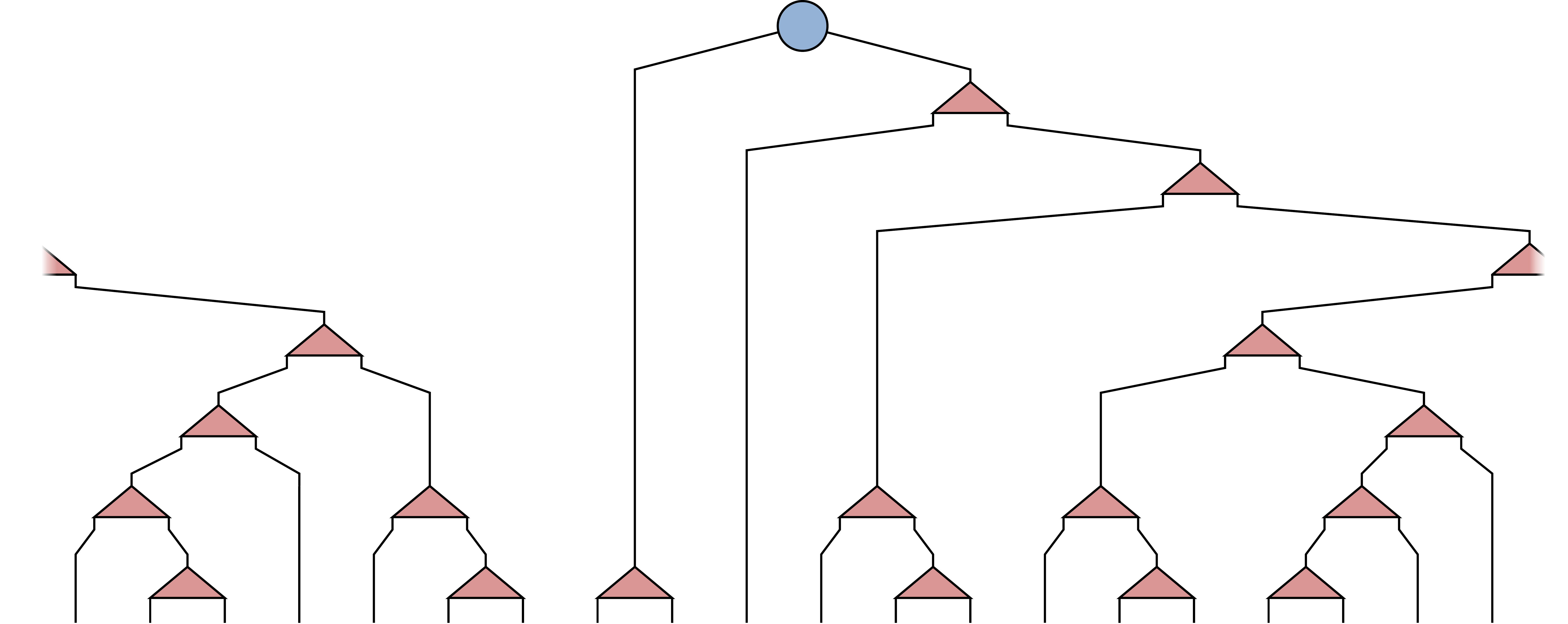} \vspace{0.4cm}\\
    (c)\includegraphics[width=0.93\columnwidth]{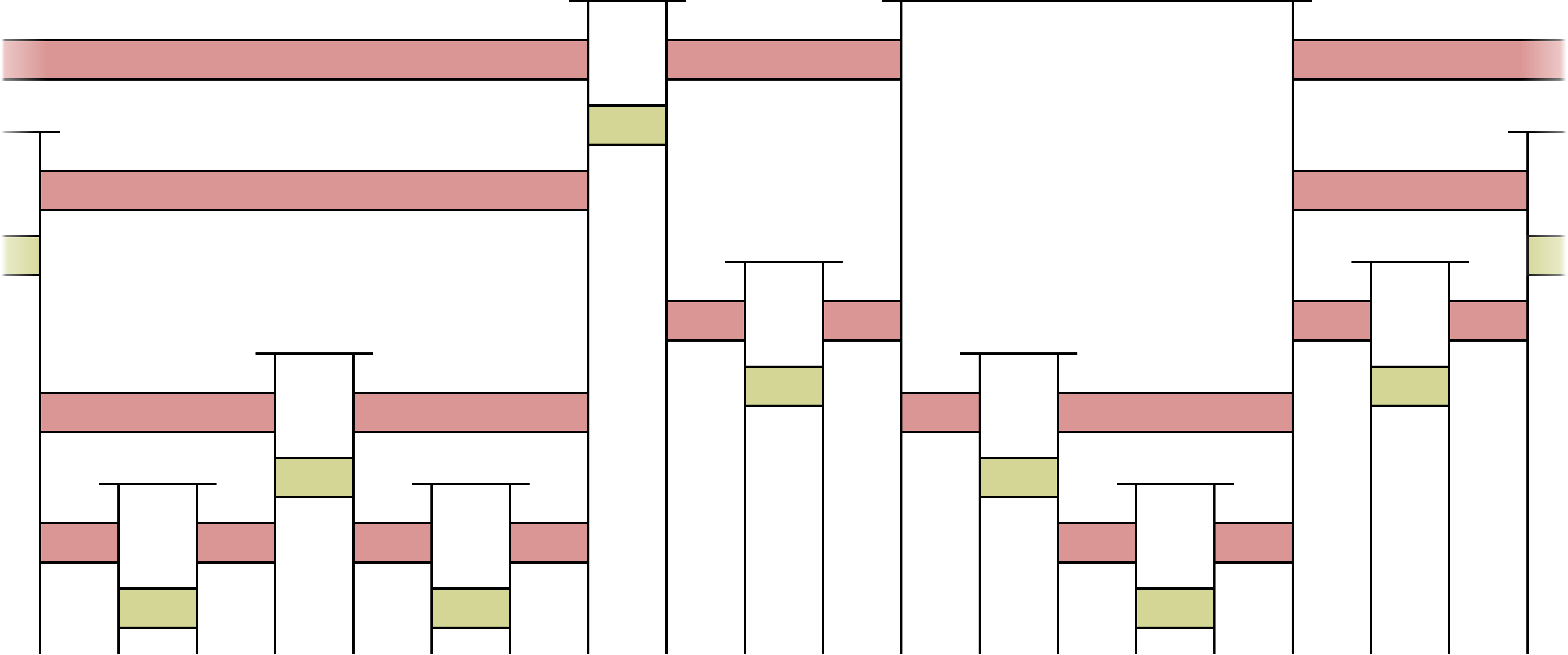}
    \caption{Examples of wavefunctions generated from various methods applied to a disordered system of 20 sites with periodic boundaries. The same set of couplings assumed for all three instances. (a) The SDRG method produces a product of singlets, denoted by horizontal lines as in Fig.\ \ref{fig:dMERA3}(a), where the numbers show the order in which the singlets are created. (b) An inhomogeneous tree tensor network, composed of isometries that each map two sites to one, generated from the tSDRG approach of Ref.\ \cite{GolR14}. (c) An instance of a disordered MERA, created through composition of the coarse-graining transformation from Fig.\ \ref{fig:dMERA3}(c).}
    \label{fig:MDH_order}
    \label{fig:tSDRG_PBC_L20}
    \label{fig:dMERA_L20}
\end{figure}

Many previous studies of the disordered Heisenberg model have used variants of the SDRG \cite{Fis92, Fis95, WesFSL95, AltKPR04, IglM05}, the archetypal method for the antiferromagnetic case was set out by Ma, Dasgupta and Hu \cite{MaDH79,DasM80}. In this procedure the pair of spins with the strongest coupling $J_{i}$ are approximated as a singlet, and then second order perturbation theory is used to find an effective coupling $\tilde{J}_{i}$ between the spins either side, as illustrated in Fig.\ \ref{fig:MDHRG}(a). This step is repeated to produce a \emph{random singlet phase}, an example is given in Fig.\ \ref{fig:MDH_order}(a), where singlets created later on in the renormalization scheme can potentially span large distance scales. When averaged over many realizations of randomly chosen couplings $J_{i}$, the spin-spin correlation function is known to obey an inverse-square power law decay and the entanglement entropy scales logarithmically in a manner consistent with an effective conformal field theory (CFT) of central charge $c=1$ \cite{RefM04, RefM07}.

\begin{figure}
    (a)\includegraphics[scale=0.22]{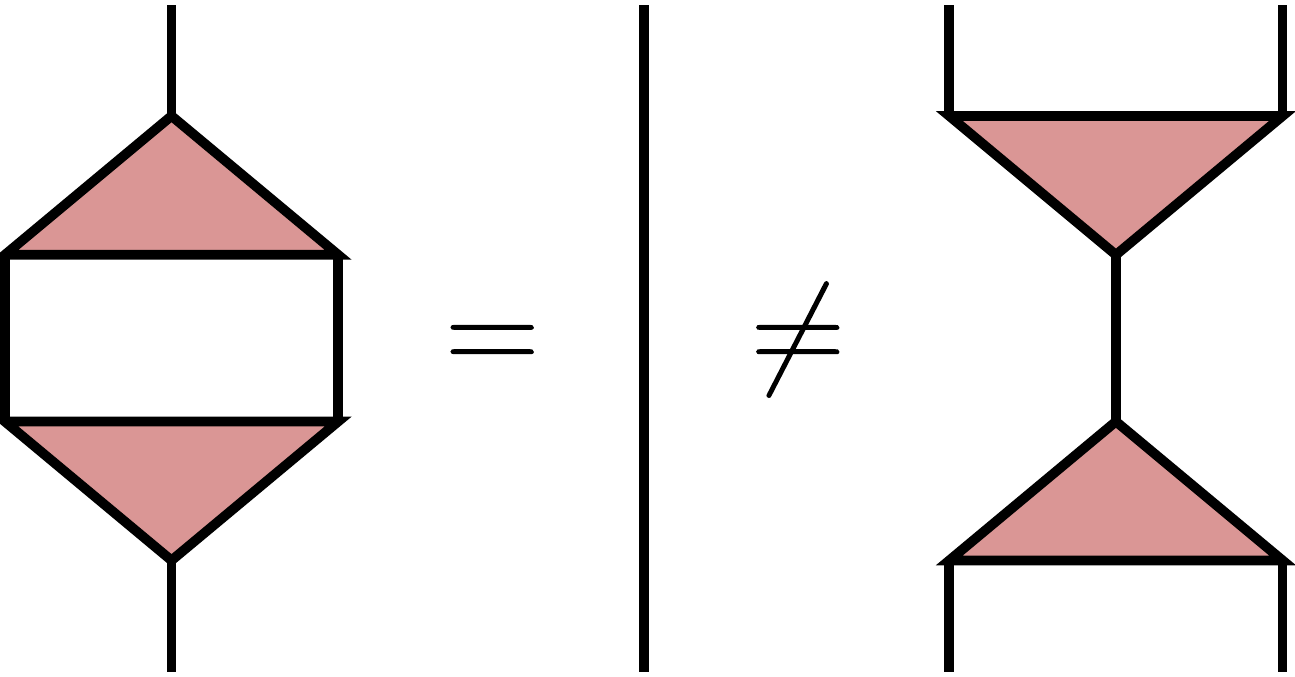} \hspace{0.8cm}
    (b) \includegraphics[scale=0.22]{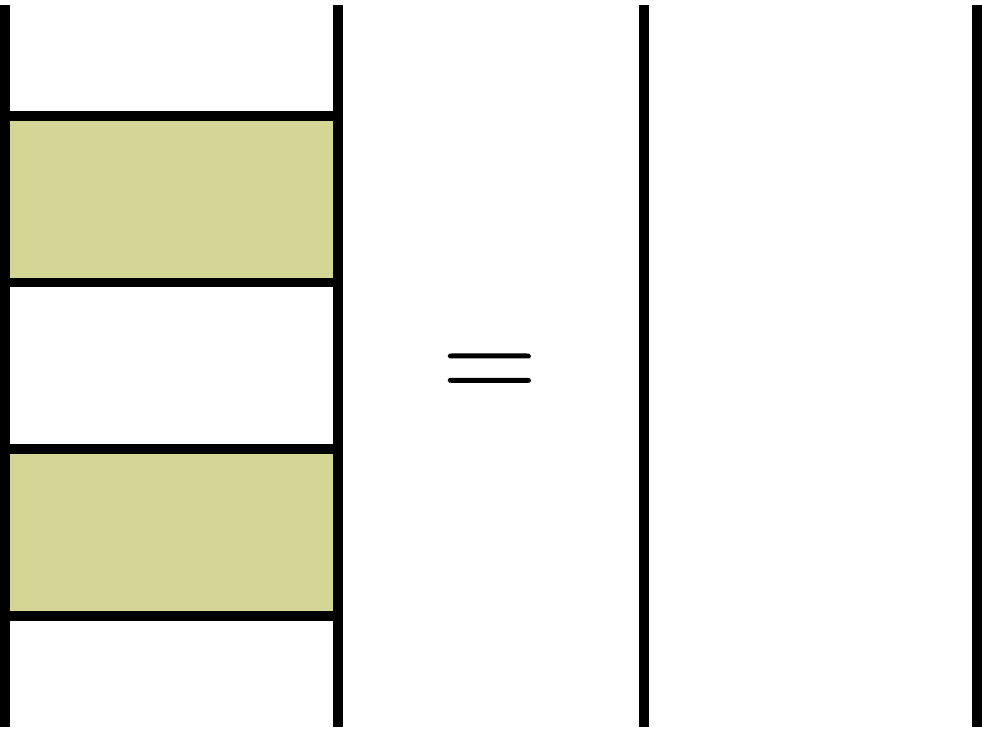}
    \caption{
        Defining properties of (a) isometric tensors or \emph{isometries} ($v v^{\dagger} = \openone \neq v^{\dagger}v$) and (b) unitary tensors or \emph{disentanglers} ($u u^{\dagger} = \openone \otimes \openone = u^{\dagger} u$).
    }
    \label{fig:iso}
    \label{fig:unitary}
\end{figure}

More recent developments have included higher order components to increase the accuracy of observables such as the spin-spin correlation functions \cite{HikFS99}, which is effectively an application of Wilson's numerical renormalization group (NRG) \cite{Wil75} to a system with many impurities. Later it was shown in Ref.\ \onlinecite{GolR14} that this form of SDRG has an efficient tensor network representation in terms of an inhomogeneous TTN, where the structure is set by the renormalization order of the couplings, which we refer to as the \emph{tSDRG} method. An example of a wavefunction produced from tSDRG is given in Fig.\ \ref{fig:tSDRG_PBC_L20}(b). The basic building block of these wavefunctions are isometries $v$, which annihilate to an identity with their conjugates $v v^{\dagger} = \openone$ as also depicted in Fig.\ \ref{fig:iso}(a), and are responsible for coarse-graining two sites into a single effective site. In the tSDRG approach the isometries are found by diagonalizing the block Hamiltonian supported on the pair of sites with the strongest coupling. Whilst this method was shown to reproduce qualitative features of the correlation functions and entanglement entropy, the ground state energy was not given with high precision \cite{GolR14}. Although the accuracy of the energy could be improved through the subsequent use of a variational sweep to minimize the energy of the corresponding TTN, this was found to be at the expense of accuracy in the other features of the state. In particular, the energy minimization update favoured short-range correlations and ignored the long-range singlets, which are key to the proper characterization of the disordered ground states.

\section{\label{sec:TN}Tensor networks and disorder}
The goal of this manuscript is to design a tensor network that can properly capture the structure of entanglement in a disordered ground state, in particular the pattern of long range singlets. 
Whereas the tSDRG approach constructs the TTN and sets the values in the tensors once, in this section we propose a more sophisticated network that allows a variational update whilst still capturing the long range entanglement.
It follows the ideas of entanglement renormalization, employing unitary \emph{disentanglers} in such a way as to modify the random singlet phase of SDRG to be relevant for finite systems with finite disorder.

Each step of the SDRG can be interpreted as a coarse-graining transformation that maps a lattice $\L$ to a new lattice $\L'$ of two fewer sites by fixing two neighbouring spins, specifically those which possess the strongest coupling $J_i$, in a singlet state as depicted in Fig.\ \ref{fig:MDHRG}(a). Here we propose two generalizations of this transformation, each a mapping from $\L$ to $\L'$. The first, presented in Fig.\ \ref{fig:dMERA2}(b), enacts a unitary disentangler $u$ across the coupling $J_i$ to be renormalized before employing isometries $v_L$ and $v_R$ that each coarse-grain two sites into one. The second, presented in Fig.\ \ref{fig:dMERA3}(c), also enacts a unitary disentangler $u$ across the coupling $J_i$ before employing additional unitary gates $w_L$ and $w_R$ with the boundary spins. The sites at the location of coupling $J_i$ are then fixed in an entangled singlet state, similar to SDRG. 
One should notice that if the spins associated to $J_{i}$ in the second scheme, Fig.\ \ref{fig:dMERA3}(c), were instead fixed as a product state then the first scheme of Fig.\ \ref{fig:dMERA2}(b) is recovered.
This follows as unitary gates $w_L$ and $w_R$ with a fixed (product state) input index can be simplified to isometries $v_L$ and $v_R$, a property that allows the MERA to be interpreted both as a tensor network and as a quantum circuit \cite{Vid08_2}.
As with other tensor network methods, we set an upper bound to the index dimensions of the tensors in order to truncate of the Hilbert space dimension for the coarse grained sites.
This is known as setting the \emph{bond dimension} $\chi$.

These coarse-graining schemes satisfy two important properties. Firstly, it is easy to see that there exist choices of tensors in both schemes for which the singlet state of Fig.\ \ref{fig:MDHRG}(a) is reproduced, thus they should at least match the effectiveness of the standard SDRG. Secondly it should be noted that these schemes preserve locality: any operator supported on two contiguous sites of the initial lattice $\L$ is mapped to an operator supported on two contiguous sites of the coarser lattice $\L'$. Hence it also follows that a nearest-neighbour Hamiltonian on $\L$ is mapped to a nearest-neighbour Hamiltonian on $\L'$. Preservation of locality is important in order to produce a computationally viable numerical algorithm as, in general, the cost of representing a local operator grows exponentially with the size of its support.

The proposed lattice transformations of Fig.\ \ref{fig:dMERA2}(b-c) can be seen to yield a class of tensor network ansatz, just as entanglement renormalization yields the class of tensor network known as the MERA. For a lattice $\L$ of $N$ sites, we consider the class of tensor network that encompasses all possible ways of applying $N/2$ of the transformations from either Fig.\ \ref{fig:dMERA2}(b) or Fig.\ \ref{fig:dMERA3}(c) in order to reach a coarse-grained lattice of $O(1)$ sites. An example of a tensor network from this class is depicted in Fig.\ \ref{fig:dMERA_L20}(c). Importantly, one should notice that a standard MERA arises as a specific instance in this more general class. More precisely, the modified binary MERA, as introduced in \cite{EveV13}, can be understood as a particular (uniform) arrangement of the unit cell of Fig.\ \ref{fig:dMERA2}(b). Accordingly, we refer to class of tensor network that generalizes the MERA for disordered systems as a disordered MERA or \emph{dMERA}. Notice that, as with the standard MERA, all instances of dMERA possess a bounded causal structure, due to the preservation of locality in the underlying coarse-graining transformation, and are thus efficiently contractible for expectation values of local observables and for correlation functions. 

\section{\label{sec:algorithm} Algorithm}
In this section we discuss an algorithm for how the proposed coarse-graining transformations can be implemented for the study of disordered systems. This algorithm consists of a procedure for first determining the overall geometry of the dMERA tensor network, before then applying variational energy minimization to optimize the tensors. 
Here we use the original SDRG algorithm \cite{MaDH79} to determine the network geometry, then build the tensor network by patterning the unit cell of isometric and unitary gates, as depicted in Fig.\ \ref{fig:dMERA2}(b-c), to match the order of singlets. 
An example of a random singlet state produced by SDRG and the corresponding dMERA is presented in Fig.\ \ref{fig:MDH_order}(a,c). 
In general, we employ the transformation of Fig.\ \ref{fig:dMERA3}(c) throughout, apart from when we wish to increase the bond dimension $\chi$ of the network (usually during the earlier RG steps).
The reason is that, despite the block in Fig.\ \ref{fig:dMERA2}(b) reducing the geometric distance between any two sites connected by a singlet \cite{EveV11}, the update process still tends to omit long range correlation.
When the singlets are an explicit part of the coarse-graining block as in Fig.\ \ref{fig:dMERA3}(c), the long range properties are preserved.

Once the structure of the dMERA has been set, we then adapt the standard MERA energy minimization algorithm \cite{EveV09,EveV13} in order to optimize the tensors. Here we provide only an outline of the algorithm, the full details of which are provided in appendix \ref{sec:alg_details}. 
In a similar manner to standard MERA, we optimize the tensors by means of a variational sweep over the layers of the network.
Each layer is comprised of a coarse-graining block as depicted by Fig.\ \ref{fig:dMERA2}(b-c), which transforms between an initial lattice $\L$ and a coarser lattice $\L'$. Following the strategy proposed in Ref.\ \onlinecite{EveV09} the tensors within the cell are updated by first computing their corresponding linearized environments from the coarse-grained Hamiltonian on lattice $\L$ and the two-site local reduced density matrices on lattice $\L'$, themselves obtained using the appropriate ascending and descending superoperators respectively. Each of the environments is then decomposed using a singular value decomposition (SVD) in order to find the new isometry or unitary that minimizes the (linearized) energy. The update is performed for all tensors within each layer, and sequentially over all layers, and this variational sweep is iterated until the tensors converge. All the manipulations required to optimize a dMERA scale as $\mathcal{O}(\chi^{7})$ in the bond dimension $\chi$ of the network.
 
Once the tensor network state is converged we then calculate expectation values and other properties of interest. Calculating expectation values of one- and two-site operators is straightforward; one can simply use the ascending superoperators to coarse-grain the operator through the network as discussed in appendix \ref{sec:expectation}. As with optimization, this evaluation has the cost $\mathcal{O}(\chi^{7})$ in general but can be simplified to $\mathcal{O}(\chi^{6})$ for operators located on particular sites. Two-point correlation functions can also be efficiently computed by initially coarse-graining each of the operators separately and then fusing them into a single operator at the scale they meet. As with standard MERA, this may incur a higher computational cost as the fusion of the two operators may result in an operator with a larger support. As shown in appendix \ref{sec:corr}, the full cost of calculating the correlation function can vary between $\mathcal{O}(\chi^{6})$ and $\mathcal{O}(\chi^{11})$ depending on the particular sites involves and the geometry of the network. Block entanglement entropy is one of the key quantities that will be analysed in sec.\ \ref{sec:results}, as one of the aims of this work is to achieve the correct entanglement structure of the wavefunction. However, the calculation of entanglement entropy is generally computationally difficult. Fortunately, the form of the tensor network alleviates this to some degree, as discussed in appendix \ref{sec:ee}, such that the entanglement entropy can be computed from tensor network states with small $\chi$.

\section{\label{sec:results}Results}
In order to benchmark the proposed method we focus on the disordered anti-ferromagnetic XX and Heisenberg models, where $\Delta_{z} = 0, 1$ in Eq.\ (\ref{eqn:Heisenberg}) respectively. We select coupling strengths $0 < J_{i} < 1$ from the \emph{strong disorder distribution} \cite{Laf05}, which has the probability density function 
\begin{equation}
    P(J) = \frac{1}{\Delta_{J}} J^{-1 + \Delta_{J}^{-1}},
\end{equation}
where $\Delta_{J} \geq 0$ is the disorder strength.
When $\Delta_{J} = 1$ the distribution is uniform and when $\Delta_{J} \to \infty$ the distribution matches that of the infinite randomness fixed point. 
For the remainder of this paper we will concentrate on the stronger disorder regime, when $\Delta_{J} \geq 1$, as this tends to be more challenging for numerical tenchniques.
Throughout this section we compare the efficacy of the dMERA with various other numerical methods, including the tSDRG algorithm \cite{GolR14} and DMRG, which is performed using the ITensor libraries \cite{Itensor023}. 
It may be possible to obtain more accurate DMRG results by, for example, changing the algorithm to take into consideration inhomogeneities of the system \cite{JuoCR97,JuoUCR99} or performing measurements at multiple $\chi$ and extrapolating to the $\chi \to \infty$ limit.
Here we concentrate on a standard DMRG implementation without extrapolating so we can compare the finite $\chi$ performance of the algorithms.
Finally we note that, as the XX model can be mapped to a free fermion model via a Jordan-Wigner transformation \cite{HenG98, Laf05}, we can obtain exact results to benchmark against, even for large system sizes.

\subsection{\label{sec:XX_results}The disordered XX model}
We begin by applying a dMERA, built from composition of the unit cell depicted in Fig.\ \ref{fig:dMERA3}(c), of bond dimension $\chi = 2$ to find the ground state of the (disordered) XX model.
{We observe that after optimization with any disorder strength, the unitaries $w_{L,R}$, when viewed as a matrix between incoming and outgoing indices, may be written as
\begin{equation}
w_{L,R} = \begin{pmatrix} 1 & 0 & 0 & 0 \\ 0 & \text{cos}\theta_{L,R} & -\text{sin}\theta_{L,R} & 0 \\ 0 & \text{sin}\theta_{L,R} & \text{cos}\theta_{L,R} & 0 \\ 0 & 0 & 0 & 1 \end{pmatrix}.
\label{eqn:XX_w_matrix}
\end{equation}
with $\theta_{L,R}$ a free variational parameter. 
The disentanglers $u$ converge to a fixed matrix,
\begin{equation}
u = \begin{pmatrix} 1 & 0 & 0 & 0 \\ 0 & 0 & -1 & 0 \\ 0 & -1 & 0 & 0 \\ 0 & 0 & 0 & -1 \end{pmatrix},
\label{eqn:XX_u_matrix}
\end{equation}
that does not contain any free variational parameters.
Therefore, in this setting, the $\chi=2$ dMERA is fully specified by $L - 3$ parameters, with $L$ the system size. The optimization of this network can also be simplified from the general algorithm introduced in Sect. \ref{sec:algorithm}, as discussed in appendix \ref{sec:XX_update}. 
Furthermore, one should notice that if all the angles are set to $\theta_{L} = \theta_{R} = 0$, then the $w_{L,R}$ become identity matrices and the the random singlet phase of SDRG is recovered.
The reason that this parameterization arises is not clear to us, and a proof using analytic methods is beyond the scope of this paper.
In the absence of a proof, the numerical results in the rest of the section are performed using a full optimization, but the simplified update reproduces the results in all test cases.

\begin{figure}
    \includegraphics[width=\columnwidth]{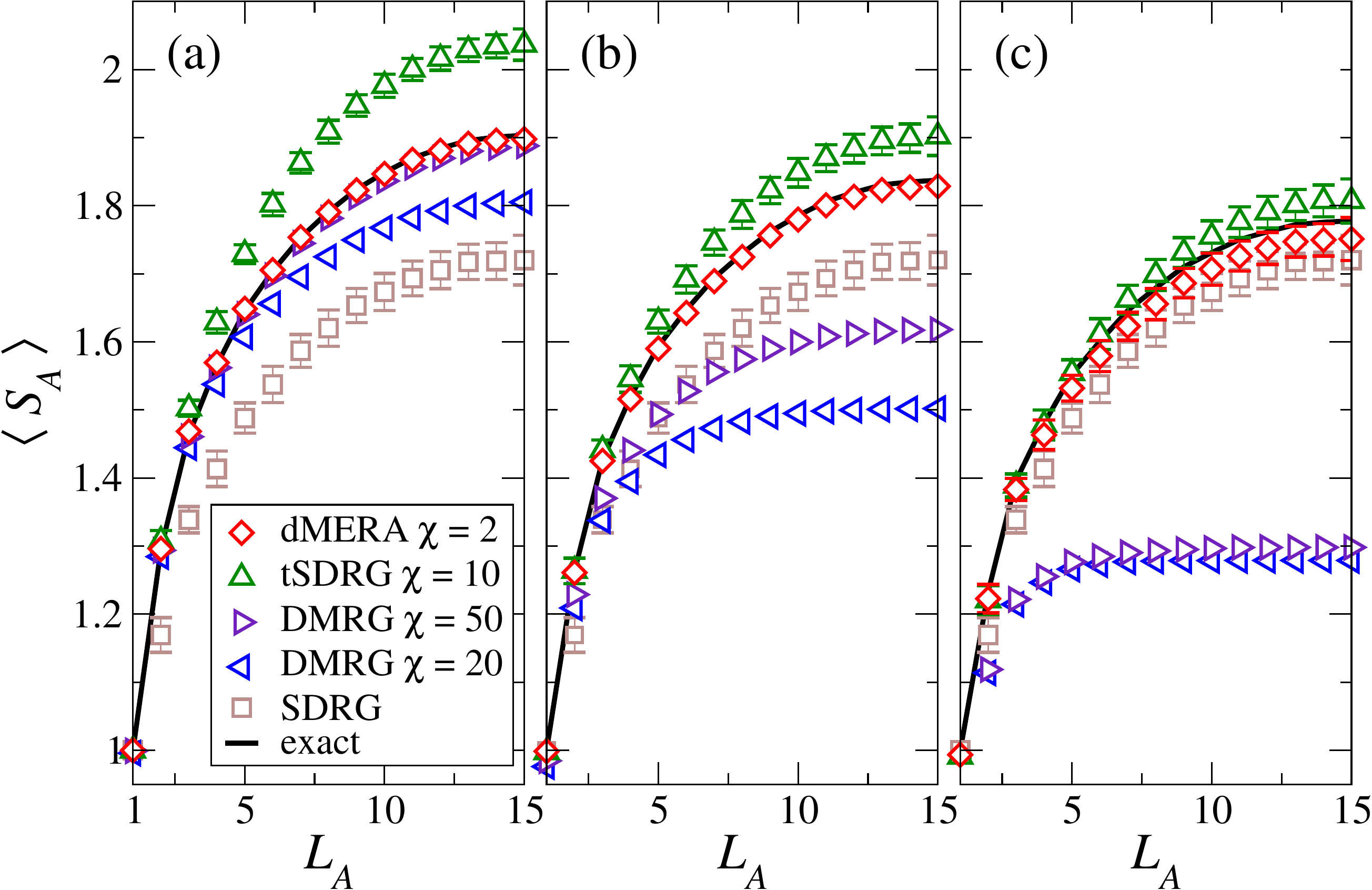}
    \caption{
        Average entanglement entropy $\langle S_{A} \rangle$ for the disordered XX model with
        (a) $\Delta_{J}$ = 1,
        (b) $\Delta_{J}$ = 2, and
        (c) $\Delta_{J}$ = 4.
        The system size is 30 sites and the results are averaged over all possible subsystems of size $L_{A}$ for 50 disorder configurations.
        Errors are within symbol size unless plotted.
    \label{fig:ee_30_XX}
    }
\end{figure}

Results comparing how the average ground state entanglement entropy $\langle S_{A} \rangle$ scales with block size $L_{A}$ for disorder strengths $\Delta_{J} = 1, 2, 4$ are presented in Figure \ref{fig:ee_30_XX}. Note that we average both over all possible blocks of size $L_{A}$ and 50 different disorder configurations. The results demonstrate that dMERA, even when restricted to bond dimension $\chi=2$, reproduces the ground state entanglement accurately over the range of disorder strengths considered. In comparison the tSDRG algorithm has a tendency to over estimate the entanglement entropy, particularly when the disorder is small, yet still reproduces the expected logarithmic scaling. 
For small disorder, $\Delta_{J} = 1$, we see that DMRG requires bond dimension $\chi = 50$ in order to rival the accuracy of the $\chi=2$ dMERA, while for stronger disorder DMRG is substantially worse and the entanglement quickly saturates to a constant even for large bond dimension $\chi$. 
Notice also that the results from standard SDRG become more accurate as the disorder increases, which is to be expected as the approach becomes exact in the limit of infinite disorder \cite{Fis94}.

\begin{figure}
    \includegraphics[width=\columnwidth]{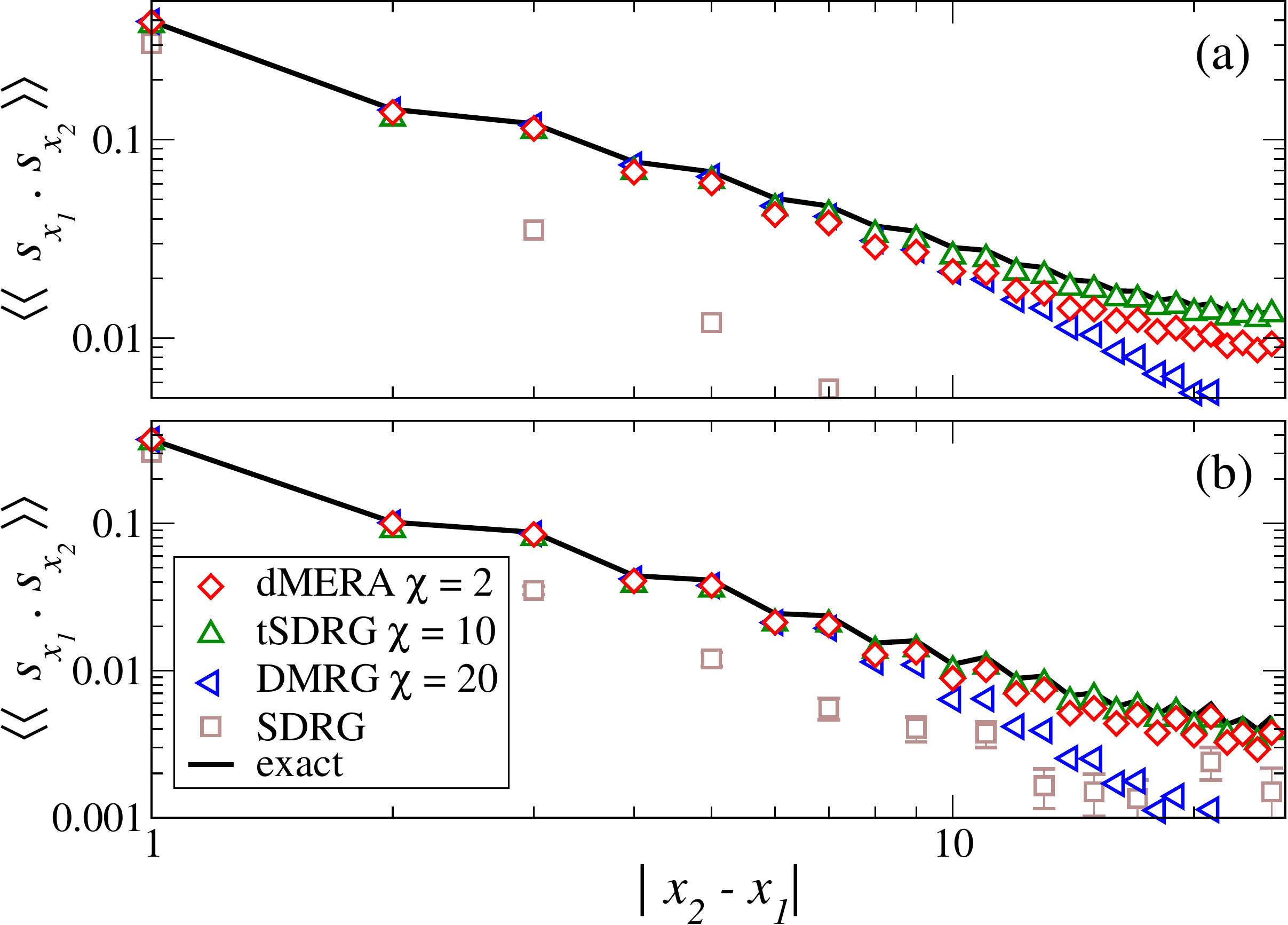}
    \caption{
        Staggered spin correlation function for the disordered XX model with
        (a) $\Delta_{J}$ = 1, and
        (b) $\Delta_{J}$ = 2.
        The system size is 50 sites and the results are averaged over all possible pairs of sites $x_{1}$ and $x_{2}$ for 100 disorder configurations. 
    \label{fig:corr_50_XX}
    }
\end{figure}

The staggered two-point correlation functions $\langle \langle \vec{s}_{x_{1}} . \vec{s}_{x_{2}} \rangle \rangle$, computed from 50 site systems, are plotted in Fig.\ \ref{fig:corr_50_XX}. These have also been averaged both over all positions and over 100 disorder configurations. For length scales smaller than half the system size, the exact results demonstrate the characteristic power law decay of the random singlet phase, with an exponent that depends on the disorder strength $\Delta_{J}$. While the $\chi=2$ dMERA does reproduce this power law decay, it tends to underestimate the correlation especially for weak disorder $\Delta_{J}=1$. However, the results are still significantly more accurate than those of the standard SDRG approach or from $\chi=20$ DMRG, where the correlations decay exponentially at larger distances.

That the $\chi=2$ dMERA tends to underestimate correlators can be understood by analysing individual disorder realizations, where it can be seen that the problem occurs when the maximum coupling $J_{i}$ is not significantly stronger than one of its neighbours, $J_{i-1}$ or $J_{i+1}$. In such instances the SDRG that is used to determine the structure of the dMERA is also less accurate; although optimization of the tensors allows the wavefunction to be improved the correlations still end up systematically smaller than they should be at larger separations. In disorder instances where the maximum coupling is much greater than its neighbours ($J_{i} \gg J_{i-1}, J_{i+1}$), which is more likely to occur at stronger disorder $\Delta_{J}$, we find that the correlation functions are faithfully reproduced by the $\chi = 2$ dMERA.

\subsection{The disordered Heisenberg model}

We now address the disordered Heisenberg model, which has $\Delta_z = 1$ in Eq.\ (\ref{eqn:Heisenberg}). This model no longer corresponds to free fermions and thus cannot be diagonalized exactly as with the XX model. Also, a dMERA of bond dimension $\chi = 2$ is no longer viable as the set of unitary gates that preserve the $SU(2)$ symmetry would be trivial; thus we increase the bond dimension to $\chi = 4, 8$. 


\begin{figure}
    \includegraphics[width=\columnwidth]{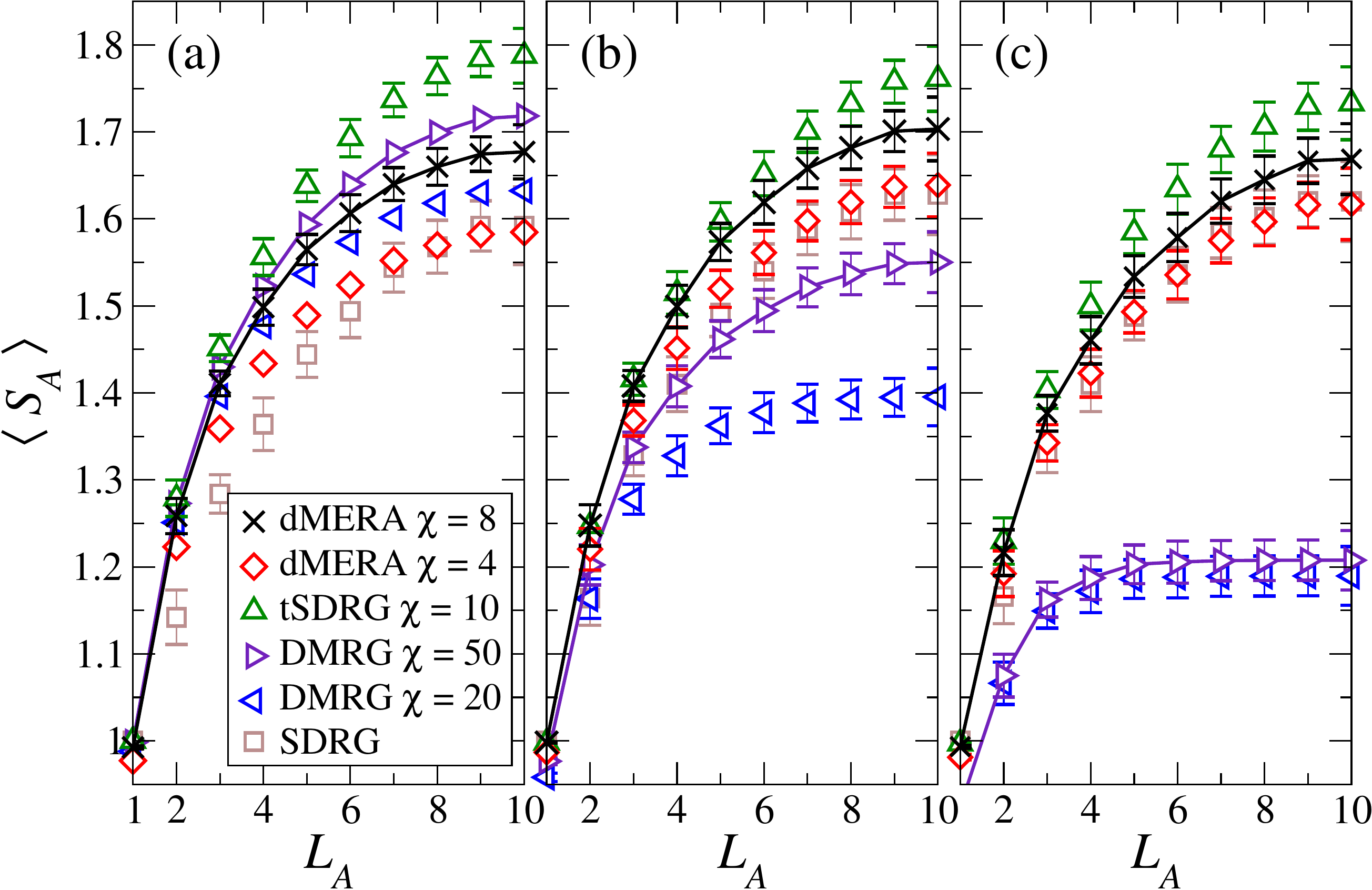}
    \caption{
        Average entanglement entropy for the disordered Heisenberg model with
        (a) $\Delta_{J}$ = 1,
        (b) $\Delta_{J}$ = 2, and
        (c) $\Delta_{J}$ = 4.
        The system size is 20 sites and the results are averaged over all possible subsystems of size $L_{A}$ for 50 disorder configurations. 
        Lines are shown as a guide to the eye for the highest $\chi$ results of dMERA and DMRG.
    \label{fig:ee_20_XXX}
    }
\end{figure}

Figure \ref{fig:ee_20_XXX} shows the average entanglement $\langle S_{A} \rangle$ for a 20 site system when averaged over 50 disorder realizations.
One can infer a rough upper bound to the expected results from tSDRG, which tends to overestimate the entanglement, and a lower bound from SDRG, which tends to underestimate the entanglement. It appears that the approaches based on energy minimization, DMRG and the dMERA, underestimate the entanglement for small bond dimension, approaching from below as the bond dimension is increased. For small disorder, $\Delta_{J} = 1$ as depicted in Fig.\ \ref{fig:ee_20_XXX}(a), the dMERA with $\chi = 8$ and $4$ appears to be slightly less accurate than DMRG with $\chi = 50$ and $20$ respectively. However in the cases with stronger disorder, $\Delta_{J} = 2,4$ as depicted in Fig.\ \ref{fig:ee_20_XXX}(b-c), the dMERA clearly reproduces the entanglement more accurately than DMRG, where the entanglement saturates to a constant for larger blocks. As expected, SDRG also becomes more accurate with stronger disorder, almost matching the entanglement from a $\chi=4$ dMERA, although still falling short of the $\chi=8$ case.

\begin{figure}
    \includegraphics[width=\columnwidth]{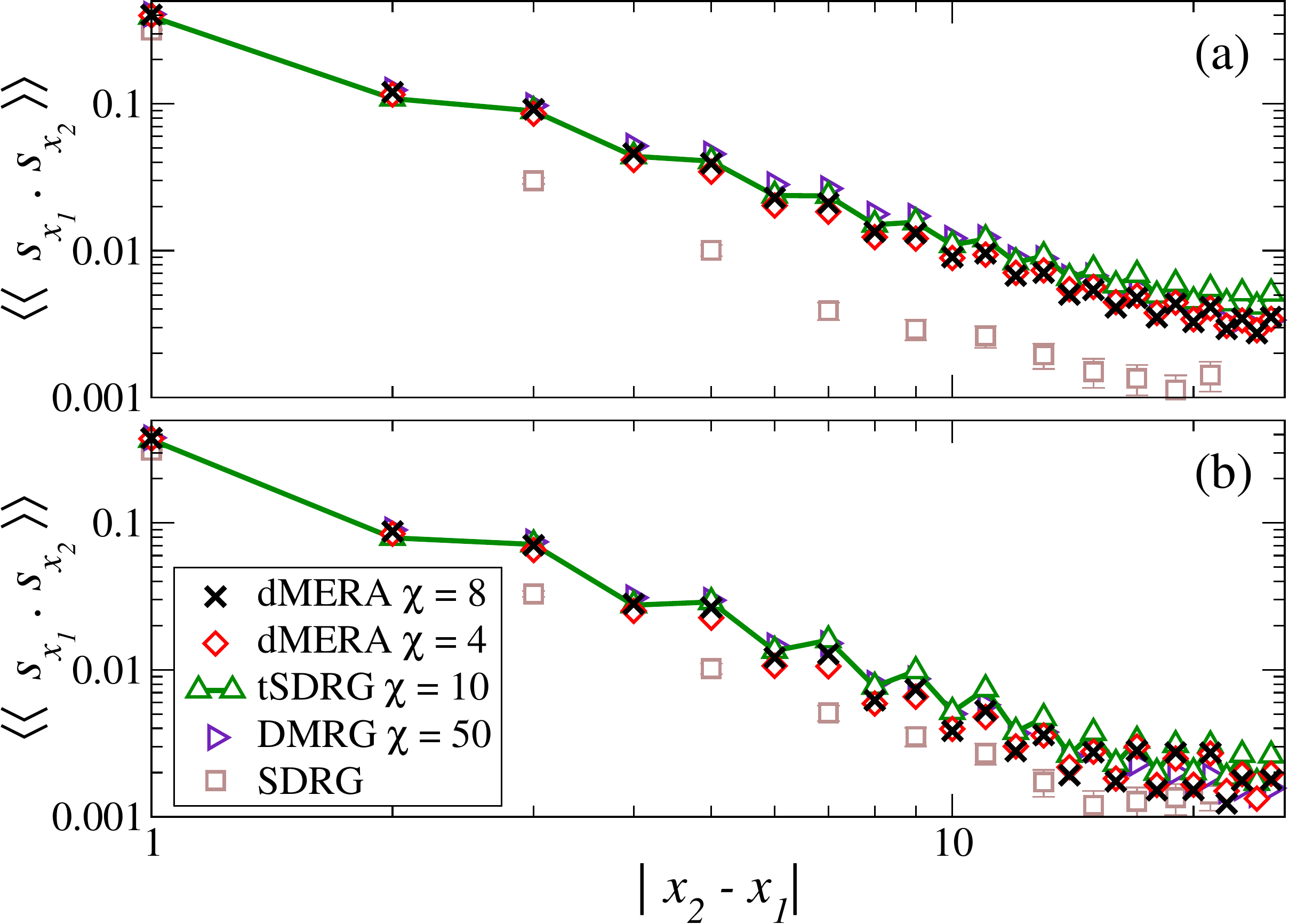}
    \caption{
        Staggered spin correlation function Entanglement entropy for the disordered Heisenberg model with
        (a) $\Delta_{J}$ = 1, and
        (b) $\Delta_{J}$ = 2.
        The system size is 50 sites and the results are averaged over all possible pairs of sites $x_{1}$ and $x_{2}$ for 200 disorder configurations.
        The line is a guide to the eye for $\chi=10$ tSDRG.
    \label{fig:corr_50_XXX}
    }
\end{figure}

The two-point spin correlation functions are plotted in Fig.\ \ref{fig:corr_50_XXX} for a system of 50 sites, averaged over 200 disorder configurations. The dMERA is seen to reproduce the expected algebraic decay of correlations, producing a pattern similar to that from tSDRG but differing in magnitude at longer distances. However, it is not clear which approach is more accurate as both methods seem quite well converged in bond dimension $\chi$.

\section{\label{sec:summary}Summary and conclusions}

We have proposed a generalization of entanglement renormalization for disordered systems, which yields a new class of efficiently contractible tensor network ansatz that we call \emph{disordered MERA}. A key feature of the dMERA is that the geometry of the network can be adjusted to the specific problem under consideration, whereas previous tensor network ansatzes typically use a fixed network geometry only adjust the content of the tensors to the specific problem. This allows much of the entanglement structure of wavefunction to be captured before the variational optimization of the tensors even begins.

Conceptually, the benefit of adjusting the network geometry can easily be understood from an example such as the \emph{concentric singlet phase} also known as the \emph{rainbow state} \cite{VitRL10,RamRS14}. This state, or something close to it, could easily be represented by a dMERA, but would require exponential growth in bond dimension with system size in order to be represented by a  conventional tensor network, such as an MPS or MERA, due to the linear growth of half-chain entanglement entropy.

In practice, the advantage of the this approach was seen in the benchmark results of Sect.\ \ref{sec:XX_results}. Here a dMERA of dimension $\chi=2$ yielded remarkably accurate results for the XX model, significantly improving over DMRG calculations with $\chi=50$. 
The dMERA also seemed to capture the entanglement structure in the disordered Heisenberg model, although a larger bond dimension was required to do so adequately. Accordingly, we expect this method may require further development and refinement before it is viable for more challenging models.

There are many possible extensions and modifications to our proposed approach. In the basic unit cell of the coarse-graining transformation we imposed that the truncated sites were either in a product state, as depicted in Fig.\ \ref{fig:dMERA2}(b), or in a singlet state, as depicted in Fig.\ \ref{fig:dMERA3}(c). In general this could be chosen as any quantum state; for instance a \emph{spin trio} type state could be used to describe the \emph{baryonic} states found in Ref.\ \onlinecite{QuiHM15}. It could also be possible to use a different metric to determine the geometry of a dMERA, rather than through use of the SDRG as was considered here, allowing the approach to be extended beyond states based on random singlet phases. For instance, one could try to use the mutual information to set the geometry, similar to the proposal in Ref.\ \onlinecite{HyaGB17}. Alternatively, it may be possible to design an algorithm that adjusts the network structure automatically during the variational optimization, which remains an interesting avenue for future work. 

\begin{acknowledgments}
    AMG would like to thank Norbert Schuch and Rudolf A. R\"{o}mer for fruitful discussions.
\end{acknowledgments}

\appendix

\section{\label{sec:alg_details}Algorithmic details}
\subsection{\label{sec:ham_shift}The Hamiltonian}
Before the update can begin, the Hamiltonian needs to have its energy \emph{shifted} so that the entire spectrum is negative. 
This is important to allow the update process to target the ground state, the details of which are discussed in sec.\ \ref{sec:update}.
The shift is performed by diagonalizing the two-site Hamiltonian operators and subtracting the identity multiplied by the largest eigenvalue $\lambda_{i,i+1}^{\text{max}}$ from each
\begin{equation}
    h^{\text{shifted}}_{i,i+1} = h_{i,i+1} - \lambda_{i,i+1}^{\text{max}} \left( \openone_{2} \otimes \openone_{2} \right).
\end{equation}
The sum of the shifts are stored and added to the final energy after minimization to obtain the true ground state energy once more.

\subsection{\label{sec:SDRG_order}The SDRG order}
The order in which the coarse-graining blocks are combined to create the full network is given by ordinary SDRG \cite{MaDH79}.
The algorithm coarse-grains in the order of strongest interaction, renormalizing neighbouring interactions via second order perturbation theory.
For more information see \cite{DasM80} and \cite{Fis94}.

The SDRG rules for the XXZ model are as follows:
Find the largest coupling $J_{i}$ in the set of coupling strengths.
The strongest coupled pair form a singlet to first order. 
Second order perturbation theory is then used to find an effective coupling $\tilde{J}$ for the spins either side
    \begin{equation}
        \tilde{J} = \frac{J_{i-1} J_{i+1}}{(1 + \Delta_{z}) J_{i}}.
    \end{equation}
The location $i$ is stored, then the strongest coupled spins are removed, coarse-graining the chain by two sites.
This is repeated until there are no remaining sites.
The result is a list of the locations of the singlets at each level of coarse-graining as shown in Fig.\ \ref{fig:MDH_order}(a), which can then be used to set the geometry of the tensor network.

\subsection{\label{sec:initial}The initial tensors}
The order of coarse-graining and the maximum allowed $\chi$ is used to set the initial state of the tensors.
If $\chi = 2$ the network is in the most simple form; all $u$ and $w$ tensors are $2 \times 2 \times 2 \times 2$ and unitary, and each coarse graining block is of the form of Fig.\ \ref{fig:dMERA3}(c).
Setting these as $\left( \openone_{2} \otimes \openone_{2} \right)$ means that the initial state is the random singlet state.
Another option is to generate random numbers to fill the tensors.

The top of the network takes a slightly different form due to the small system size and periodic boundaries. 
When the system consists of four sites it is only possible to put one coarse-graining block, then the remaining pair of sites will be connected by a singlet.
As shown in Fig.\ \ref{fig:dMERA_top}, due to the PBCs this second singlet can be considered to be part of the same coarse-graining operation.
\begin{figure}
    \includegraphics[scale=0.17]{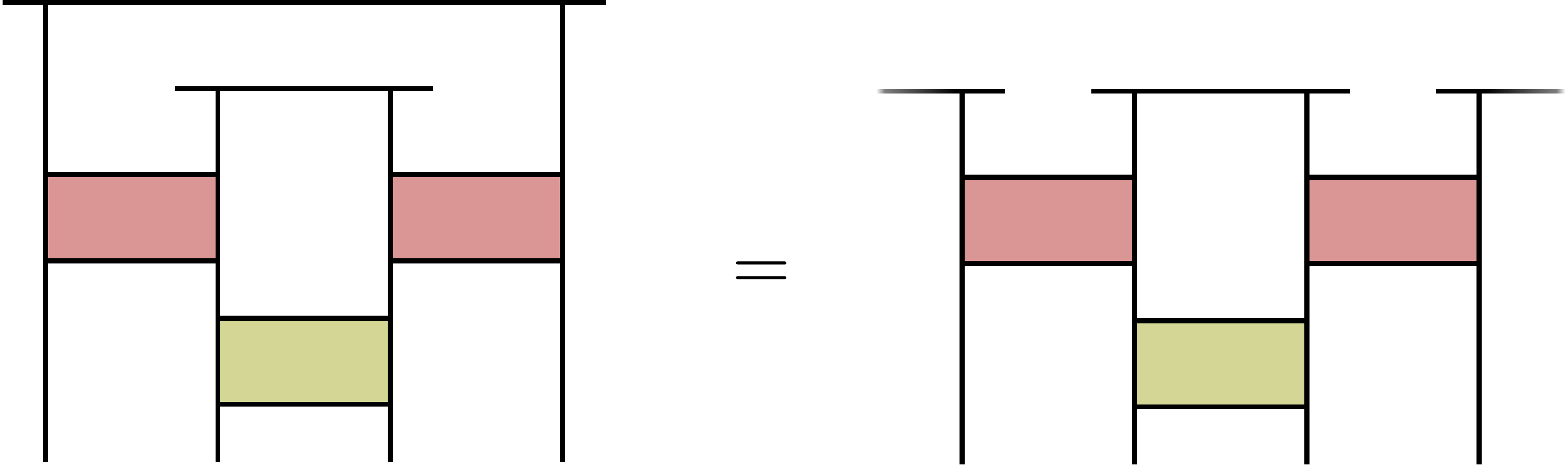} 
    \caption{
        At the top of the network, when the system is just four sites, SDRG gives two singlets.
        Due to periodic boundary conditions the final singlet can be seen as being over or round the coarse-graining block.
        Lines and symbols are as in previous figures.
    \label{fig:dMERA_top}
    }
\end{figure}
The full tensor network for a system of 20 sites with $\chi = 2$ is shown in Fig.\ \ref{fig:dMERA_L20}(c).
Notice that this is the network obtained given the order set out in Fig.\ \ref{fig:MDH_order}(a).

If $\chi > 2$ then the network needs to be altered to allow the bond dimensions to grow.
By looking at Fig.\ \ref{fig:dMERA3}(c) it is clear that if the tensors at the lower levels of the network have $\chi = 2$, then all of the tensors above must do also.
Thus the coarse-graining block of Fig.\ \ref{fig:dMERA2}(b) is used to increase the dimension of the effective sites.
Once the bond dimension is above 2 it can grow using the standard coarse-graining block of Fig.\ \ref{fig:dMERA3}(c).
The dimensions and form of the tensors is set in the initialization stage and will then not change during the algorithm.

There are many choices that can be made relating to how to implement the bond dimension increase.
We choose to only use the $\chi$-increase block [Fig.\ \ref{fig:dMERA2}(b)] when no information is to be discarded, this is because this form has been shown to be unreliable at retaining long range interactions with lower bond dimensions.
The update is performed without the $\rho$'s so that is is just a coarse-graining of the Hamiltonian.
As such, we only update it on the first sweep as the environment will not change as the algorithm progresses. 
For the rest of the blocks the dimensions of the top two legs of the unitary in the $\chi$-increase block are swapped, e.g. if the bottom two legs have dimensions $2$ and $4$ then the top legs have $4$ and $2$.
This is to be consistent with the simplified results of the XX model, which show that the unitary has properties similar to a swap tensor. 
All of these choices have been seen to be more effective as well as being conceptually preferable.

\subsection{\label{sec:desc}The descending superoperators}
\begin{figure}
    \centering
    (a)\includegraphics[scale=0.17]{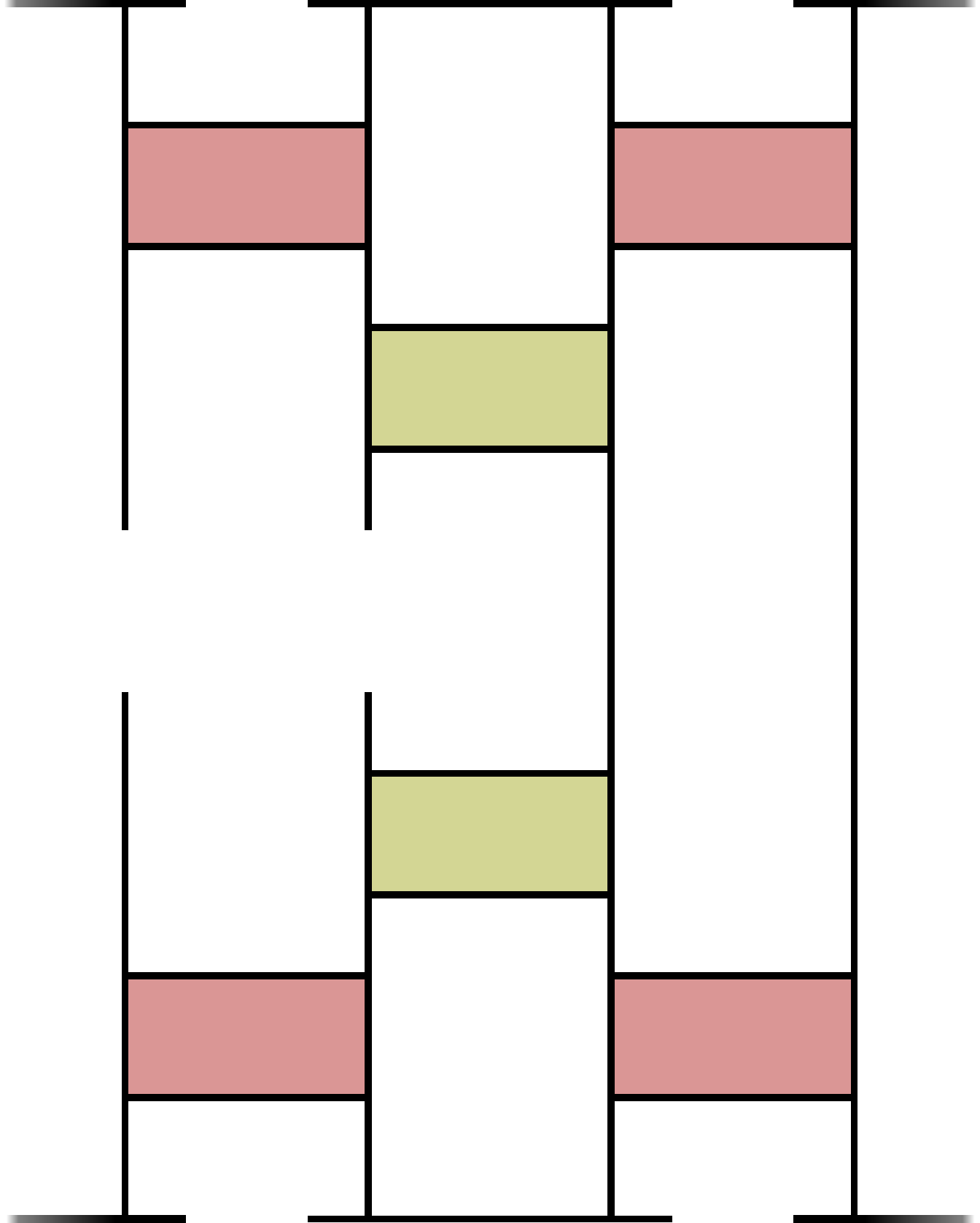} \quad 
    (b)\includegraphics[scale=0.17]{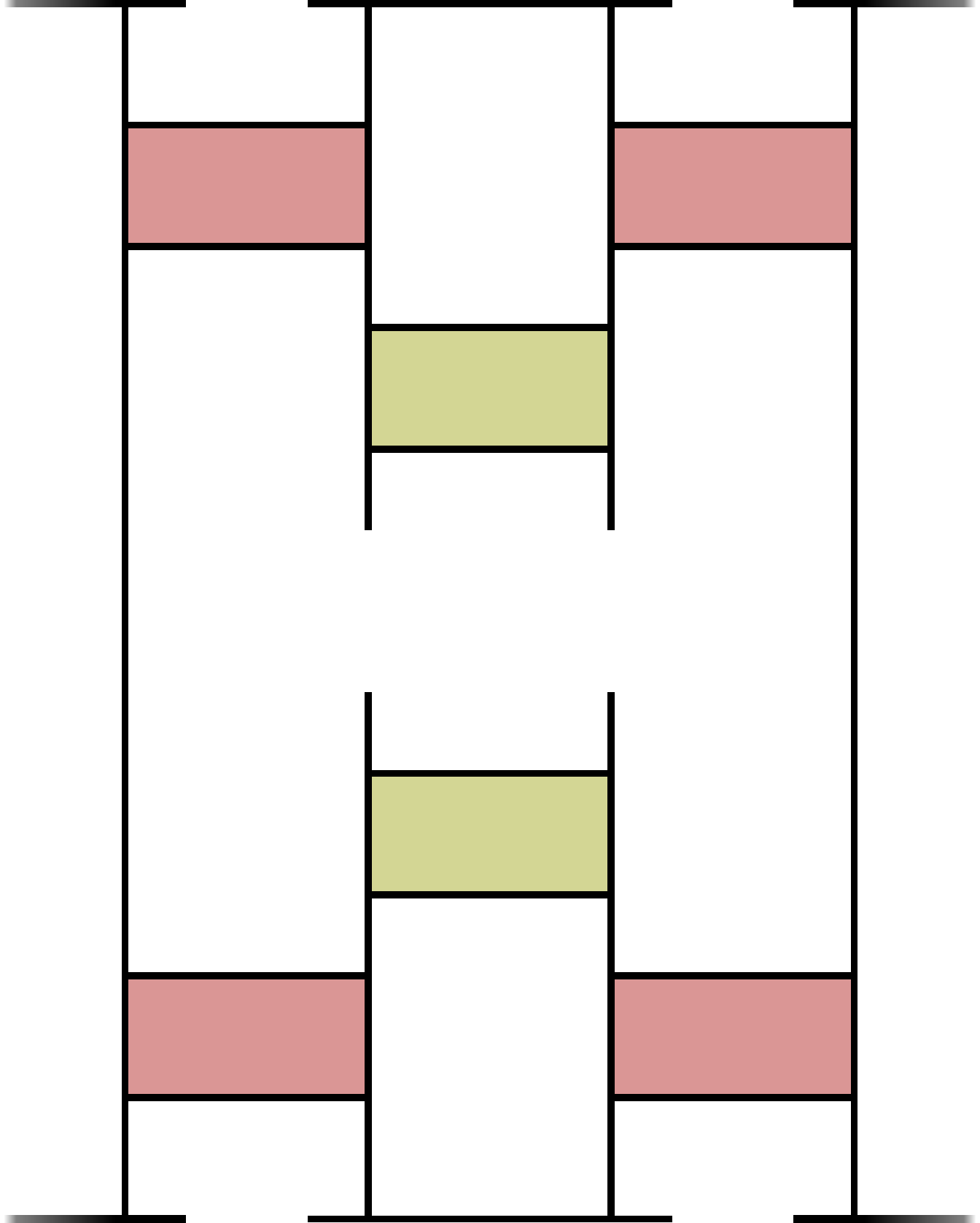} \vspace{0.5cm} \\
    (c)\includegraphics[scale=0.17]{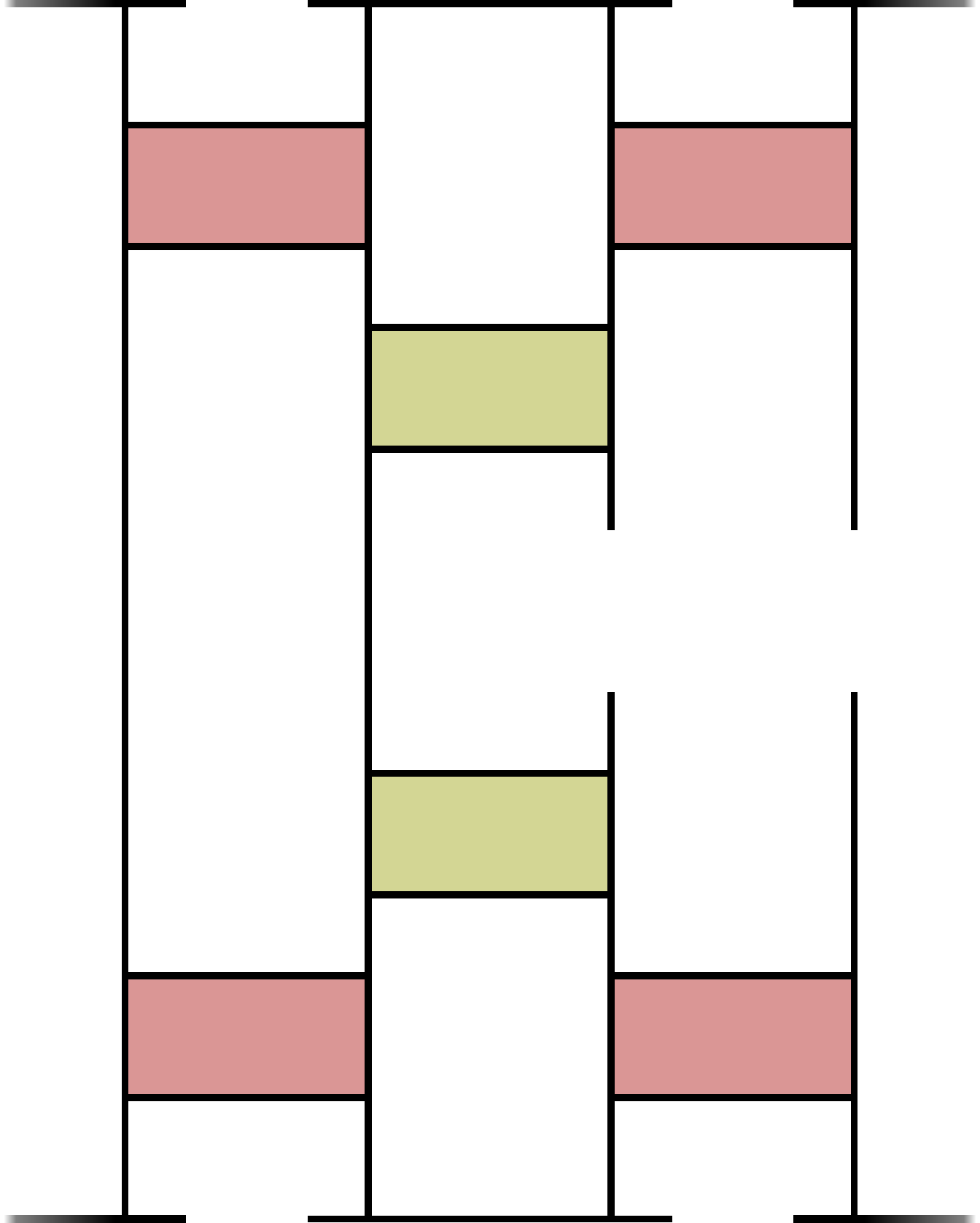} \quad
    (d)\includegraphics[scale=0.17]{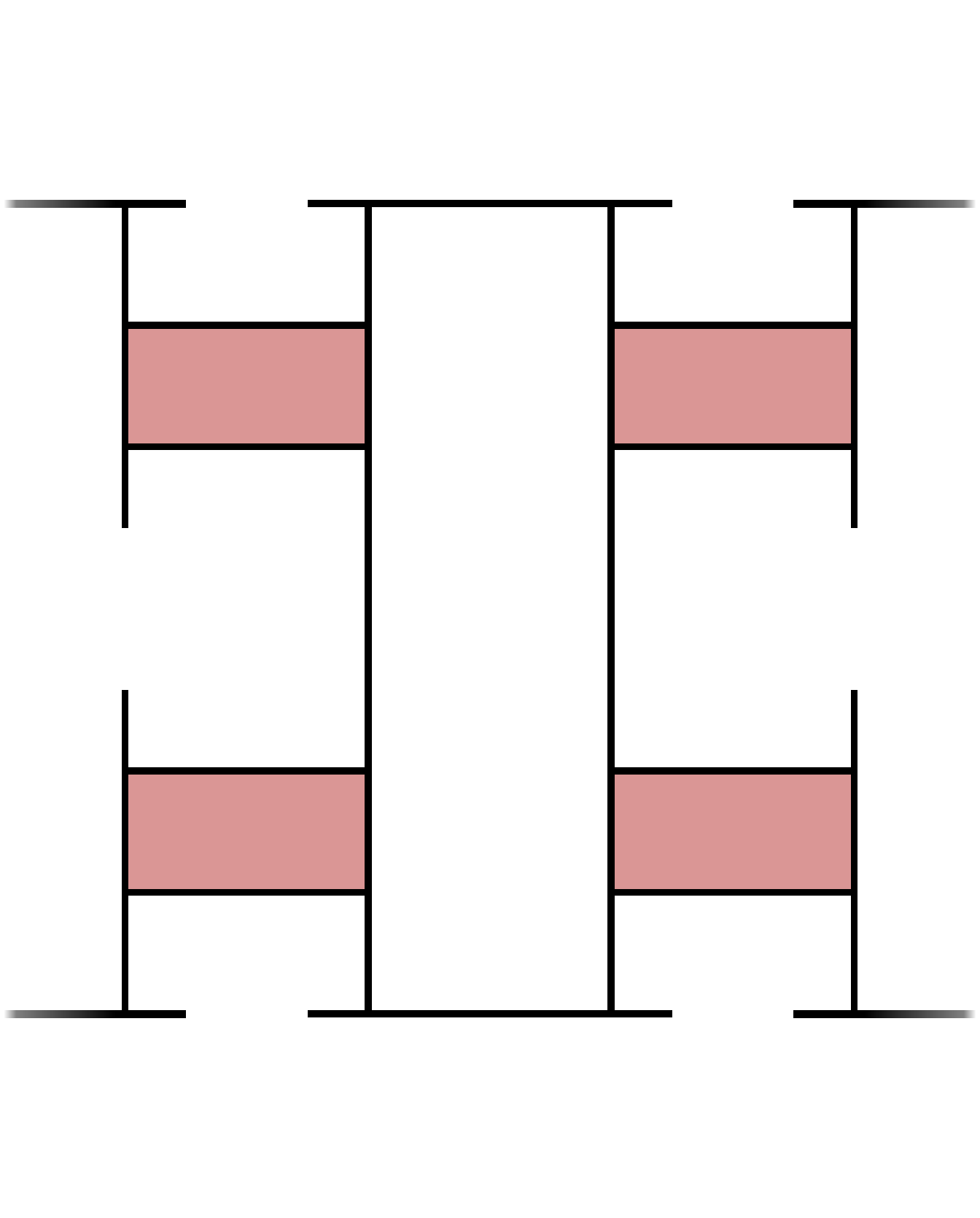}  
    \caption{
        The four reduced density matrices $\rho$ for the top of the network for positions:
        (a) $i-1$, 
        (b) $i$, 
        (c) $i+1$, and
        (d) $i+2$.
        (a - c) have leading order cost $\mathcal{O}(\chi^{6})$, (d) is $\mathcal{O}(\chi^{4})$.
    \label{fig:rho_top1}
    \label{fig:rho_top2}
    \label{fig:rho_top3}
    \label{fig:rho_top4}
    }
\end{figure}
The two-site local reduced density matrices $\{\rho\}$ contain all of the tensors that are \emph{above} the tensor to be updated so that the process minimizes the energy of the network as a whole.
To create the set of $\rho$'s, start at the top and contract down.
The tensors of the coarse-graining block make up the \emph{descending superoperators}, mapping a density matrix on level $\L'$ to one at level $\L$.
The highest level $\rho$'s consist of the four ways of contracting the top block and its conjugate as shown in Fig.\ \ref{fig:rho_top1}, where $i$ is the position of maximum coupling from SDRG when the system is four sites.
Here and in the following diagrams the optimal contraction order is found using the \emph{netcon} function \cite{PfeHV14} and the contraction itself performed using \emph{ncon} \cite{PfeESV14}.

For the next level down there are six sites on the lattice and the $\rho$'s are created by contracting the coarse-graining block to the correct $\rho$ at the level above (labelled as $\rho^{\prime}$ in the figures).
There are three sites ($i-1$, $i$ and $i+1$) that are below $\rho^{\prime}_{i-1}$ shown in Figs.\ \ref{fig:rho_m1}(b,c,d).
The sites either side of the main coarse-graining block, $\rho_{i-2}$ and $\rho_{i+2}$ are below $\rho^{\prime}_{i-2}$ and $\rho^{\prime}_{i}$ respectively as shown in Figs.\ \ref{fig:rho_l}(a,e).
Any sites that are not affected by the coarse-graining block have $\rho$'s copied down from the level above. 
This is repeated until the bottom of the network creating the necessary $\rho$'s for the updates of all of the tensors.
\begin{figure}
    (a)\includegraphics[scale=0.15]{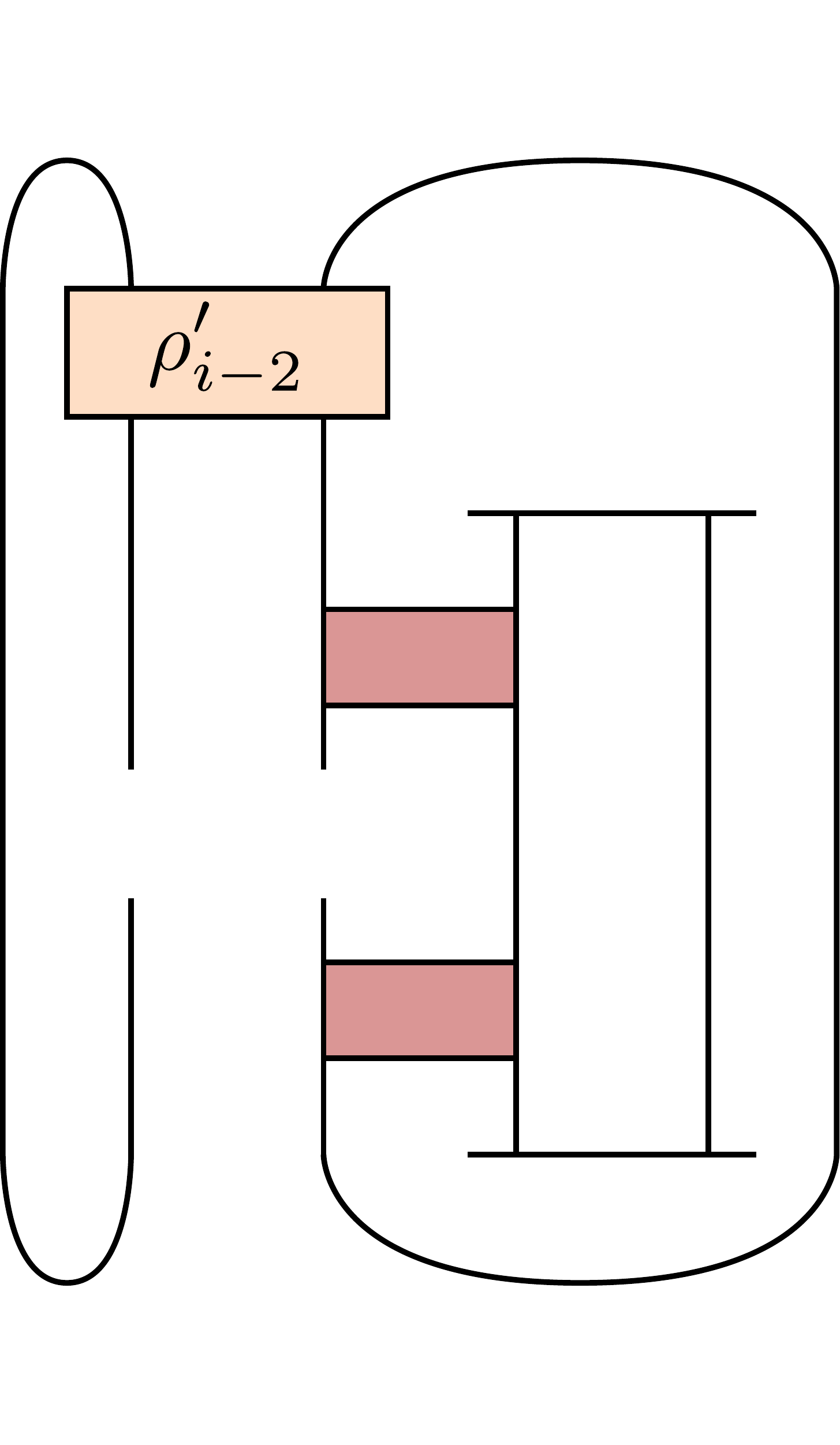}  
    (b)\includegraphics[scale=0.15]{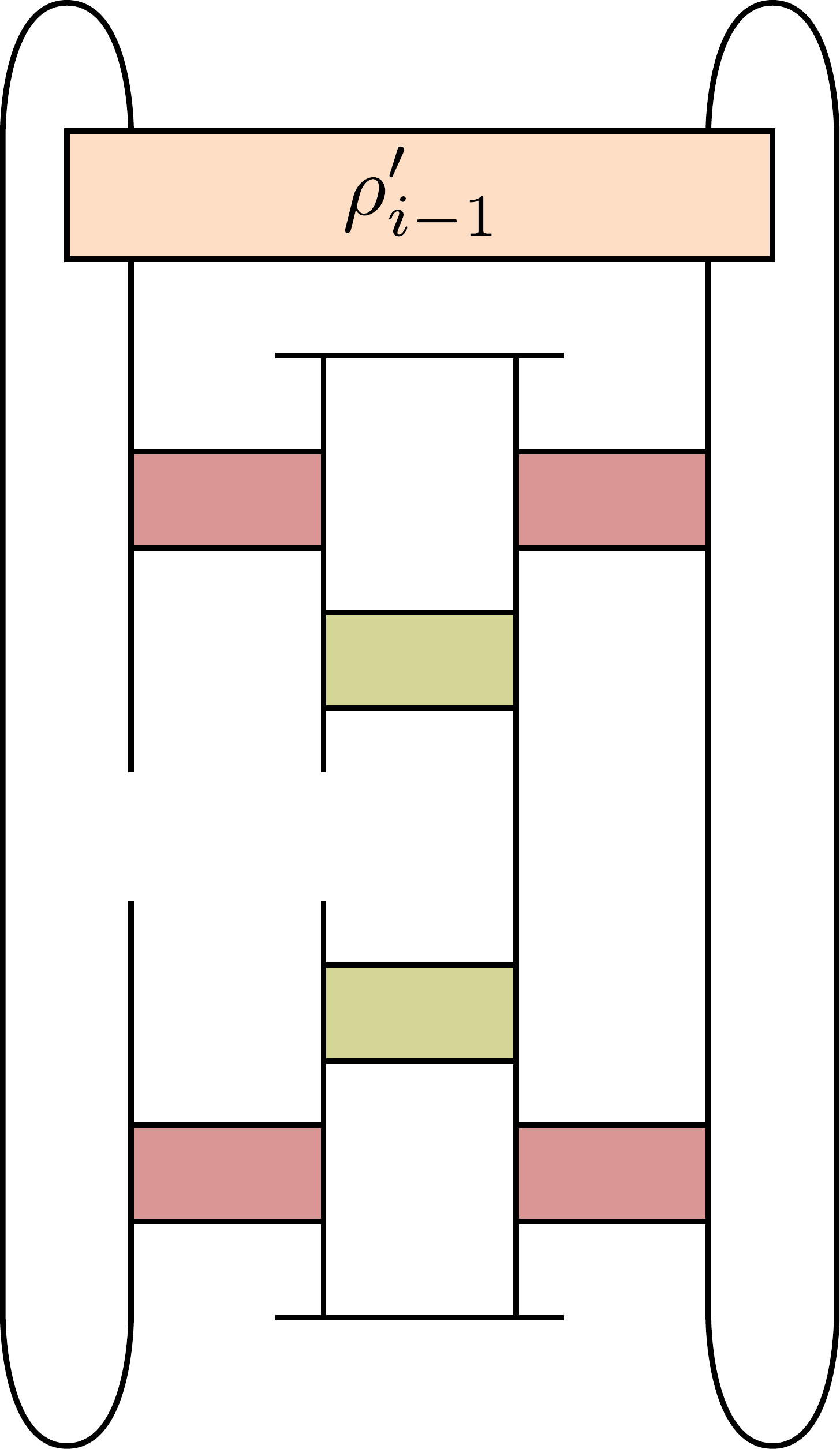} 
    (c)\includegraphics[scale=0.15]{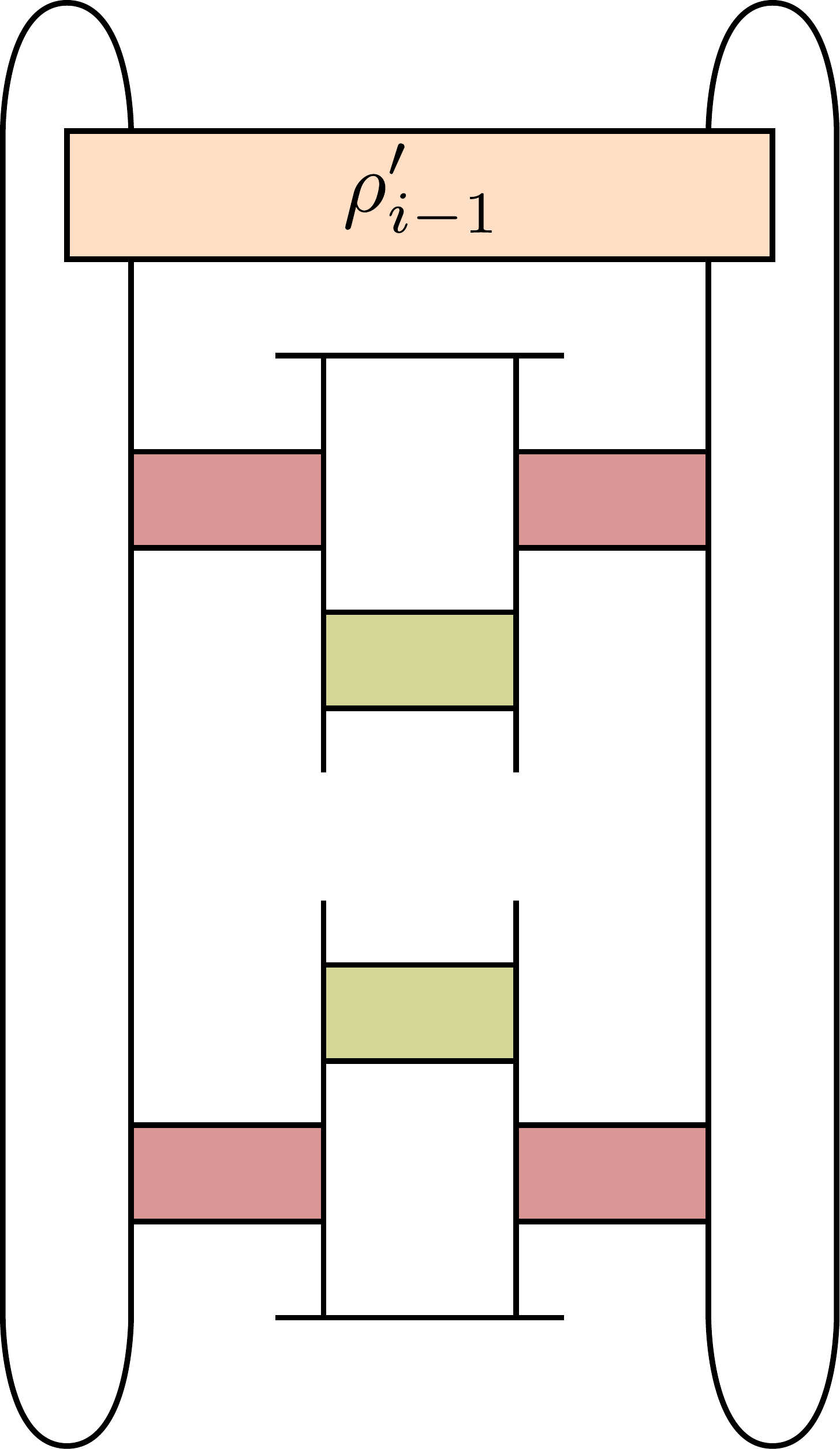} \vspace{0.2cm} \\ 
    (d)\includegraphics[scale=0.15]{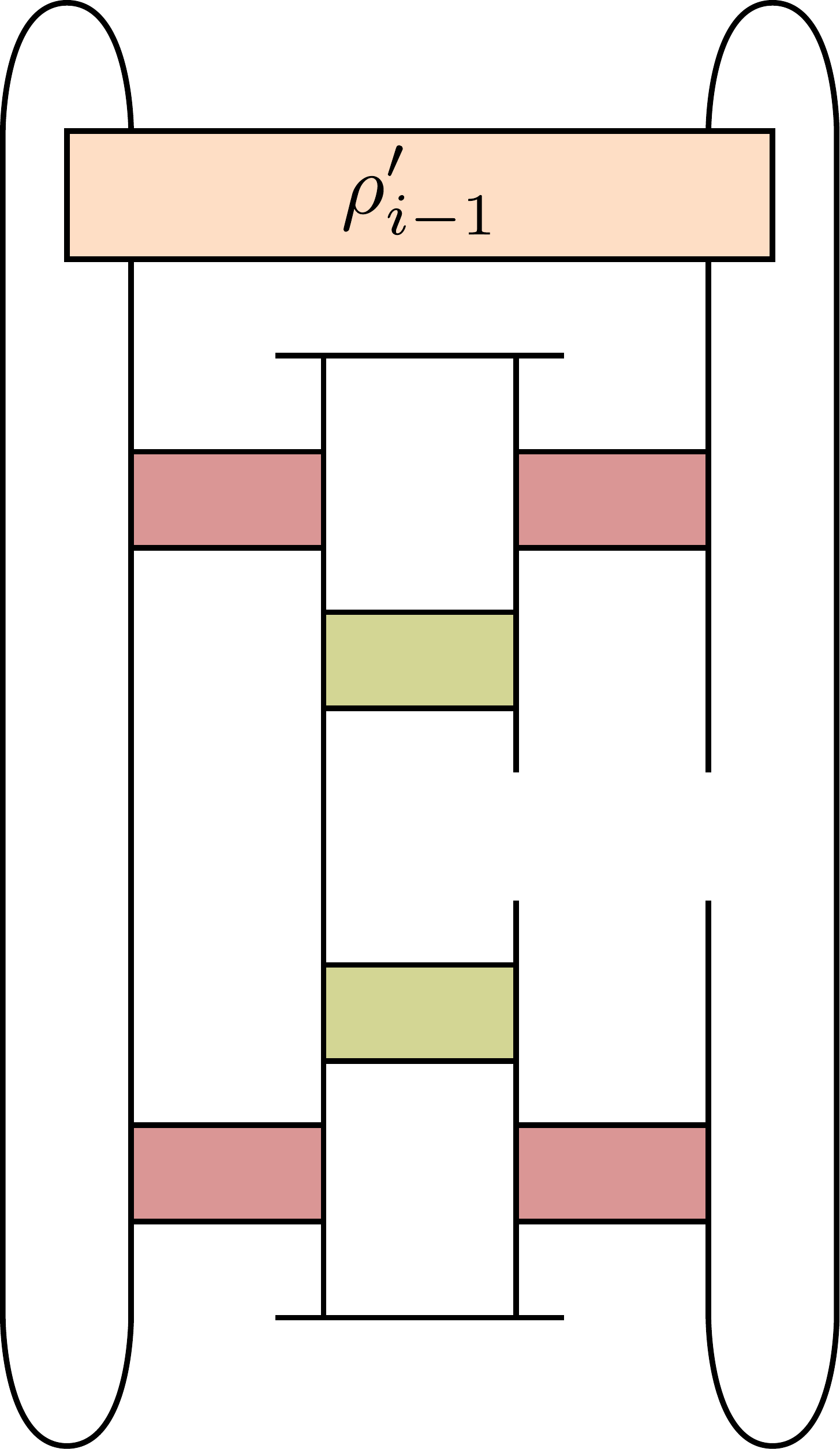} \quad
    (e)\includegraphics[scale=0.15]{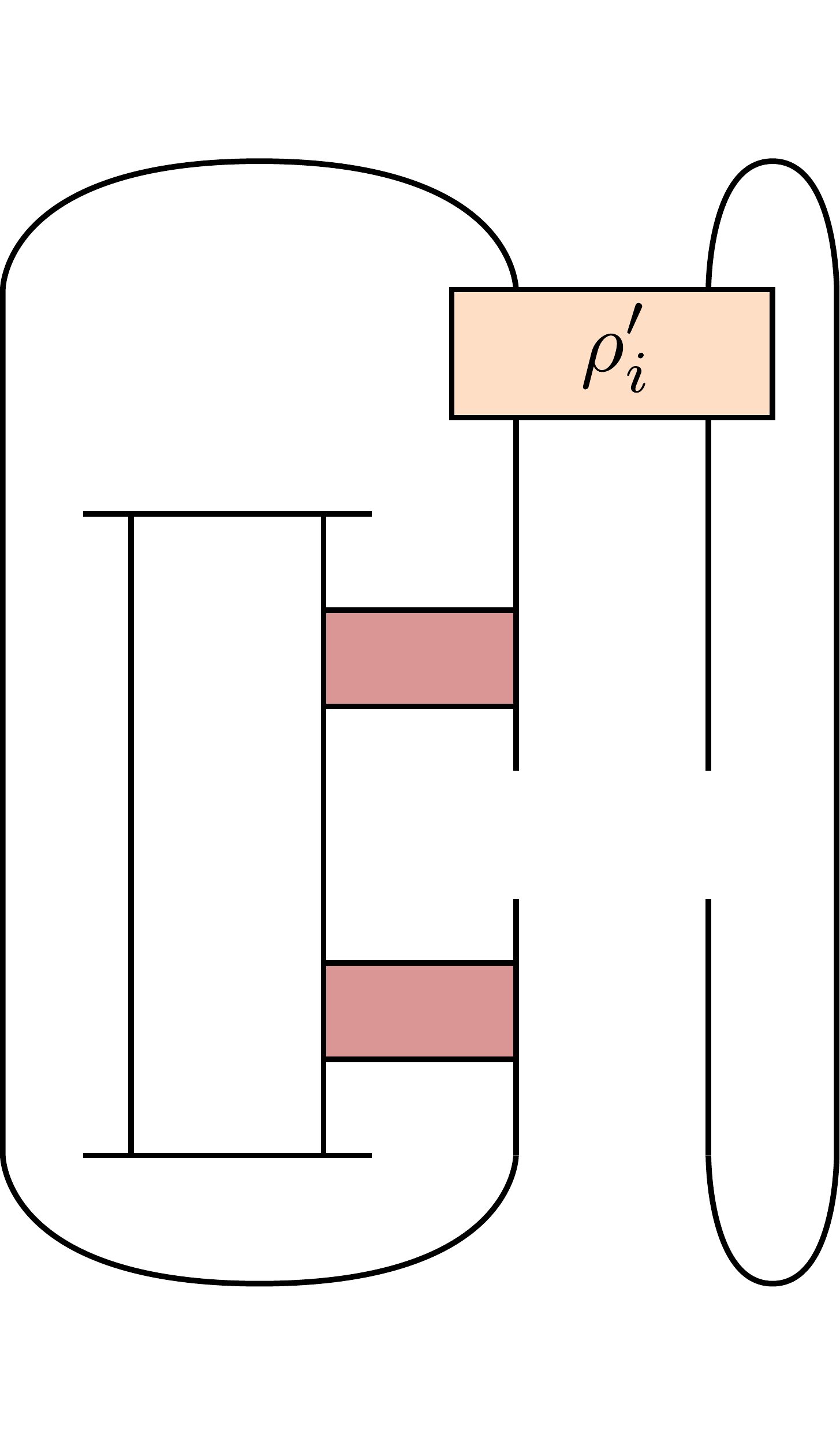}  
        \caption{
            Reduced density matrices $\rho$ for positions 
            (a) $i-2$, 
            (b) $i-1$, 
            (c) $i$,
            (d) $i+1$, and 
            (e) $i+2$,
            where $i$ is the position of maximum $J$ at that level.
            The peach coloured boxes at the top are the reduced density matrices from the level above $\rho^{\prime}$, otherwise boxes and lines are as in previous figures.
            The cost of (a, c, e) is $\mathcal{O}(\chi^{6})$, and (b, d) is $\mathcal{O}(\chi^{7})$ to leading order.
        \label{fig:rho_l}
        \label{fig:rho_m1}
        \label{fig:rho_m2}
        \label{fig:rho_m3}
        \label{fig:rho_r}
    }
\end{figure}

\subsection{\label{sec:update}The update}
The process is much the same as for standard MERA \cite{EveV09}.
The update of the unitary and isometries is \emph{quadratic} in the sense that we want to find, for example, the optimal $u$ and $u^{\dagger}$ at the same time.
Unfortunately there is no exact way to perform a general quadratic update, thus the problem is \emph{linearized}.
Linearization here means setting $u^{\dagger}$ as the hermitian conjugate of the \emph{current} $u$, so the problem now is just to update $u$ on its own.
Once the new tensor has been found the $u^{\dagger}$ is replaced with the new $u$ and the process is repeated until convergence.

The update is the process of finding the tensor, here $u$, that minimizes the energy of the network as a whole.
We call the rest of the network the \emph{environment} of $u$, $\Upsilon_{u}$.
The fact that the Hamiltonian tensors contain all of the tensors below the coarse-graining block and the $\rho$'s contain all of the network above makes the contraction of $\Upsilon_{u}$ simply the sum of the three diagrams containing the Hamiltonian components below the $u$, as shown in Fig.\ \ref{fig:env_u}.
Notice that the linearization is accomplished by contracting $u^{\dagger}$ into the environment.
\begin{figure}
    \includegraphics[width=\columnwidth]{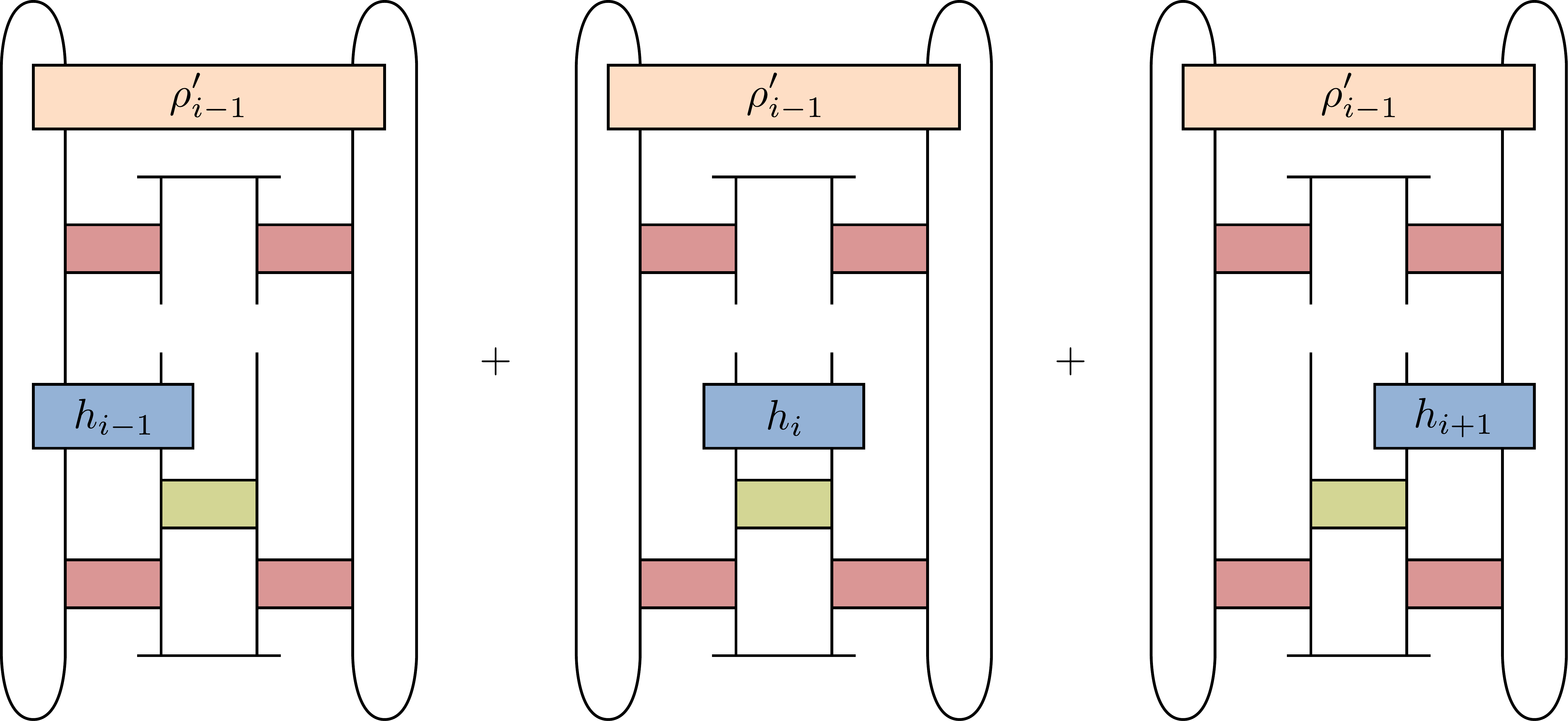} 
    \caption{
        Tensor network diagram showing contributions to the environment $\Upsilon_{u}$ of $u$, cost is $\mathcal{O}(\chi^{7})$ to leading order.   
    \label{fig:env_u}
    }
\end{figure}
Then the optimization of $u$ can be stated as
\begin{equation}
    E_{\text{min}} = \text{min}_{u \in \mathcal{U}} \text{Tr} \left[ u \Upsilon_{u} \right],
\end{equation}
where $\mathcal{U}$ is the set of unitary tensors.
Because the Hamiltonian is shifted, as mentioned in section \ref{sec:ham_shift}, the minimum energy is the negative trace of the singular values of $\Upsilon_{u}$ \cite{Vid08}.
Thus $u = -VU^{\dagger}$ where $U$ and $V$ are found by performing a singular value decomposition (SVD) of the environment ($\Upsilon_{u} = USV^{\dagger}$)
This is shown in Fig.\ \ref{fig:u_update1}, where it is clear that this choice of $u$ results in the energy being the negative trace of $S$.
\begin{figure}
    (a)\includegraphics[scale=0.2]{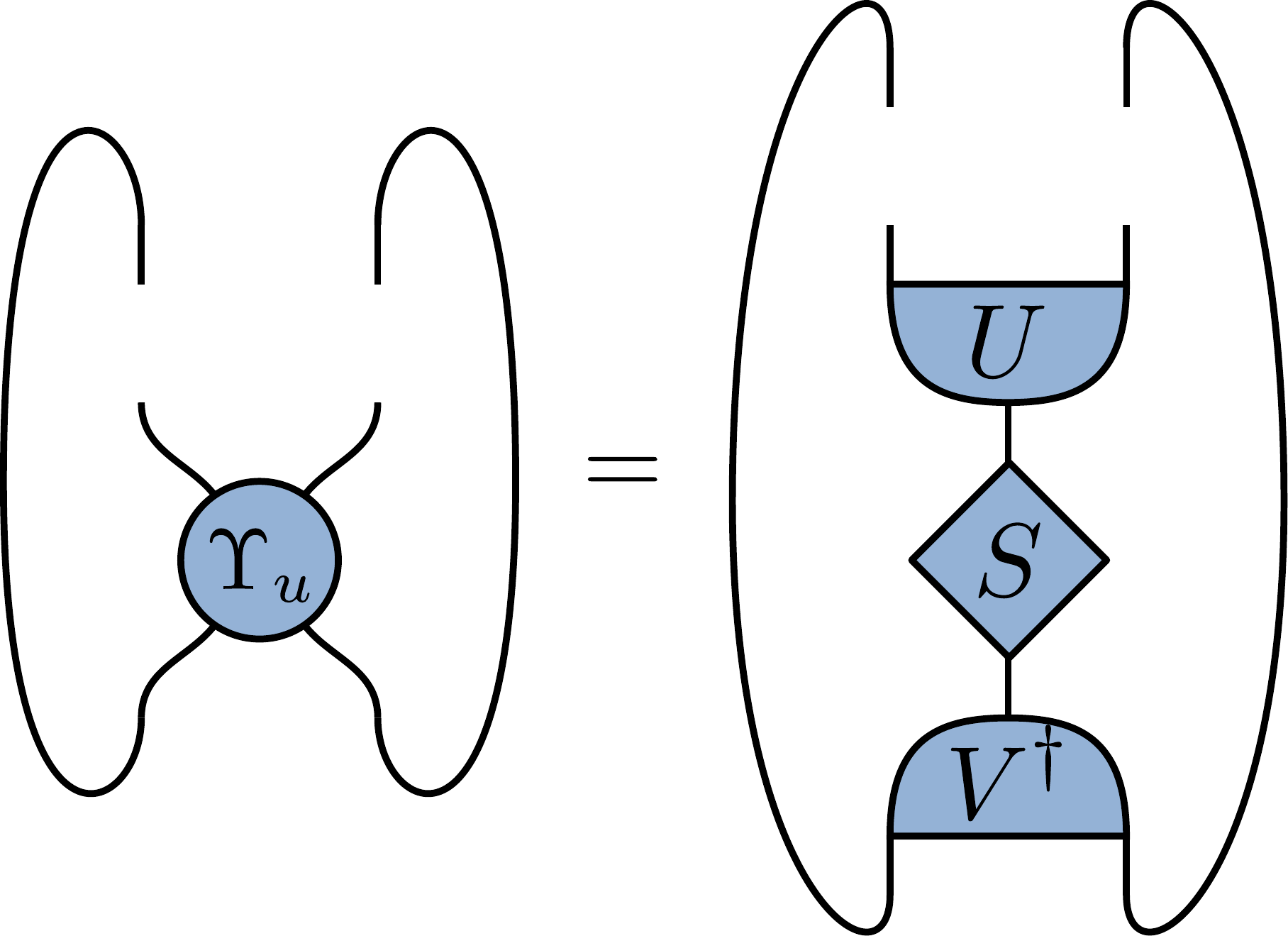} \: 
    (b)\includegraphics[scale=0.2]{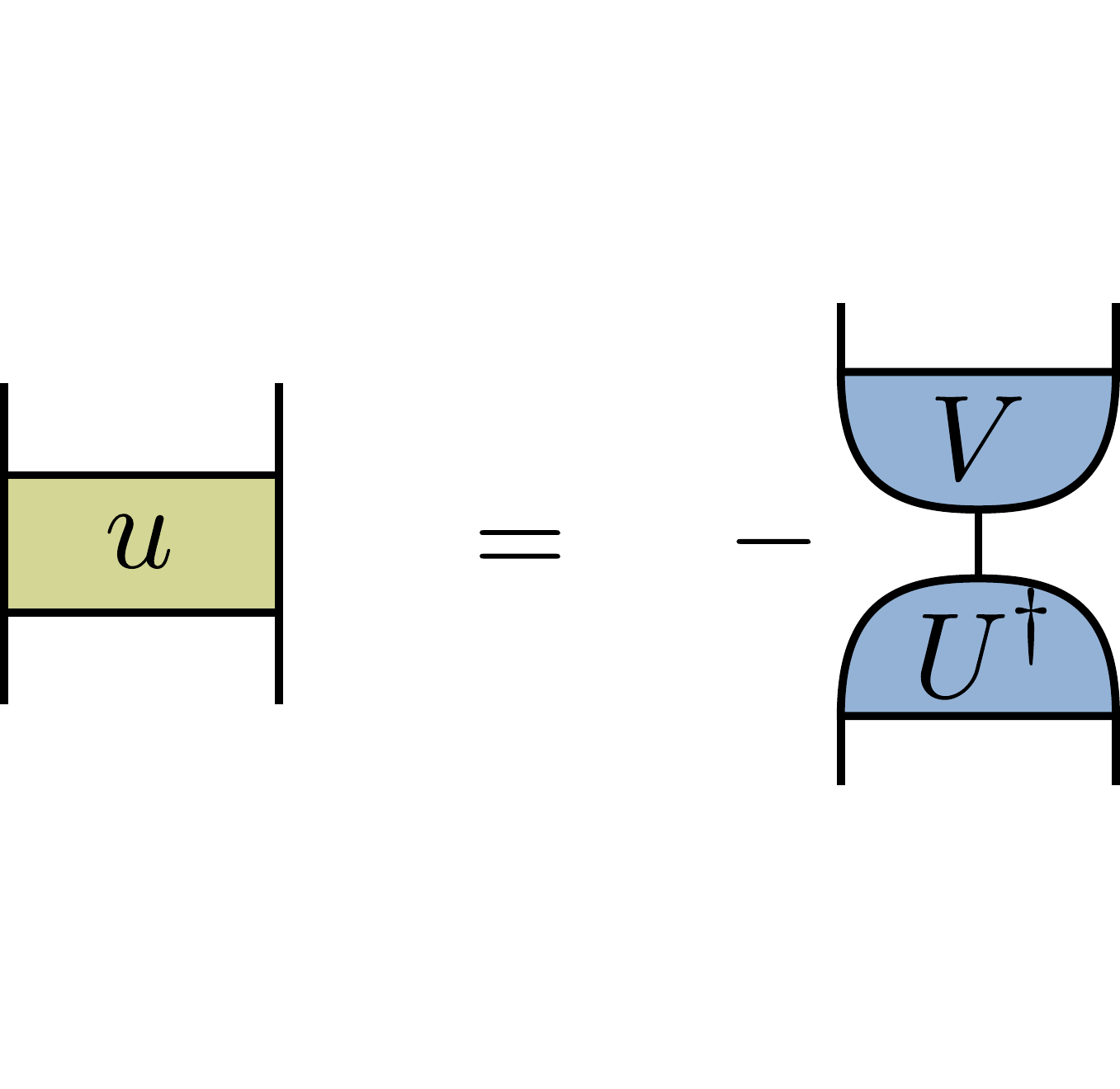}
    \caption{
        Tensor network diagrams showing 
        (a) the SVD of the environment of $u$ and 
        (b) the choice of $u = -VU^{\dagger}$ that makes the energy the negative trace over the singular values $S$.
    \label{fig:u_update1}
    \label{fig:u_update2}}
\end{figure}
The environment is then re-contracted with the new $u^{\dagger}$ and the process is repeated until convergence.
The update of the isometries is effectively the same but with different environment diagrams to be contracted.
The environment of $w_{L}$ is given in Fig.\ \ref{fig:env_wL} and the environment of $w_{R}$ constructed similarly.
The main difference to that of $u$ is the $i - 2$ term, which does not contribute to the environment of $u$.
\begin{figure}
    \includegraphics[scale=0.15]{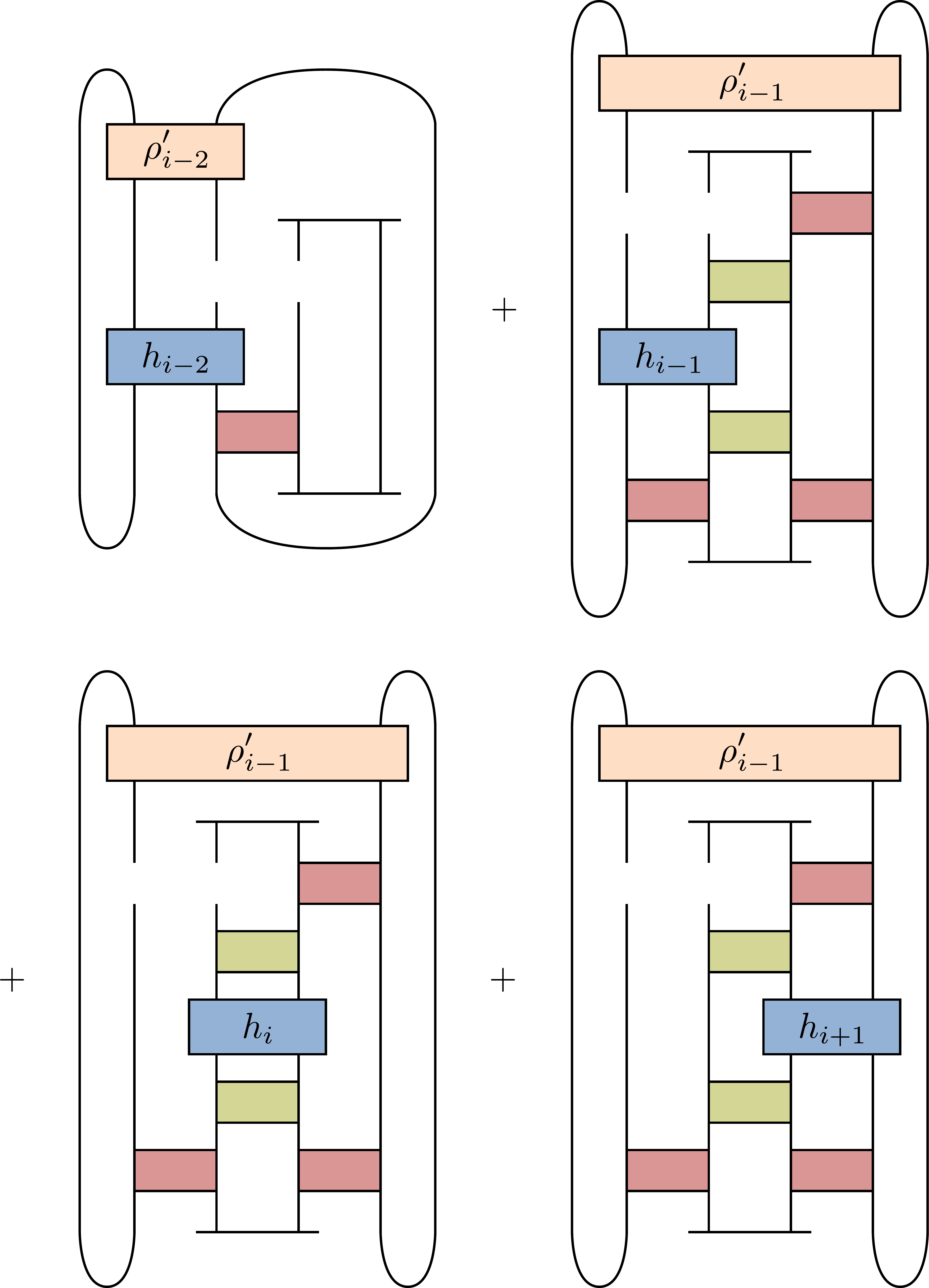}
    \caption{
        Tensor network diagram showing contributions to the environment $\Upsilon_{w_{L}}$ of left isometry $w_{L}$. 
        Cost is $\mathcal{O}(\chi^{7})$ to leading order.
    \label{fig:env_wL}}
\end{figure}

The top of the network is again slightly different as there are no $\rho$'s, just the singlets that make up the top, as shown in Fig.\ \ref{fig:env_u_top}.
This is updated by SVD in the same way as the rest of the tensors.
Diagrams for the environments of $w_{L}$ and $w_{R}$ are omitted but are straightforward to recreate.
\begin{figure}
    \includegraphics[scale=0.17]{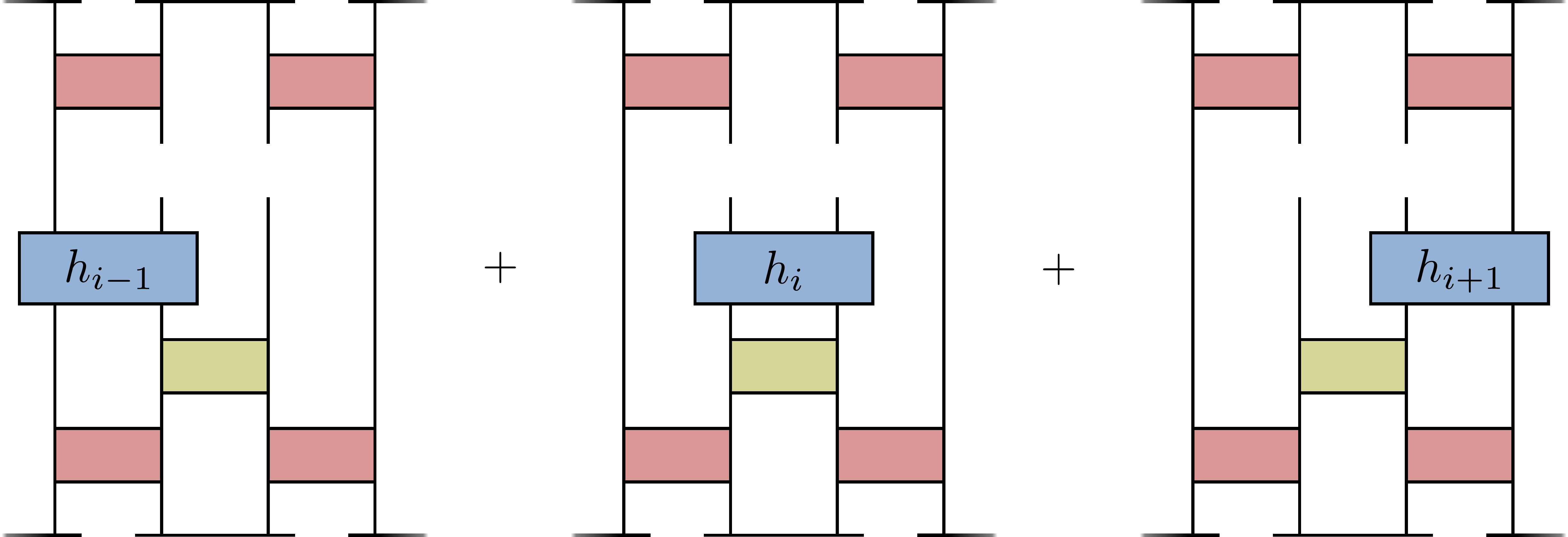}
    \caption{
        Tensor network diagram showing contributions to the environment $\Upsilon_{u}$ of disentangler $u$ at the top of the network. 
        The cost is $\mathcal{O}(\chi^{6})$ to leading order.
    \label{fig:env_u_top}}
\end{figure}

\subsection{\label{sec:raise}The ascending superoperators}
Once the tensors of a block have been updated the Hamiltonian components are \emph{raised} to the next level of coarse-graining using \emph{ascending superoperators} \cite{EveV09}.
This is accomplished by applying the block to each Hamiltonian tensor individually.
Because the centre three components get mapped onto the same site of the coarse-grained lattice, the new tensor for that site is the sum of these three terms.
The tensor network diagrams for this raising operation are shown in Fig.\ \ref{fig:ham_m}.

\begin{figure}
    (a)\includegraphics[scale=0.18]{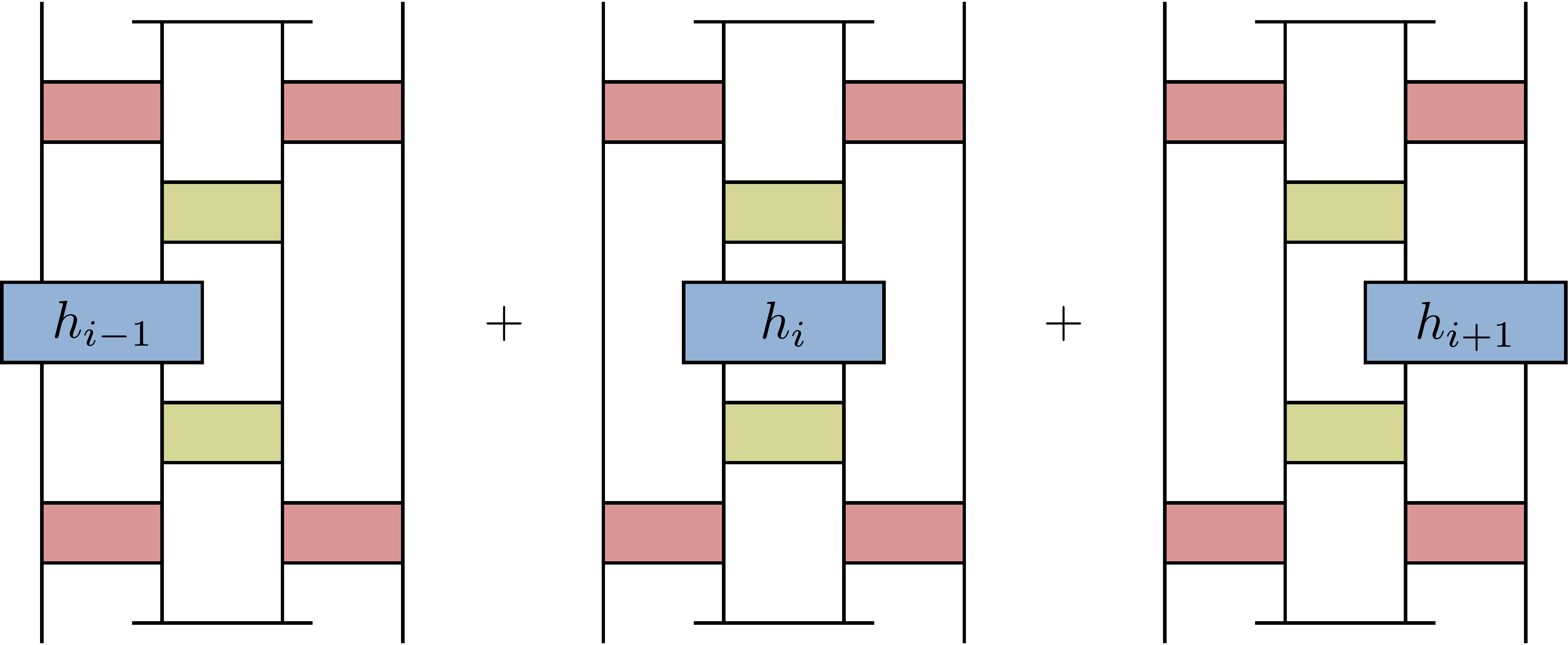} \vspace{0.2cm} \\
    (b)\includegraphics[scale=0.18]{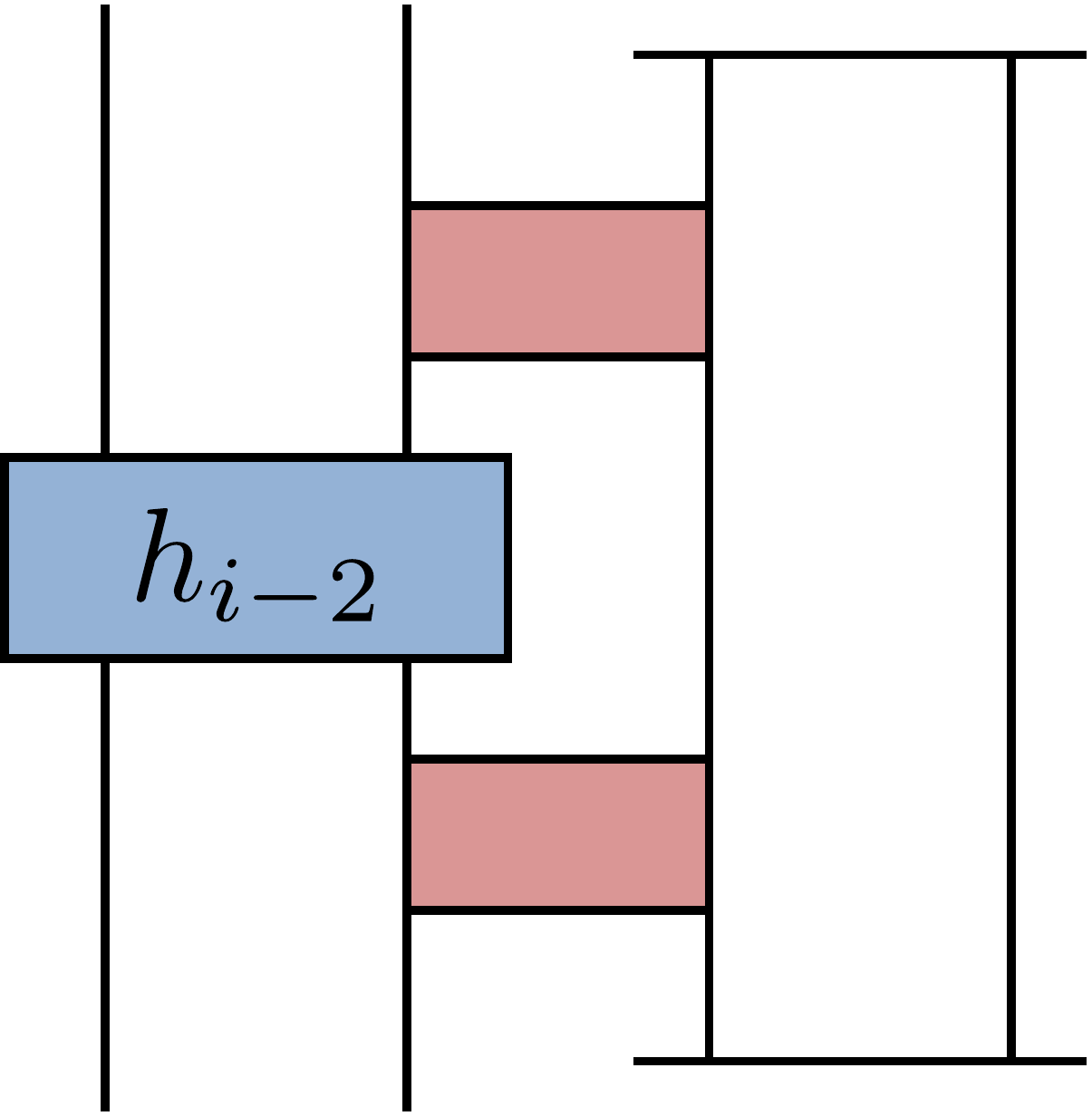} \quad 
    (c)\includegraphics[scale=0.18]{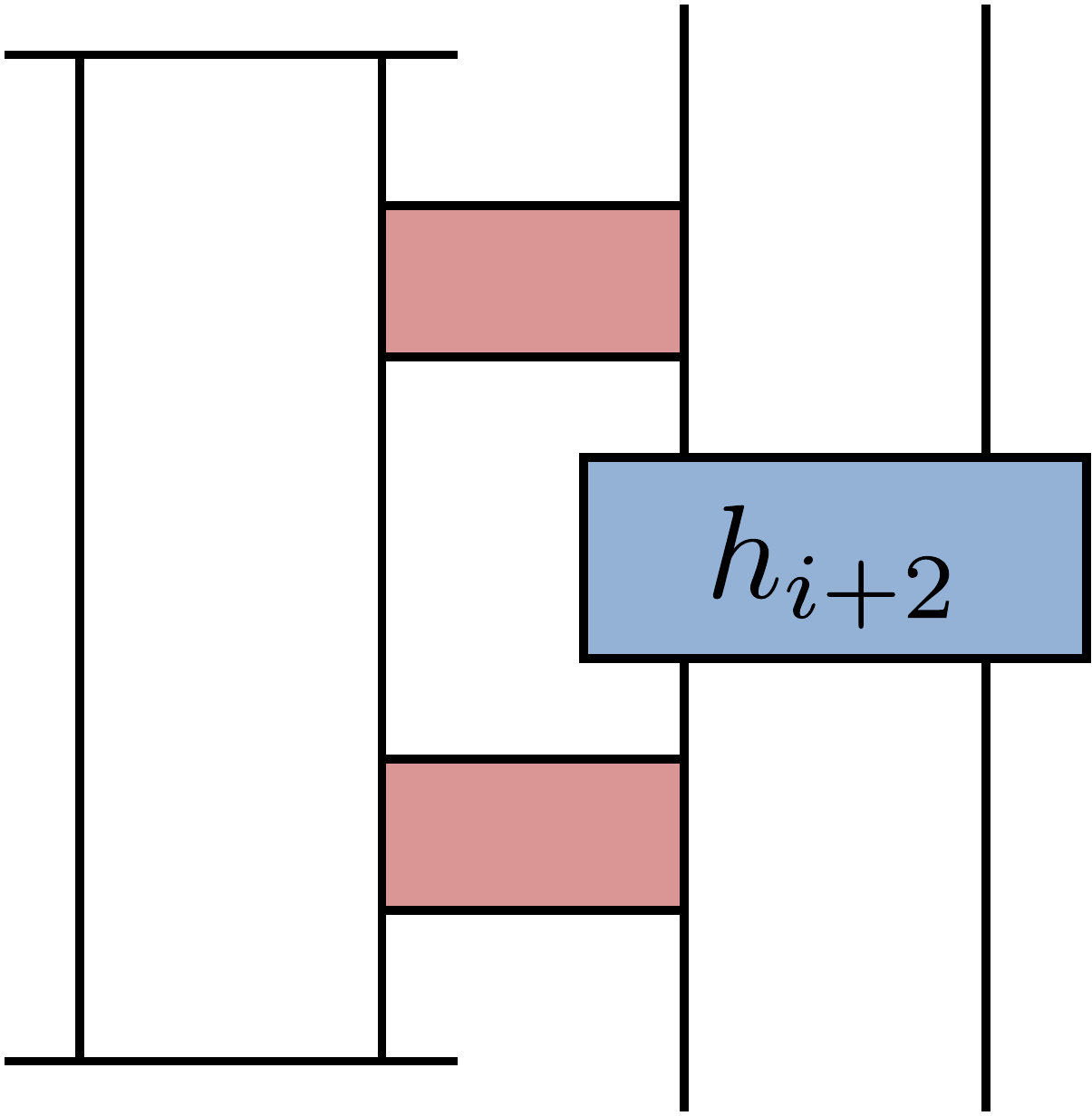} 
    \caption{
        Tensor network diagrams showing the \emph{raising} of the Hamiltonian operators to the next level of coarse-graining for 
        (a) the central components $i-1$, $i$ and $i+1$,
        (b) the left of the block $i-2$, 
        (c) the right of the block $i+2$.
        The cost is $\mathcal{O}(\chi^{7})$ to leading order.
    \label{fig:ham_m}
    \label{fig:ham_l}
    \label{fig:ham_r}
    }
\end{figure}

\section{\label{sec:observables}Observables}
\subsection{\label{sec:expectation}Single- and two-site operators}
Expectation values are calculated by applying the coarse-graining blocks on operators.
Because the block maps two-site operators on one level to two-site operators on the level above, calculating the expectation values means simply applying the raising operators on the operator in the same way as the Hamiltonian operators during the update.
However due to the unitary or isometric properties of the tensors, usually only a fraction of the tensors need to be contracted and the rest cancel to identities.
The remaining tensors are said to be in the \emph{causal cone} of the operator.
An example of this is given in Fig.\ \ref{fig:dMERA_exp_full}, which shows an operator acting on sites $11$ and $12$ in the network from Fig.\ \ref{fig:dMERA_L20}(c).
The causal cone is highlighted in blue, the remaining tensors are not involved in the expectation value.
This allows expectation values to be calculated in a highly efficient manner.
\begin{figure}
    \includegraphics[width=0.7\columnwidth]{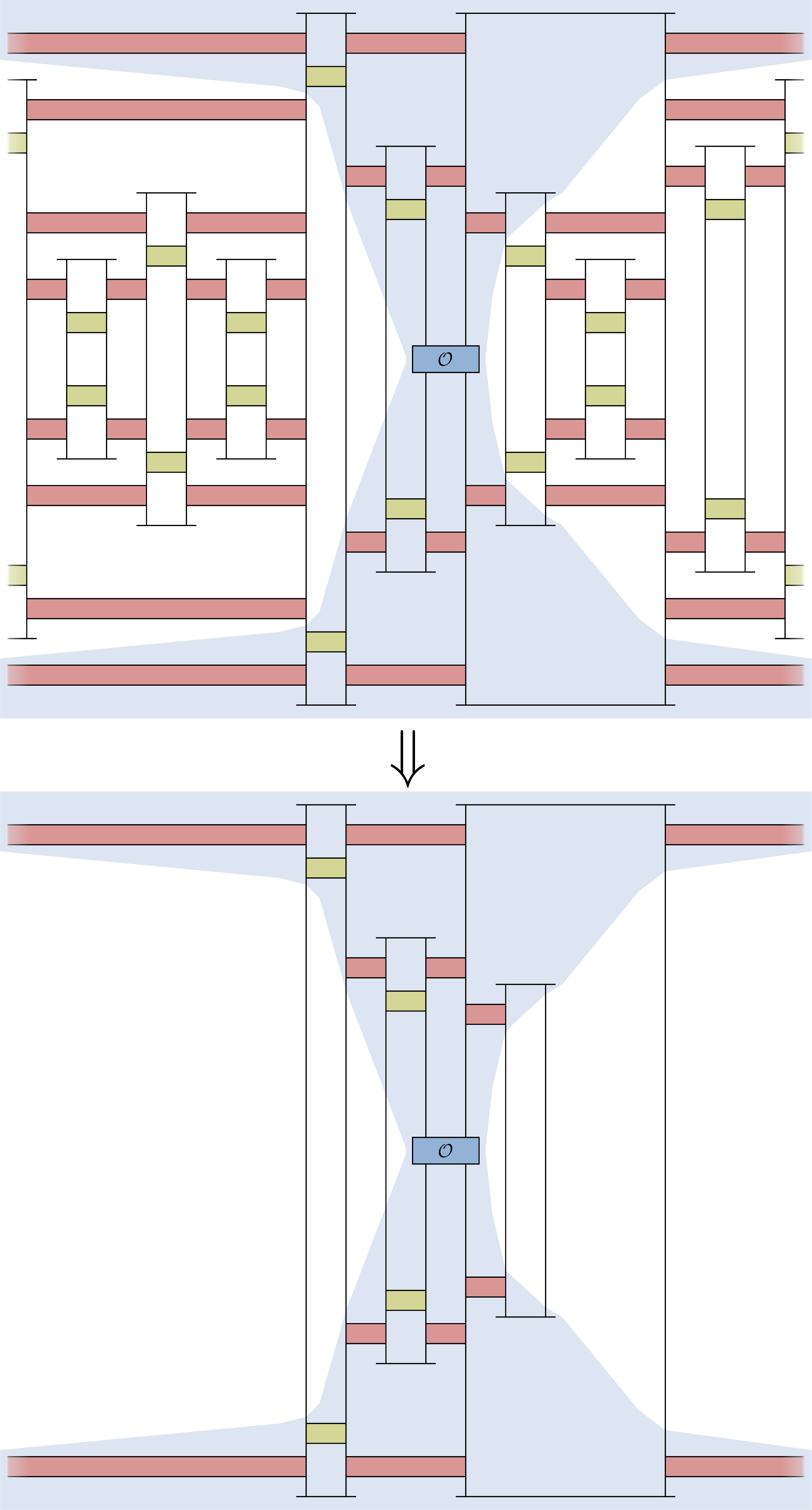}
    \caption{
        Tensor network diagram showing a full expectation value of a two-site operator acting on sites $11$ and $12$.
        Notice that due to the unitary and isometric properties of the tensors only those within the \emph{causal cone} of the operator (highlighted in blue) need to be contracted, thus reducing the computational cost.
    \label{fig:dMERA_exp_full}
    }
\end{figure}

The diagrams for the raising operation are effectively the same as those in Fig.\ \ref{fig:ham_m} except the summation of terms in (a) is unnecessary as the expectation value will only include one of the three diagrams at each level.
The numerical result of the expectation value is found when the top is contracted, as shown in Fig.\ \ref{fig:corr_h_PBC1_top}.
The leading order cost is $\mathcal{O}(\chi^{7})$, but can be $\mathcal{O}(\chi^{6})$ for operators at some sites in some network geometries.
Single-site operators can be expressed as two-site operators via tensor product with an identity. 
The expectation values of which can then be contracted using the same diagrams as before.
This is the most practical method as a single-site operator will be mapped to two sites by the coarse-graining block anyway.
\begin{figure}
    (a)\includegraphics[scale=0.18]{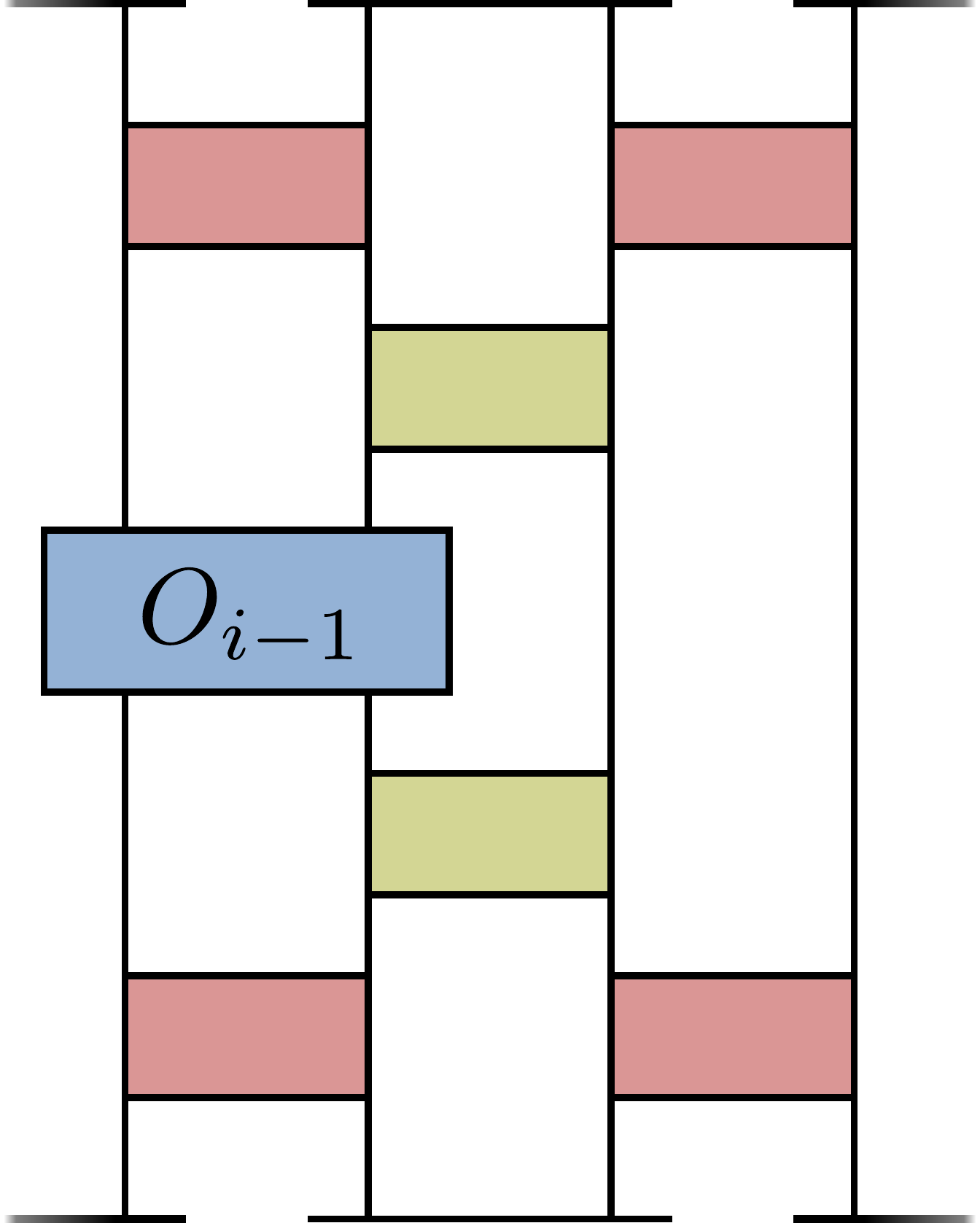} \hspace{0.05cm}
    (b)\includegraphics[scale=0.18]{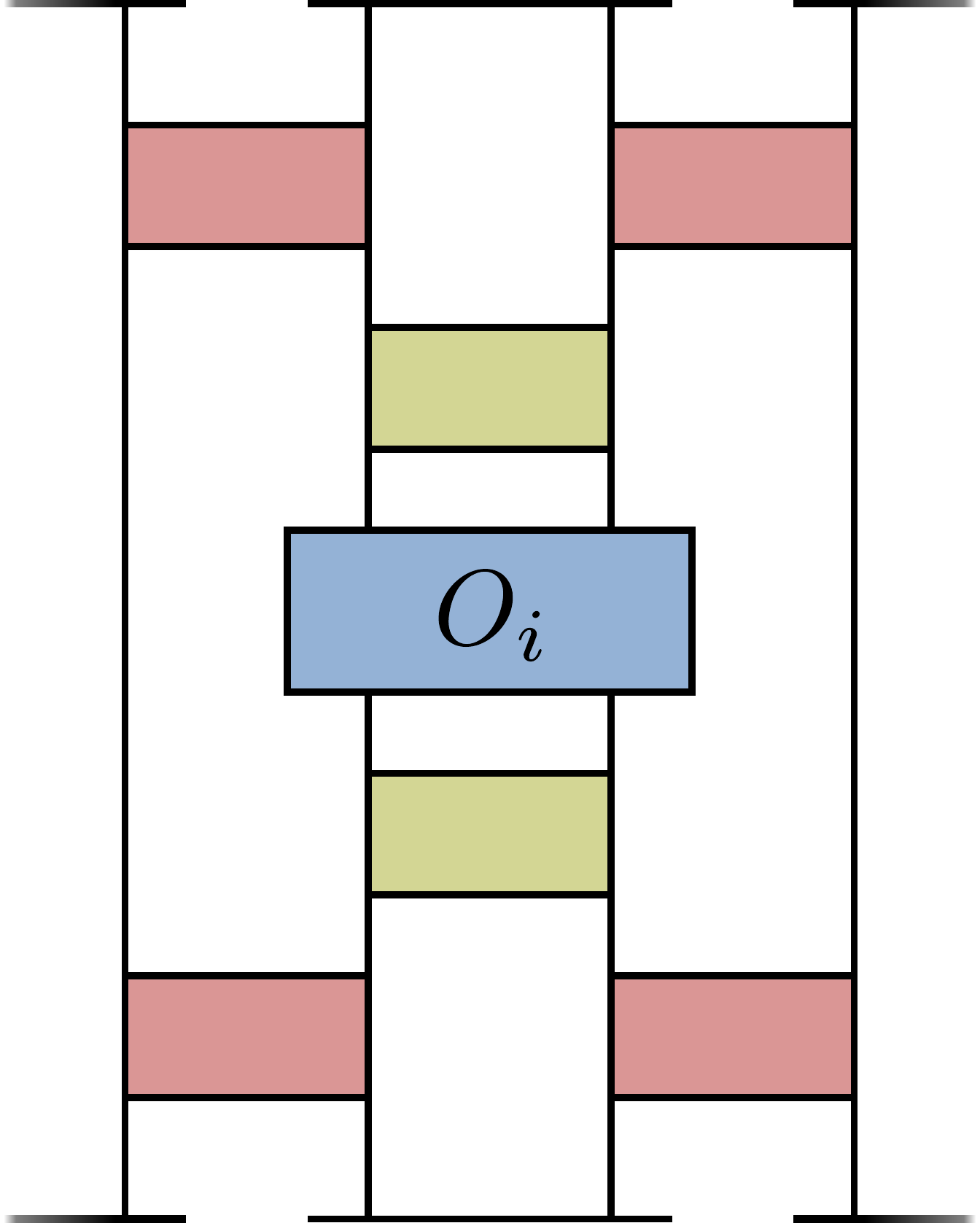} \vspace{0.4cm} \\ \hspace{0.2cm}
    (c)\includegraphics[scale=0.18]{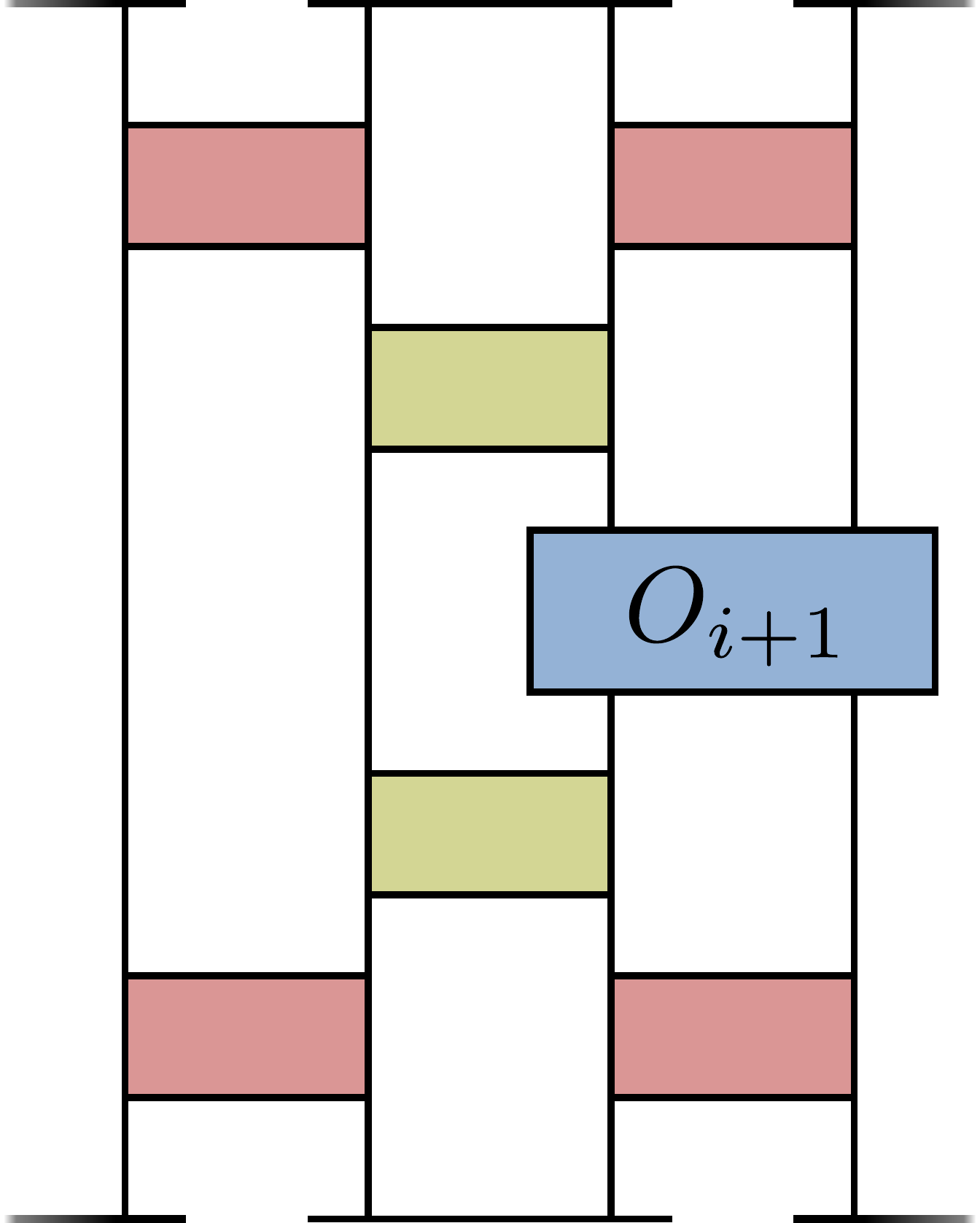} \hspace{0.05cm} 
    (d)\includegraphics[scale=0.18]{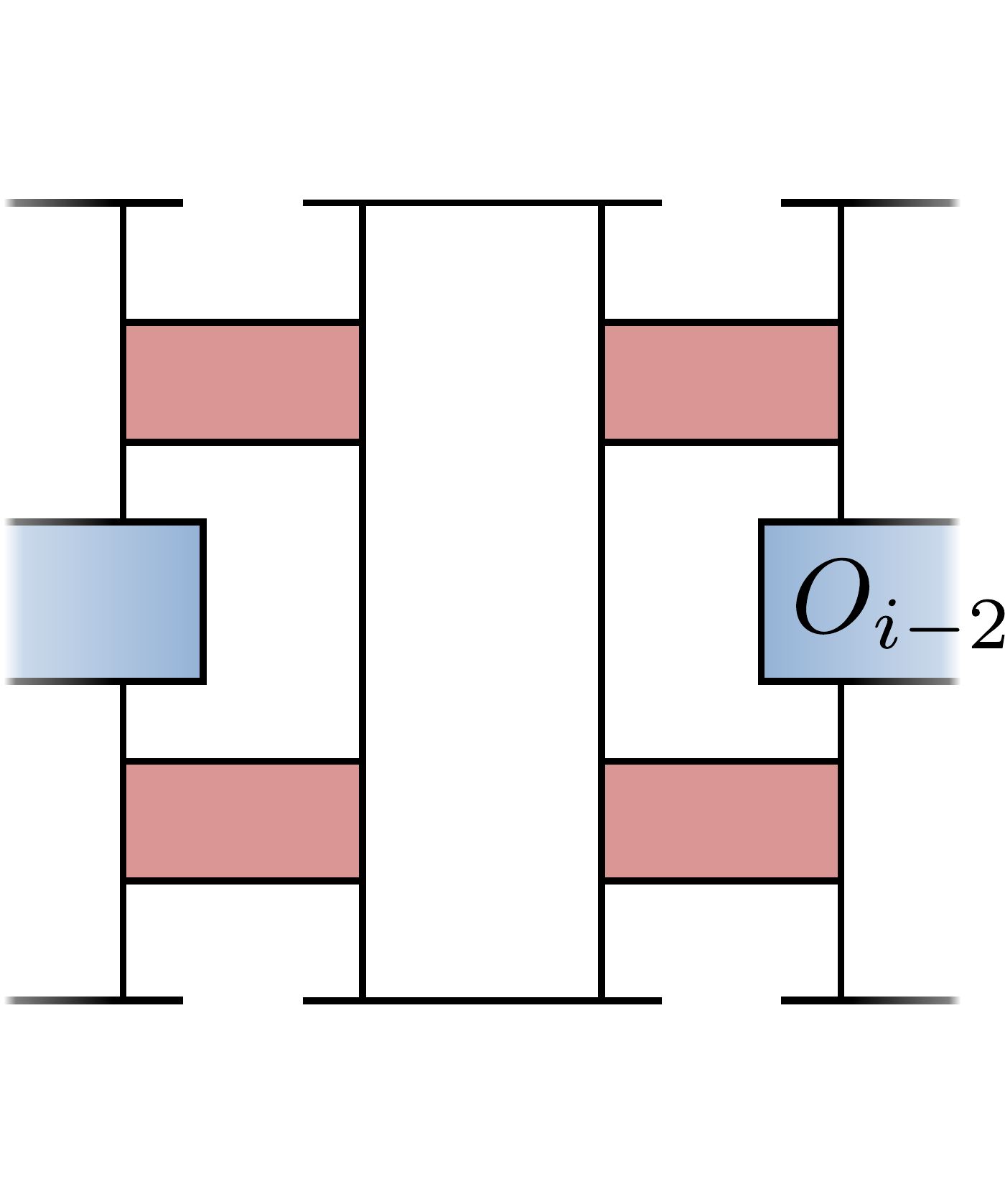}  
    \caption{
        Tensor network diagrams showing the contraction of two-site operators at the top of the network for: 
        (a) $i-1$, 
        (b) $i$, 
        (c) $i+1$, and
        (c) $i+2$.
        (a - c) have cost $\mathcal{O}(\chi^{6})$ to leading order, (d) is $\mathcal{O}(\chi^{4})$.
    \label{fig:corr_h_PBC1_top}
    \label{fig:corr_h_PBC2_top}
    \label{fig:corr_h_PBC3_top}
    \label{fig:corr_h_PBC4_top}
    }
\end{figure}

\subsection{\label{sec:corr}Two-point correlation functions}
\begin{figure}
    \includegraphics[width=0.7\columnwidth]{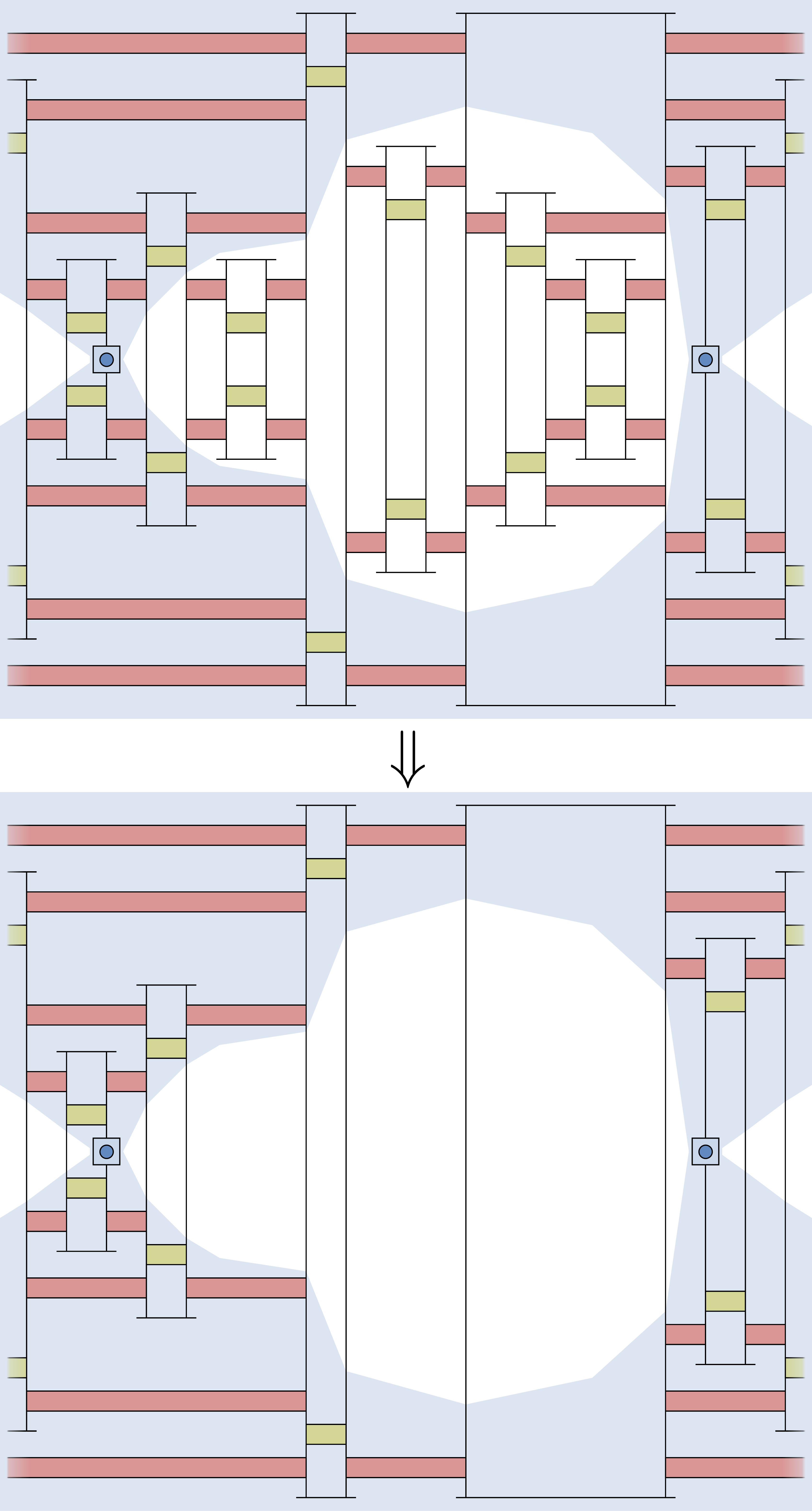}
    \caption{
        Tensor network diagram showing a two-point correlation function for sites $3$ and $18$.
        As in Fig.\ \ref{fig:dMERA_exp_full}, the causal cone is highlighted in blue.
        The two sites will be joined by a single coarse-graining block at the point when the causal cones of the two sites meet.
    \label{fig:dMERA_corr_full}
    }
\end{figure}
Correlation functions are calculated by applying the coarse-graining blocks on the two operators independently just as single- and two-site expectation values.
As before, only the tensors within the causal cone need to be contracted. 
However, at some point in the process the two sites will be joined by a single block.
This is the point at which the causal cones of the two operators join, as shown in Fig.\ \ref{fig:dMERA_corr_full}.
The joining means that the calculation of two-point correlation functions can be significantly more costly than a single-site operator.
This additional expense was also observed for standard MERA and was one of the reasons for the development of ternary MERA as opposed to the original binary variant \cite{EveV09}.
The reason is that when the two operators meet the outcome is not necessarily another two-site operator.
For the ternary MERA case however, there are choices of sites where the cost is $\mathcal{O}(\chi^{8})$.
This is when the two sites are at the centre leg of the isometry, i.e. at separations $r = 3^{q}$ with $q = 1,2,3 \dots$. 
Otherwise the cost grows beyond $\mathcal{O}(\chi^{8})$.
For dMERA the problem can not be so simply worked around.
When performing calculations for disordered systems, discarding individual correlations will result in errors in the averaging and therefore losing the characteristic power law of the random singlet phase.
It is therefore necessary to calculate correlations between all pairs of sites or discard all data for a given separation $r$ if one is too costly to compute.

When the operators in dMERA meet the possible outcomes are two-, three- and four-site operators, as shown in the tensor network diagrams of Figs.\ \ref{fig:corr_corr}(a-f).
The raising of two-site operators is the same as in sec.\ \ref{sec:expectation}, three- and four-site operators are shown in Figs.\ \ref{fig:corr_prop3_1}(g-l) and \ref{fig:corr_prop4_1}(m-t) respectively.
The expectation value of the correlation function is finally determined by contraction with the top of the network as shown in Fig.\ \ref{fig:corr_corr_PBC1_top}.
In full, the cost of a calculating a two-point correlation function can be between $\mathcal{O}(\chi^{6})$ and $\mathcal{O}(\chi^{11})$ depending on the sites and structure of the network.
As mentioned in sec.\ \ref{sec:expectation}, for algorithmic simplicity a single site-operator is encoded in a two-site operator by tensor product with an identity.
This suggests that there are two ways of encoding the each operator: $O_{i} \otimes \openone_{i+1}$ and $\openone_{i-1} \otimes O_{i}$.
Due to the fact that the network is inhomogeneous, this choice of encoding can affect the cost of contracting the correlation function. 
It is therefore important to choose the operator encoding that minimizes the cost.
\begin{figure*}
    (a)\includegraphics[scale=0.15]{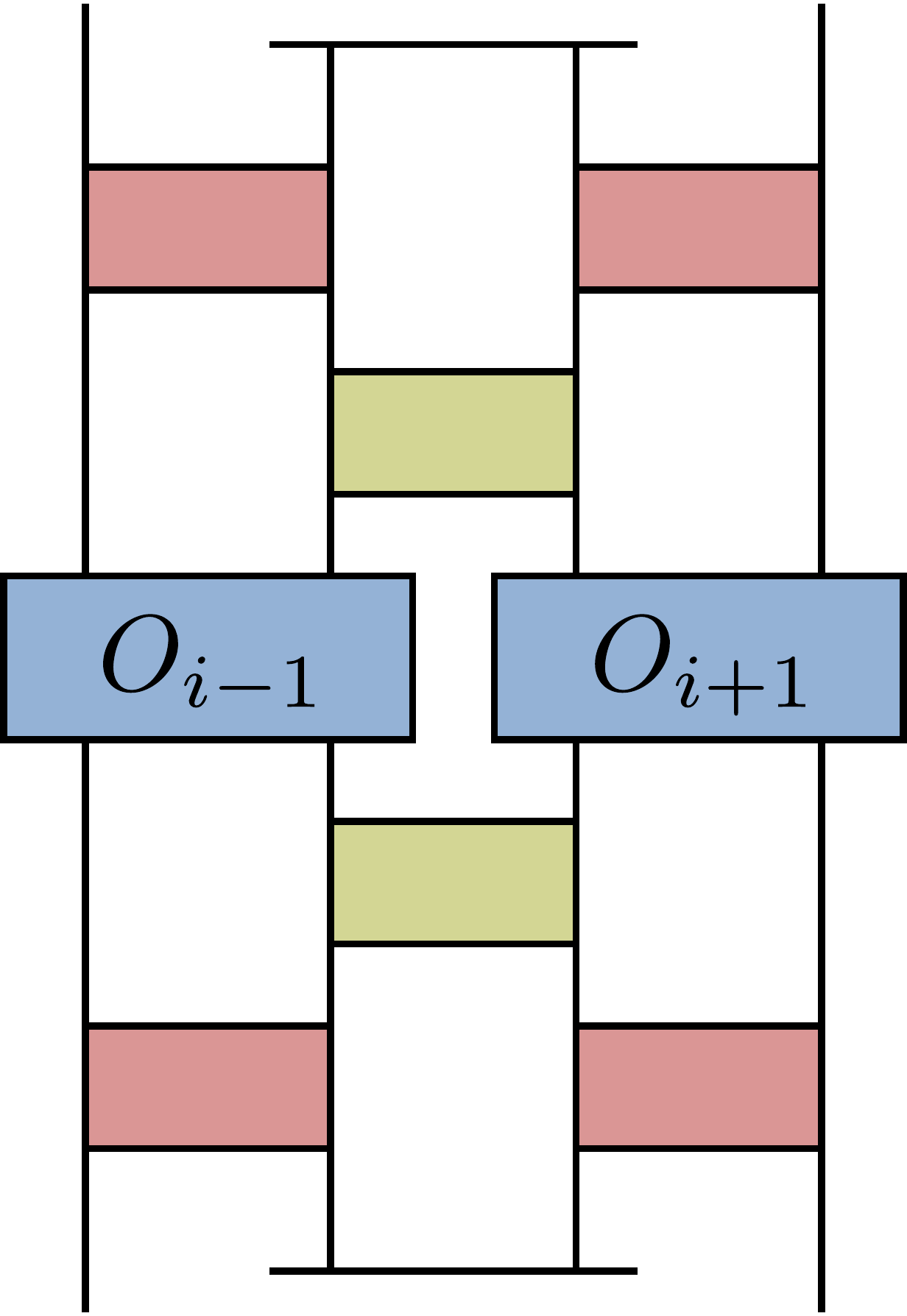} \hspace{0.05cm}  
    (b)\includegraphics[scale=0.15]{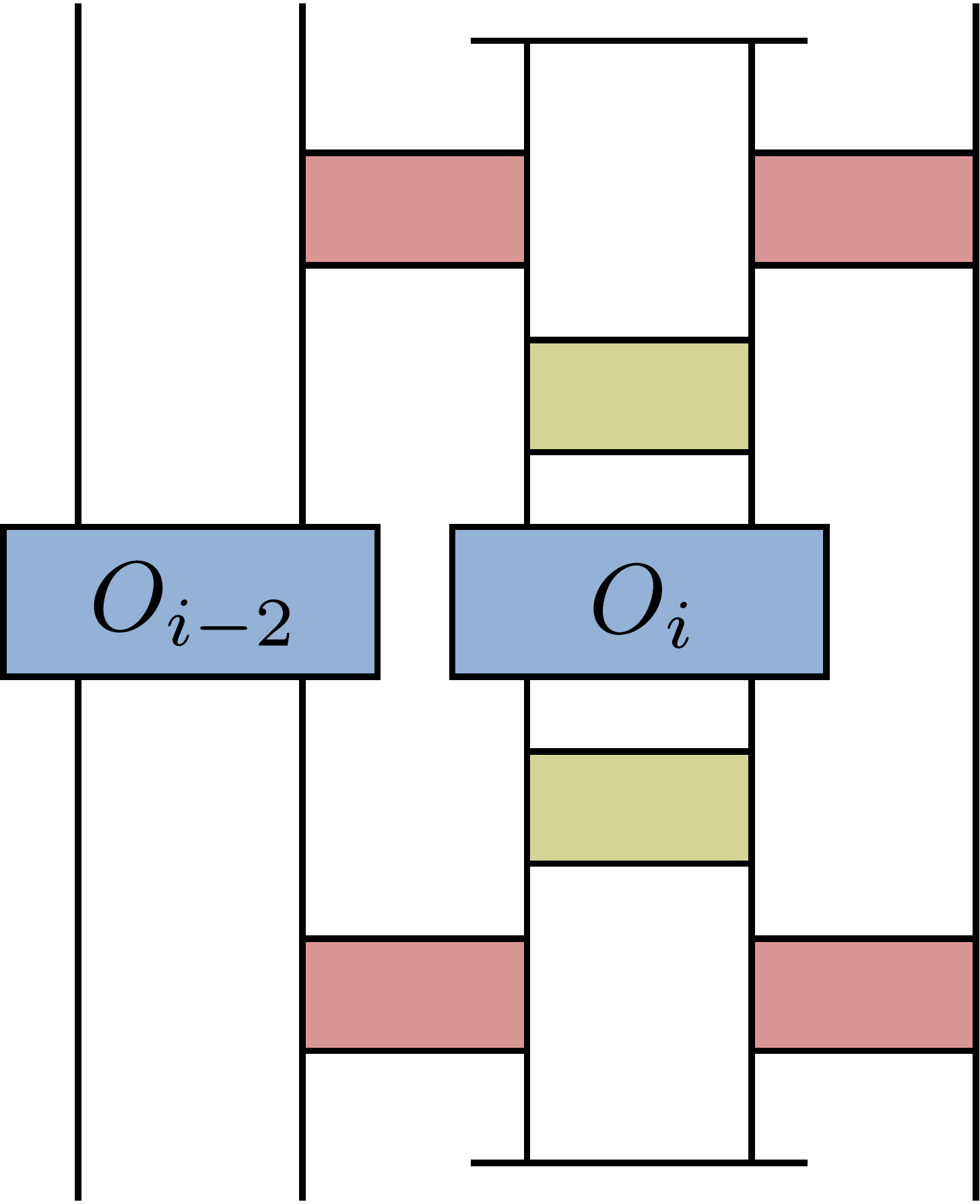} \hspace{0.05cm}  
    (c)\includegraphics[scale=0.15]{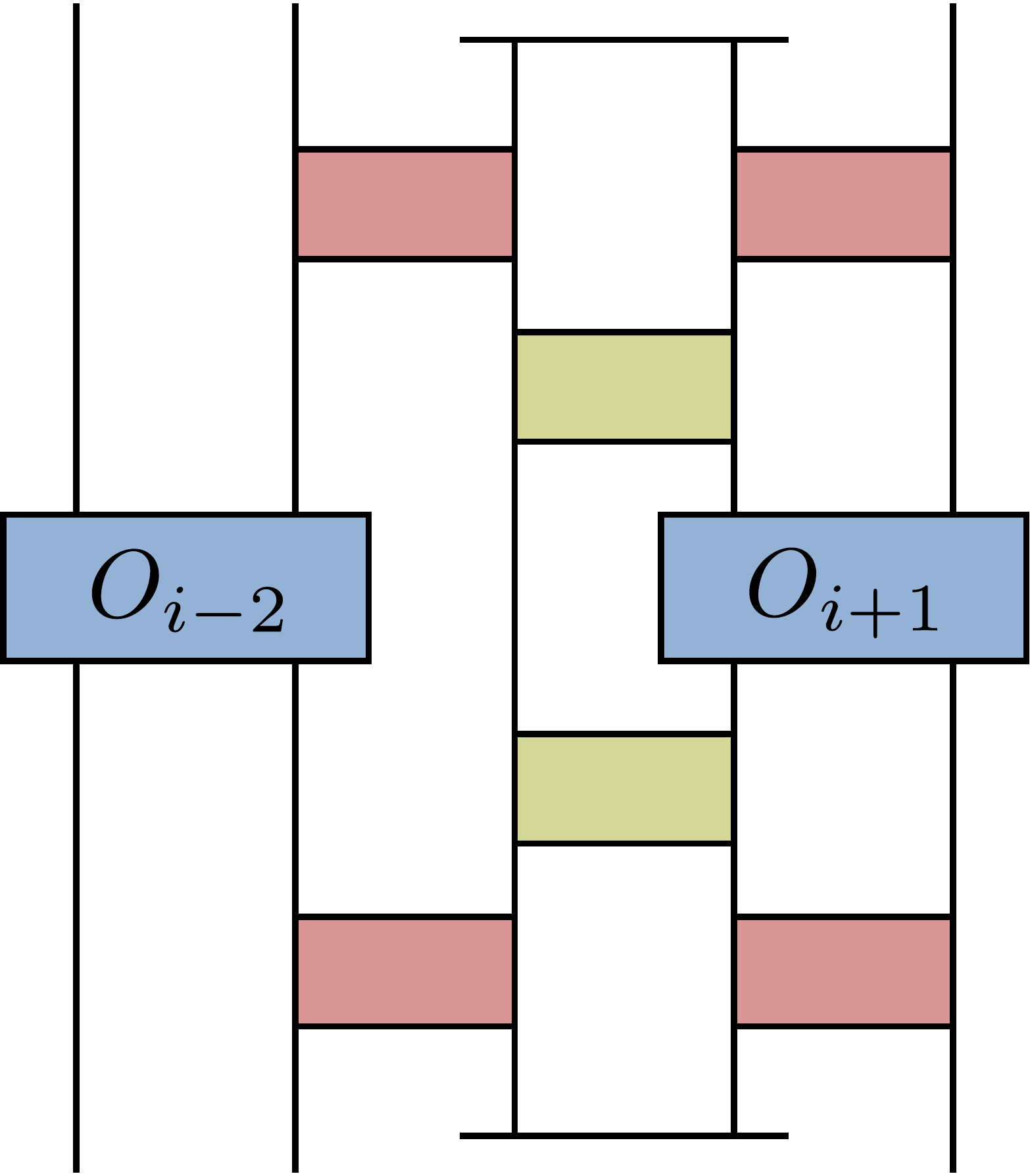} \hspace{0.05cm} 
    (d)\includegraphics[scale=0.15]{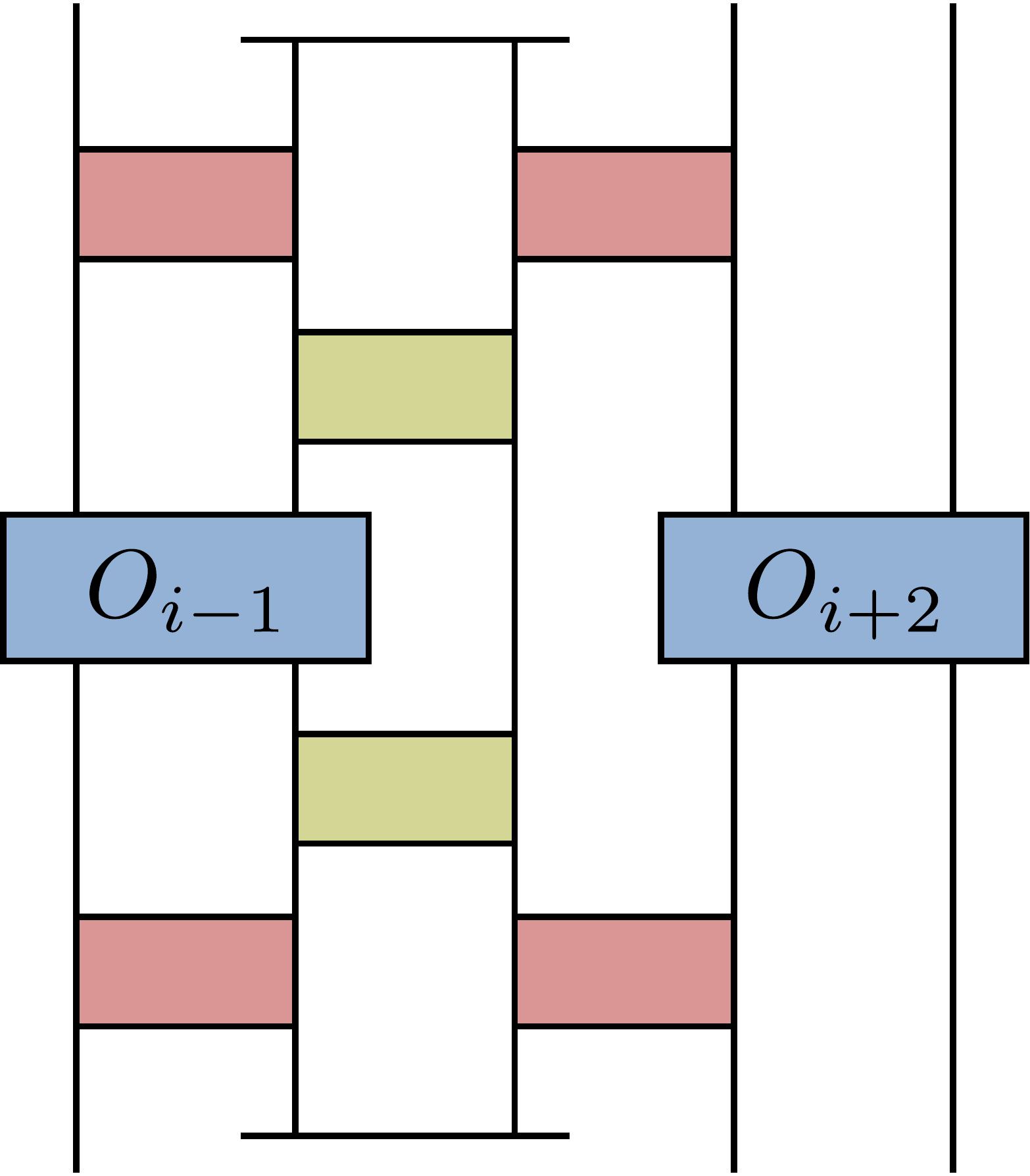} \hspace{0.05cm}  
    (e)\includegraphics[scale=0.15]{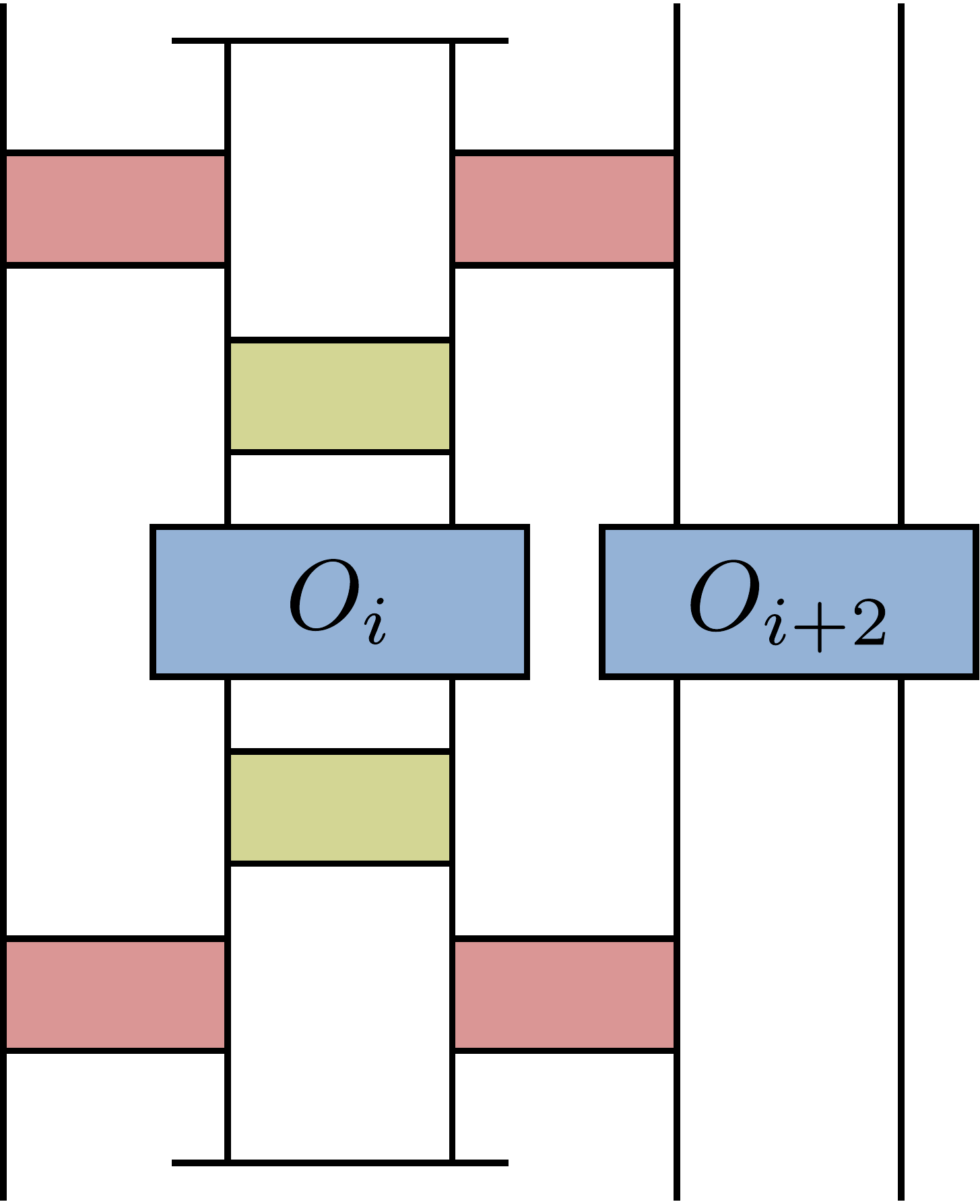} \hspace{0.05cm} 
    (f)\includegraphics[scale=0.15]{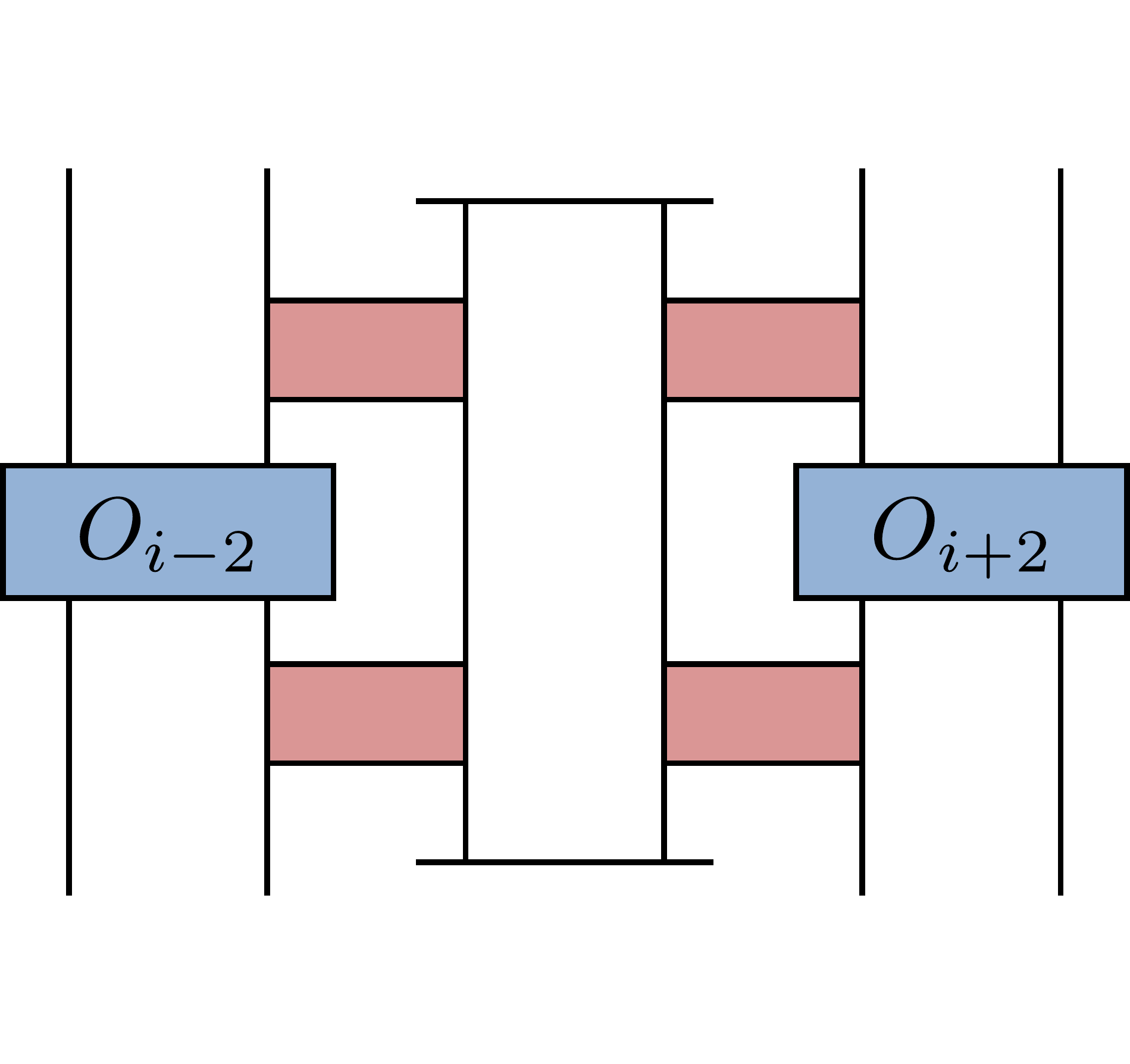} \vspace{0.2cm} \\  
    (g)\includegraphics[scale=0.15]{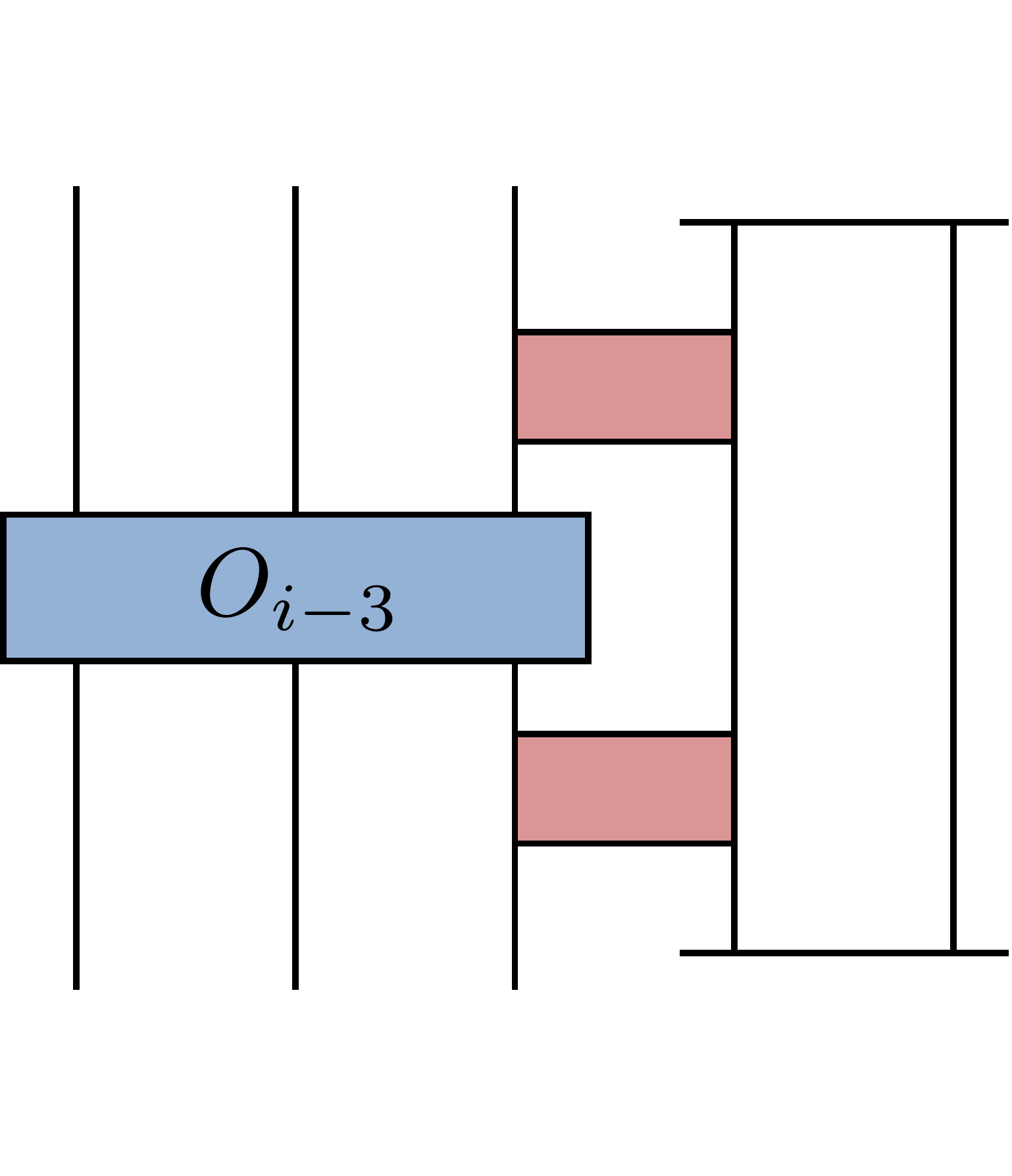} \hspace{0.05cm} 
    (h)\includegraphics[scale=0.15]{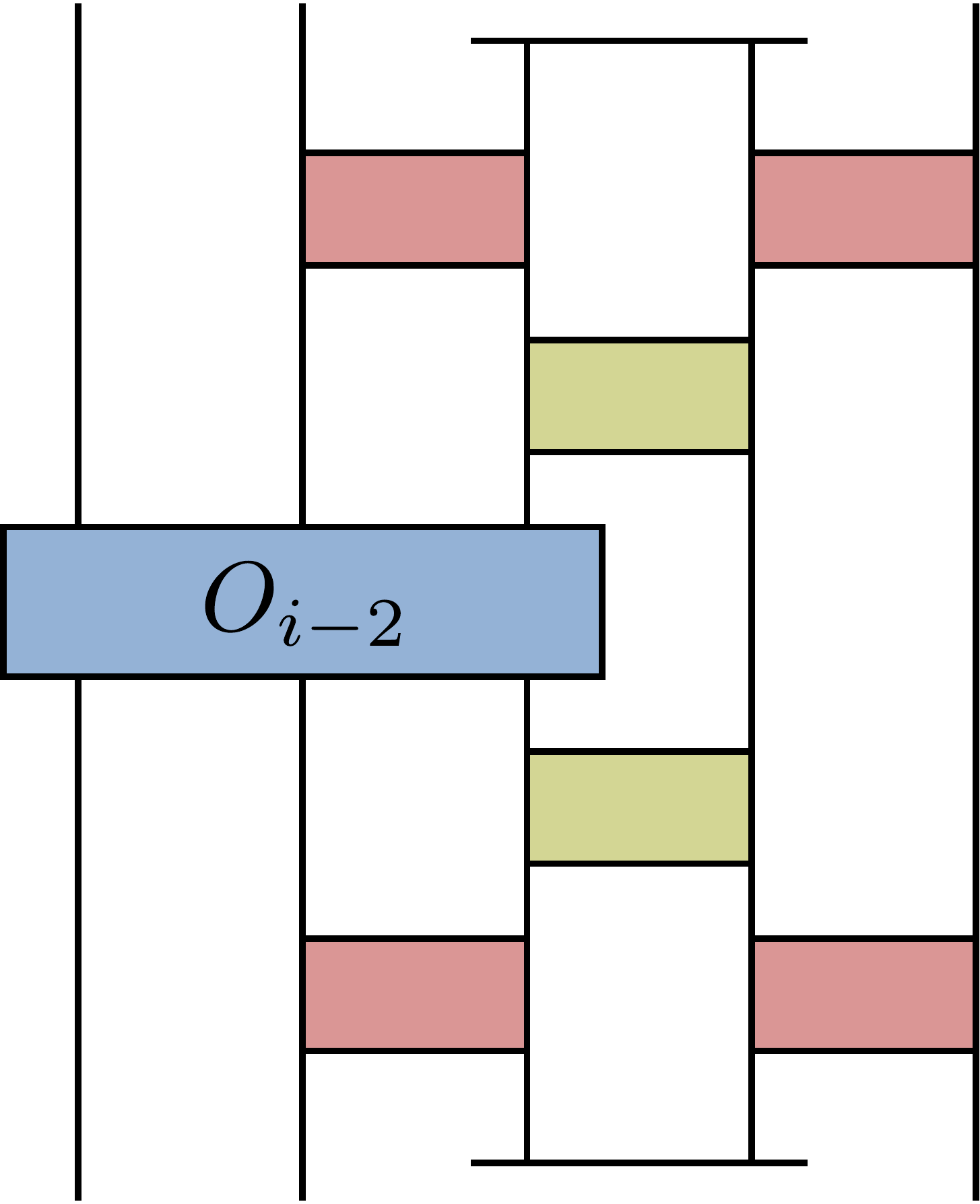} \hspace{0.2cm}  
    (i)\includegraphics[scale=0.15]{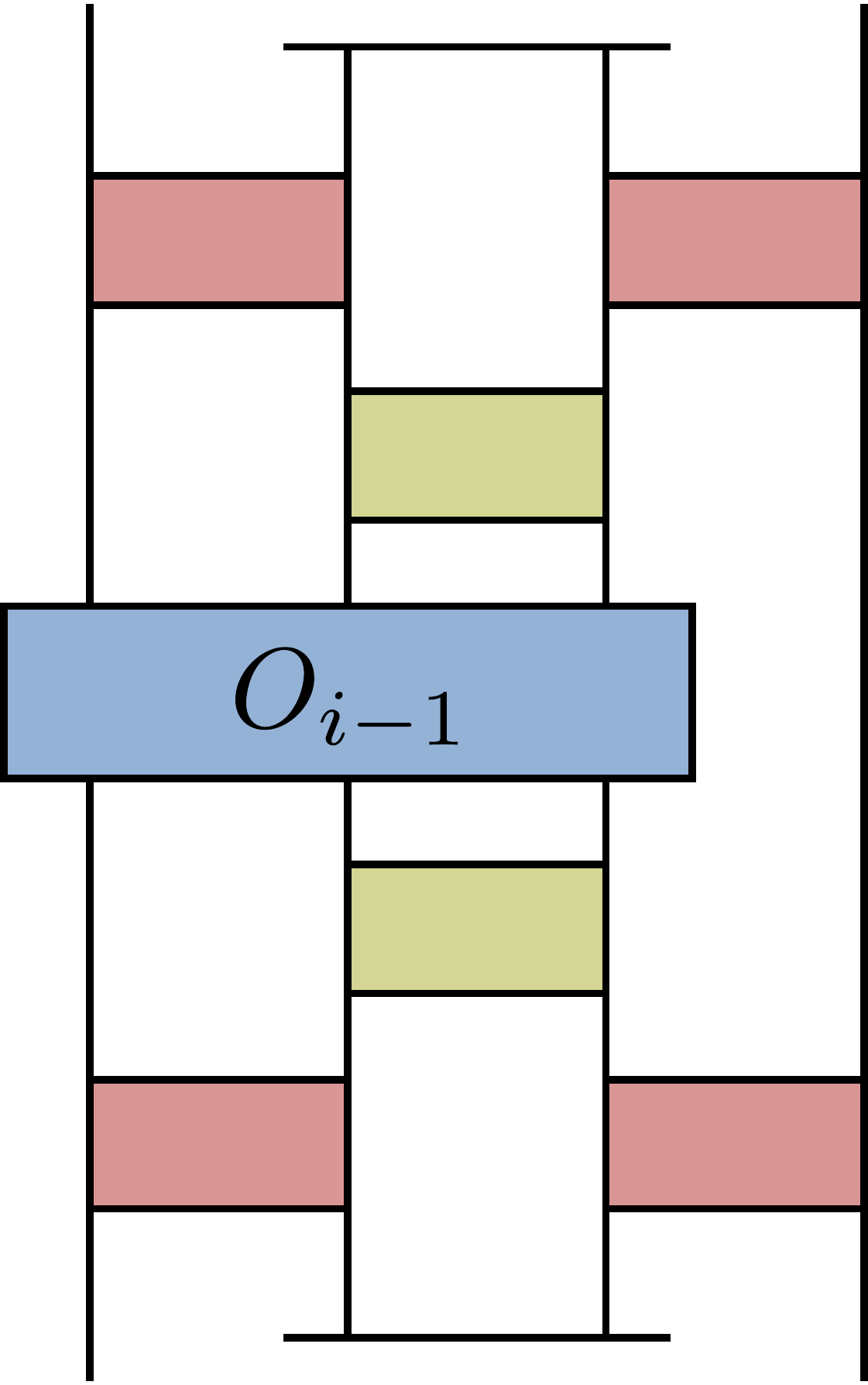} \hspace{0.2cm} 
    (j)\includegraphics[scale=0.15]{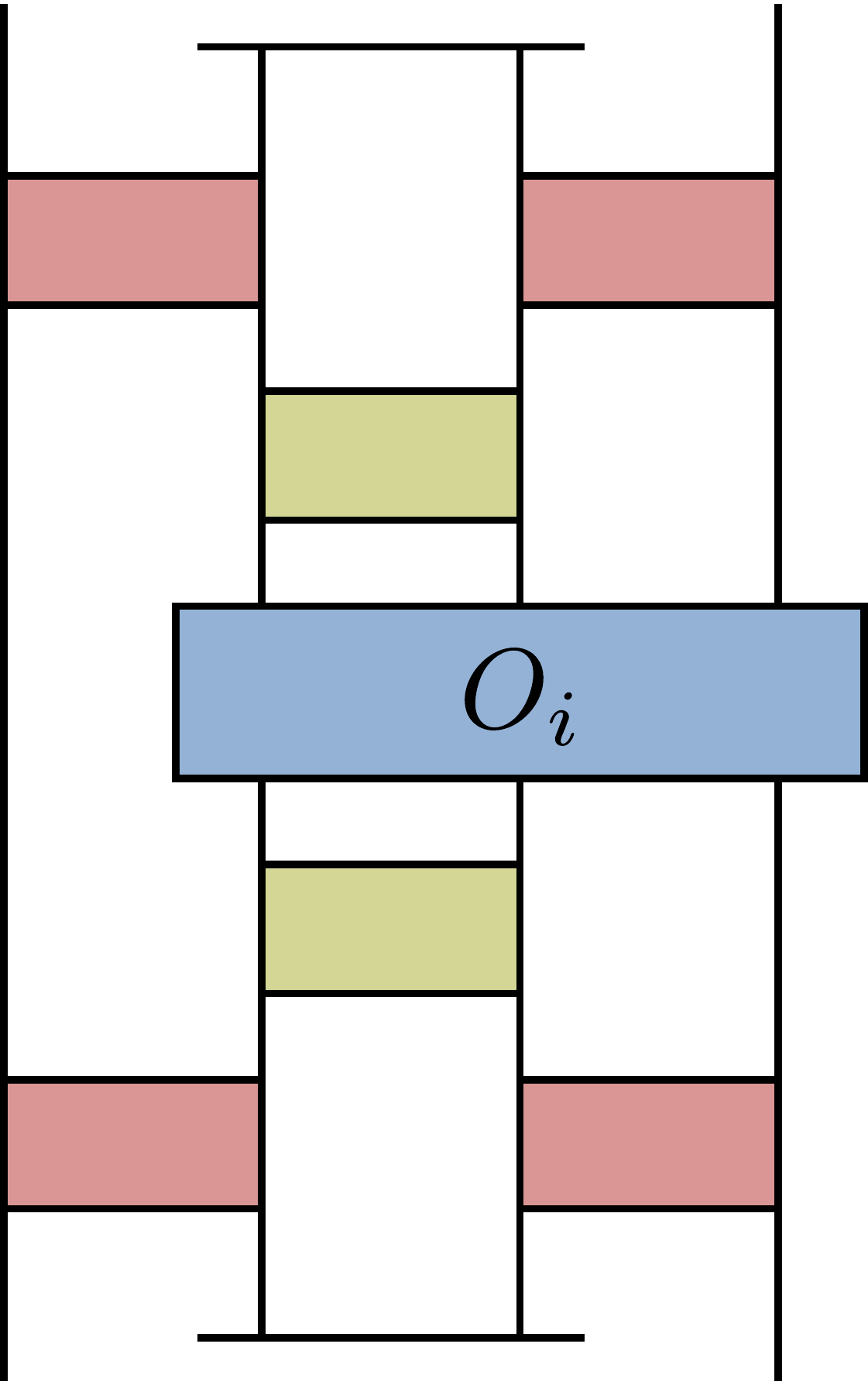} \hspace{0.05cm}  
    (k)\includegraphics[scale=0.15]{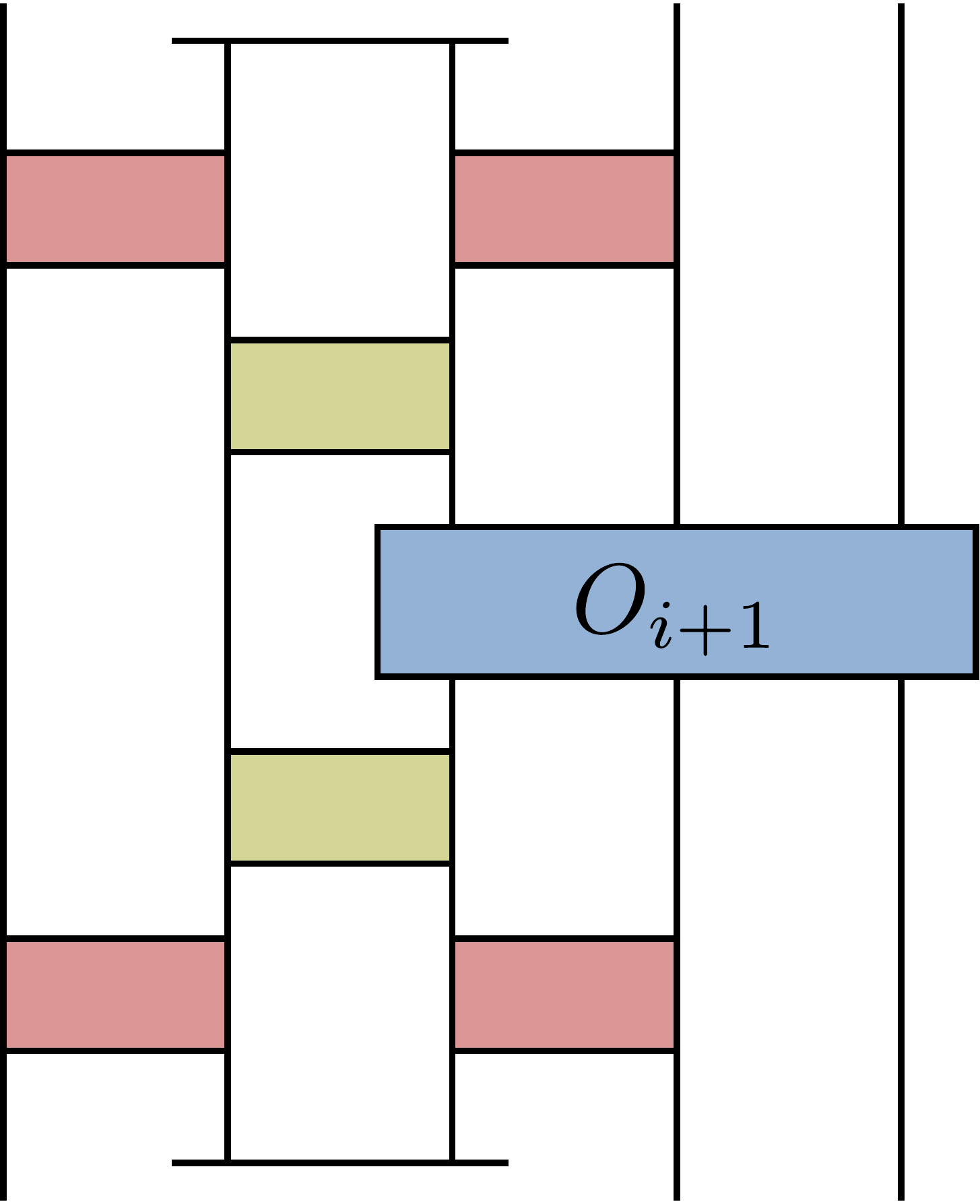} \hspace{0.05cm}  
    (l)\includegraphics[scale=0.15]{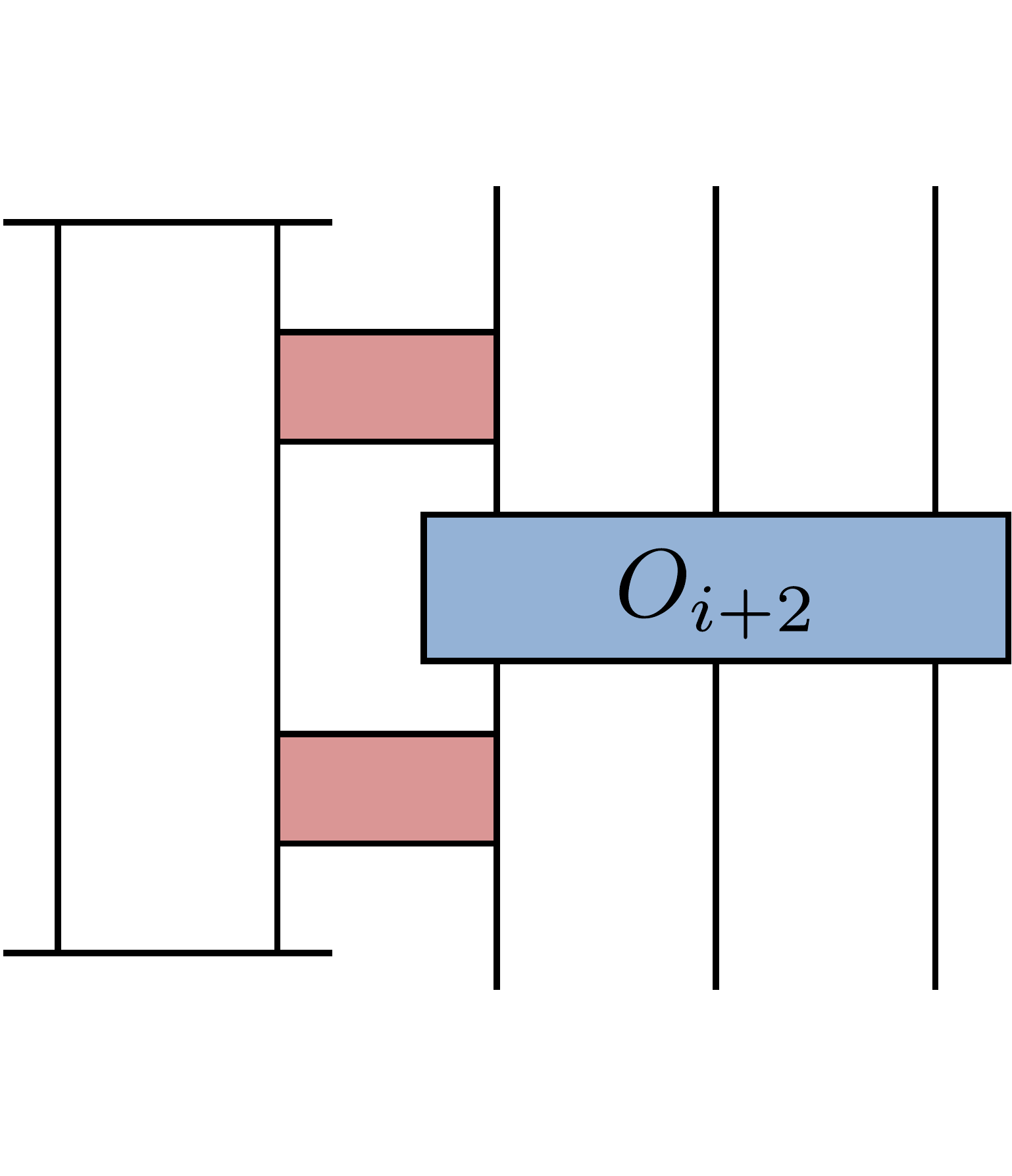} \vspace{0.2cm} \\
    (m)\includegraphics[scale=0.15]{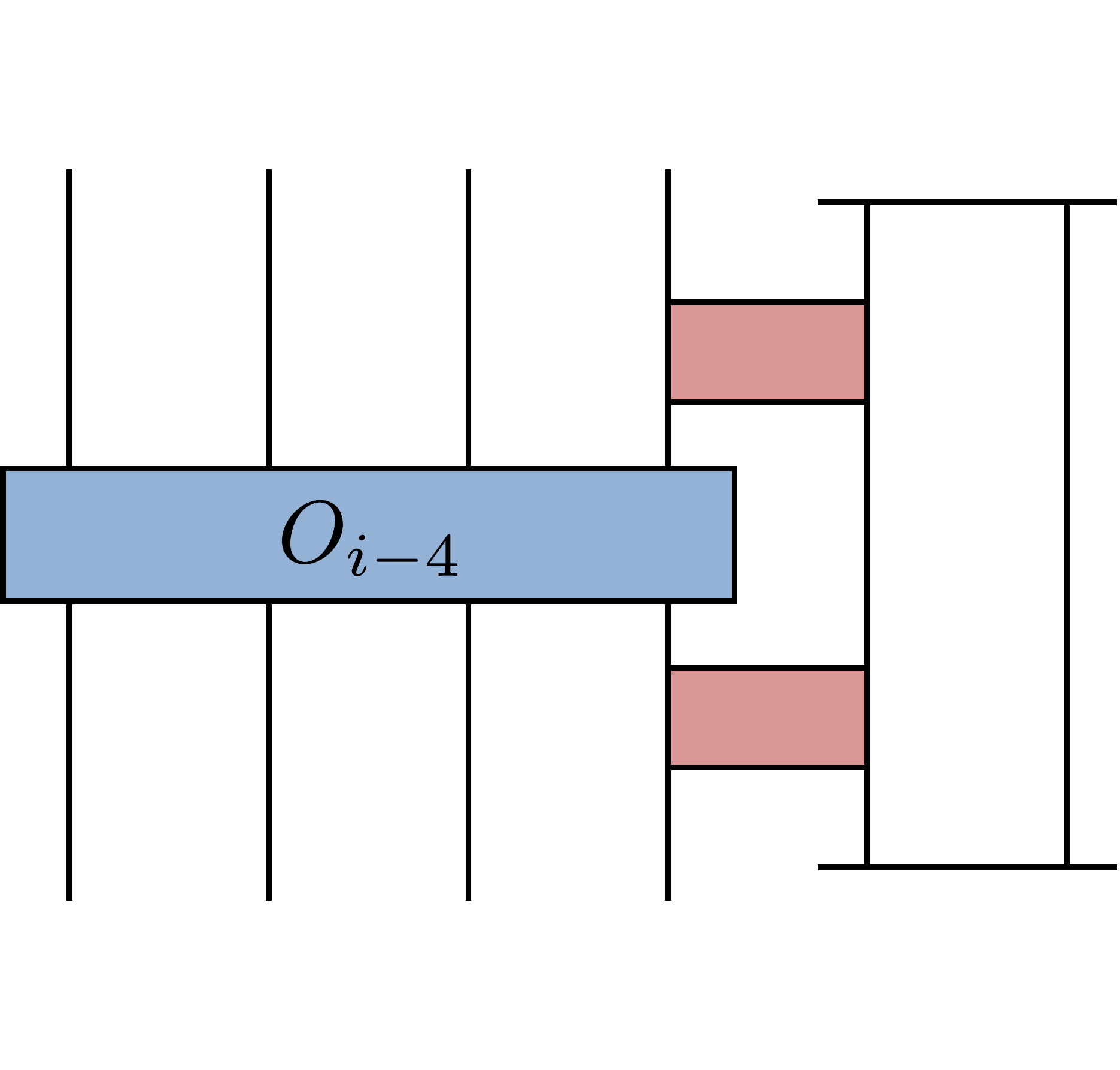} \hspace{0.05cm}
    (n)\includegraphics[scale=0.15]{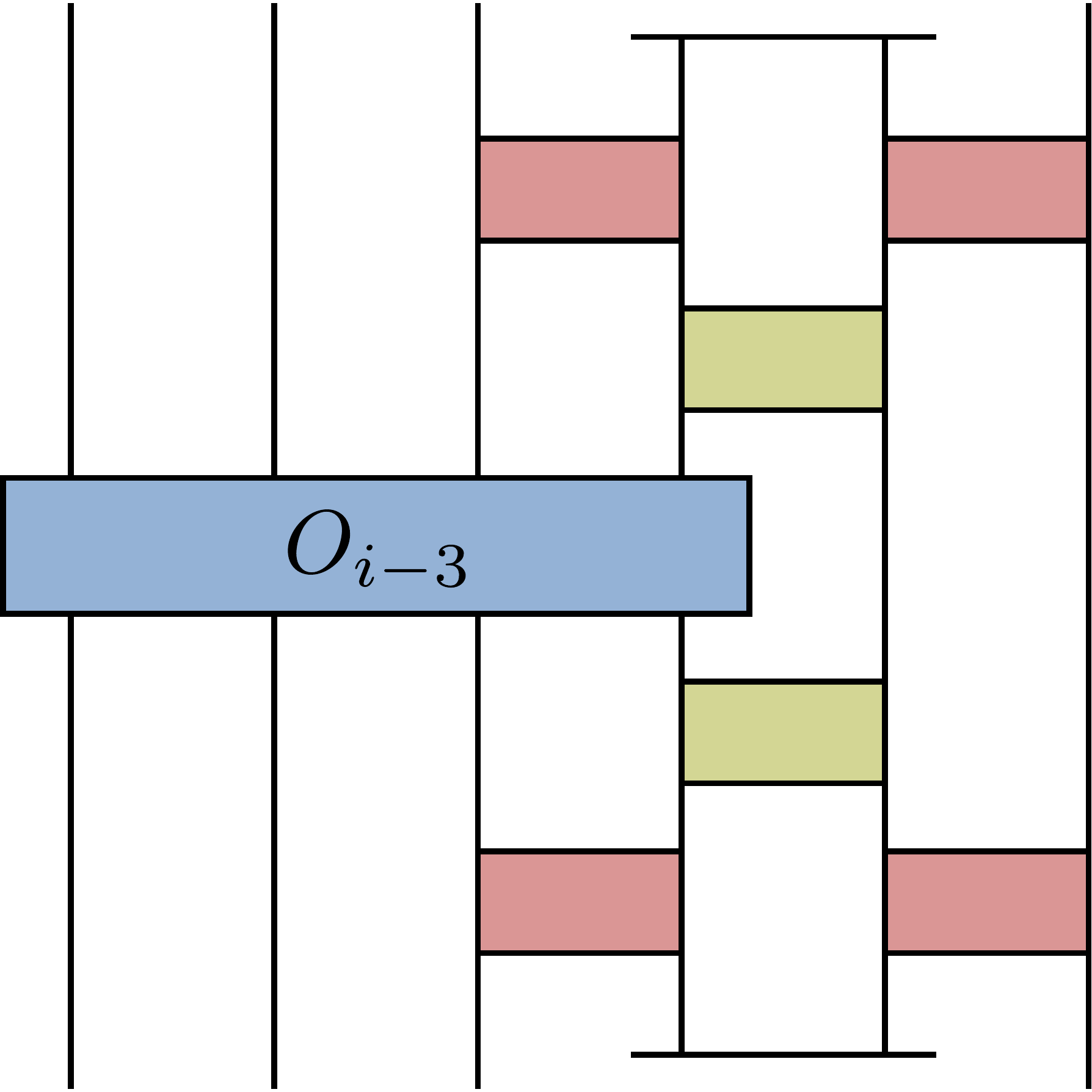} \hspace{0.05cm}
    (o)\includegraphics[scale=0.15]{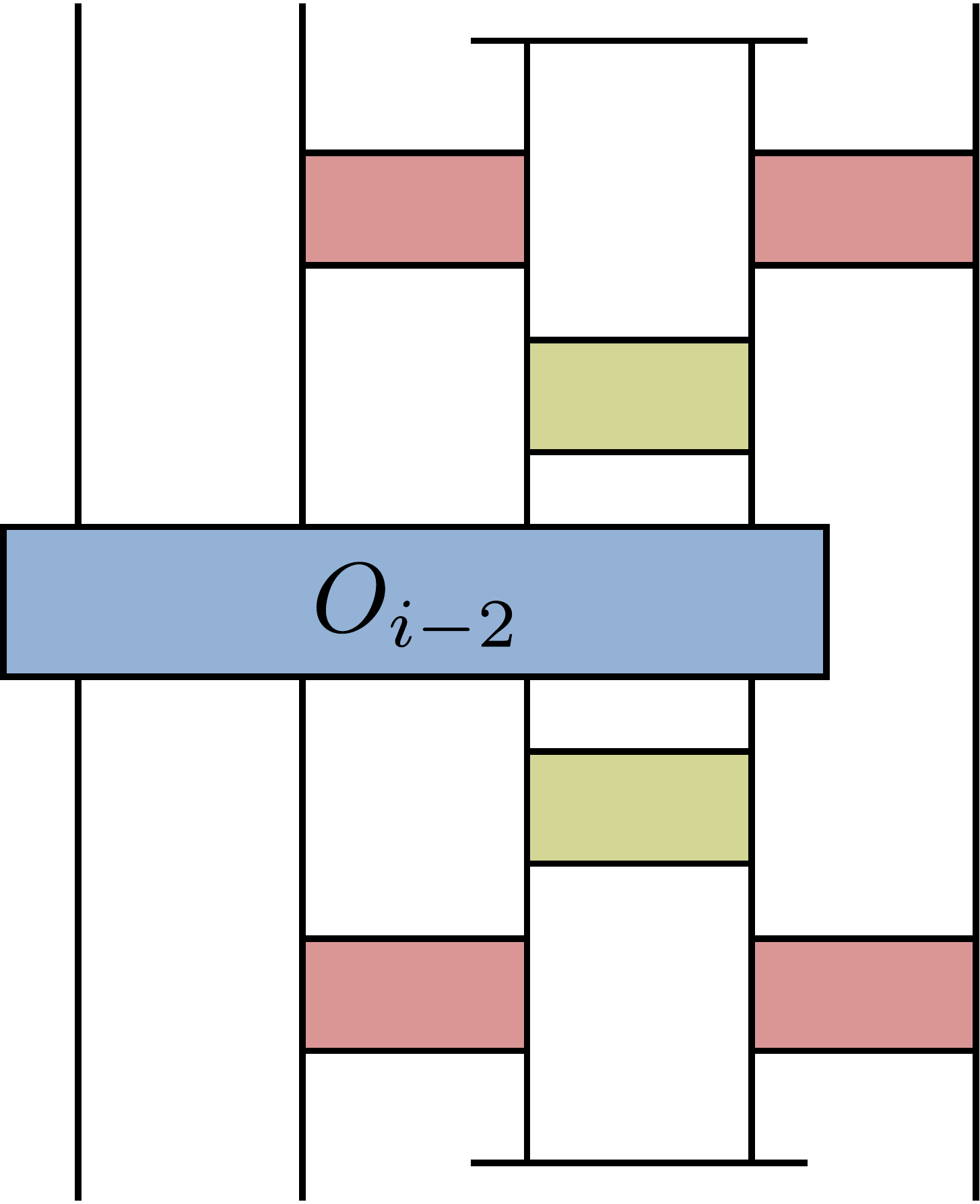} \hspace{0.05cm}
    (p)\includegraphics[scale=0.15]{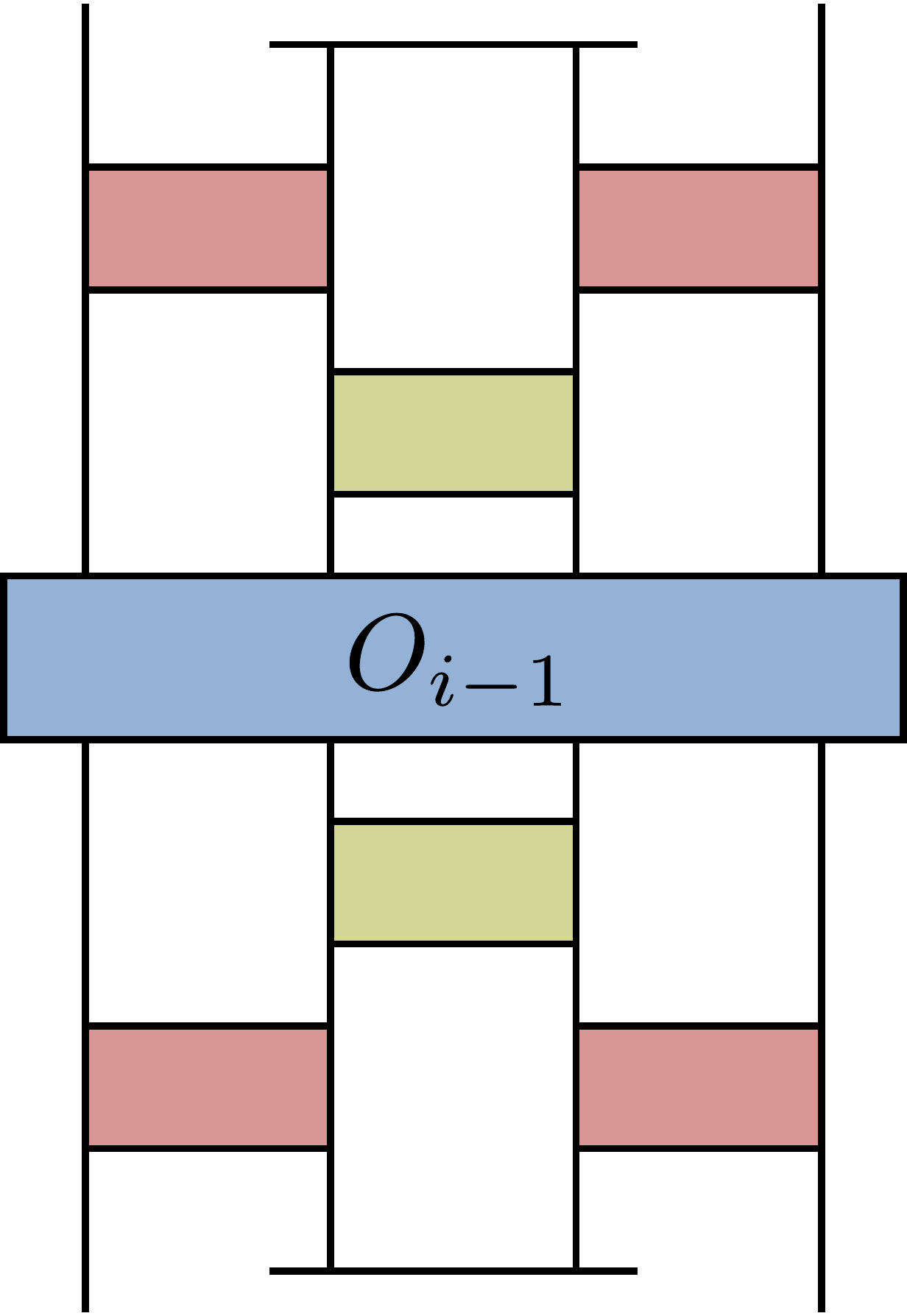} \hspace{0.05cm} 
    (q)\includegraphics[scale=0.15]{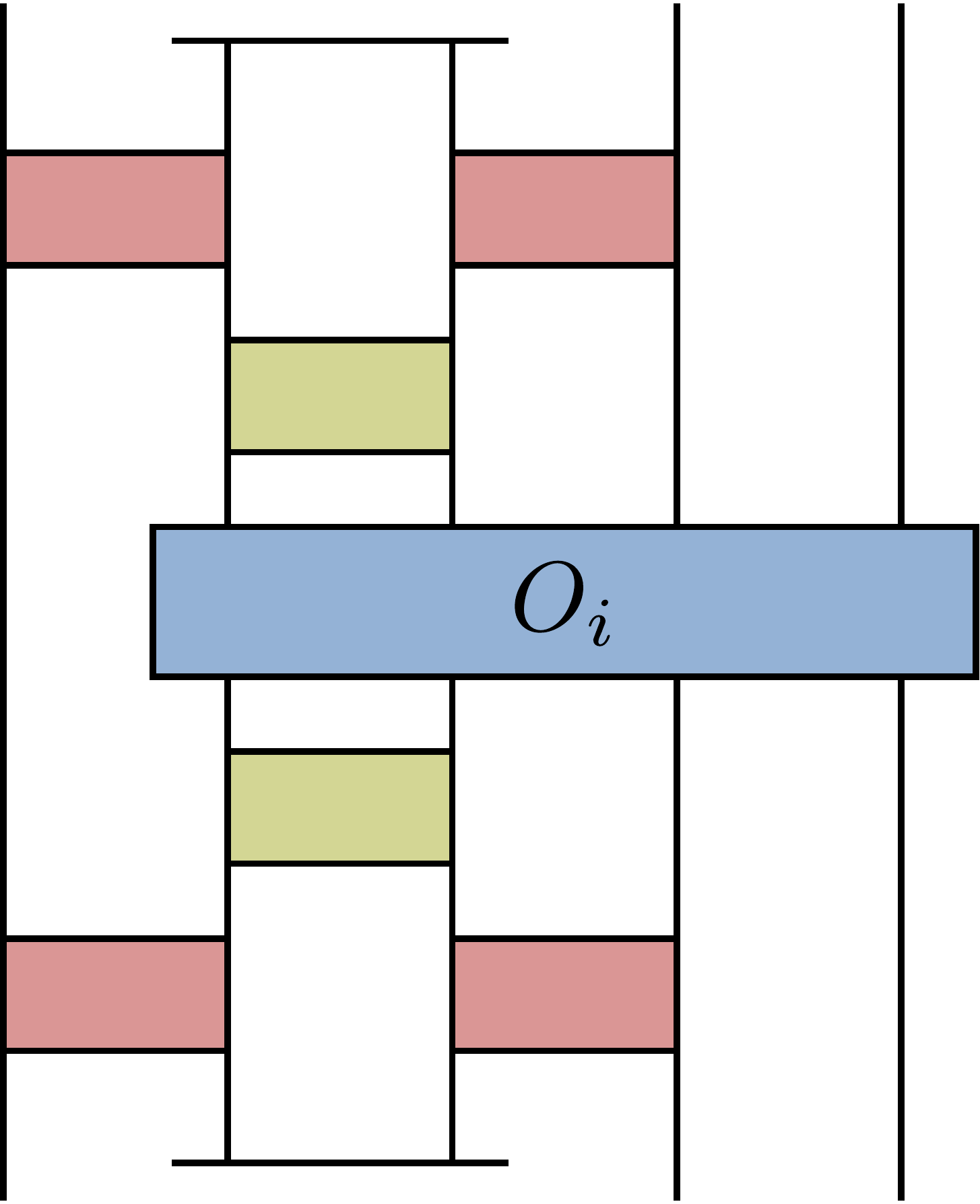} \vspace{0.2cm} \\
    (r)\includegraphics[scale=0.15]{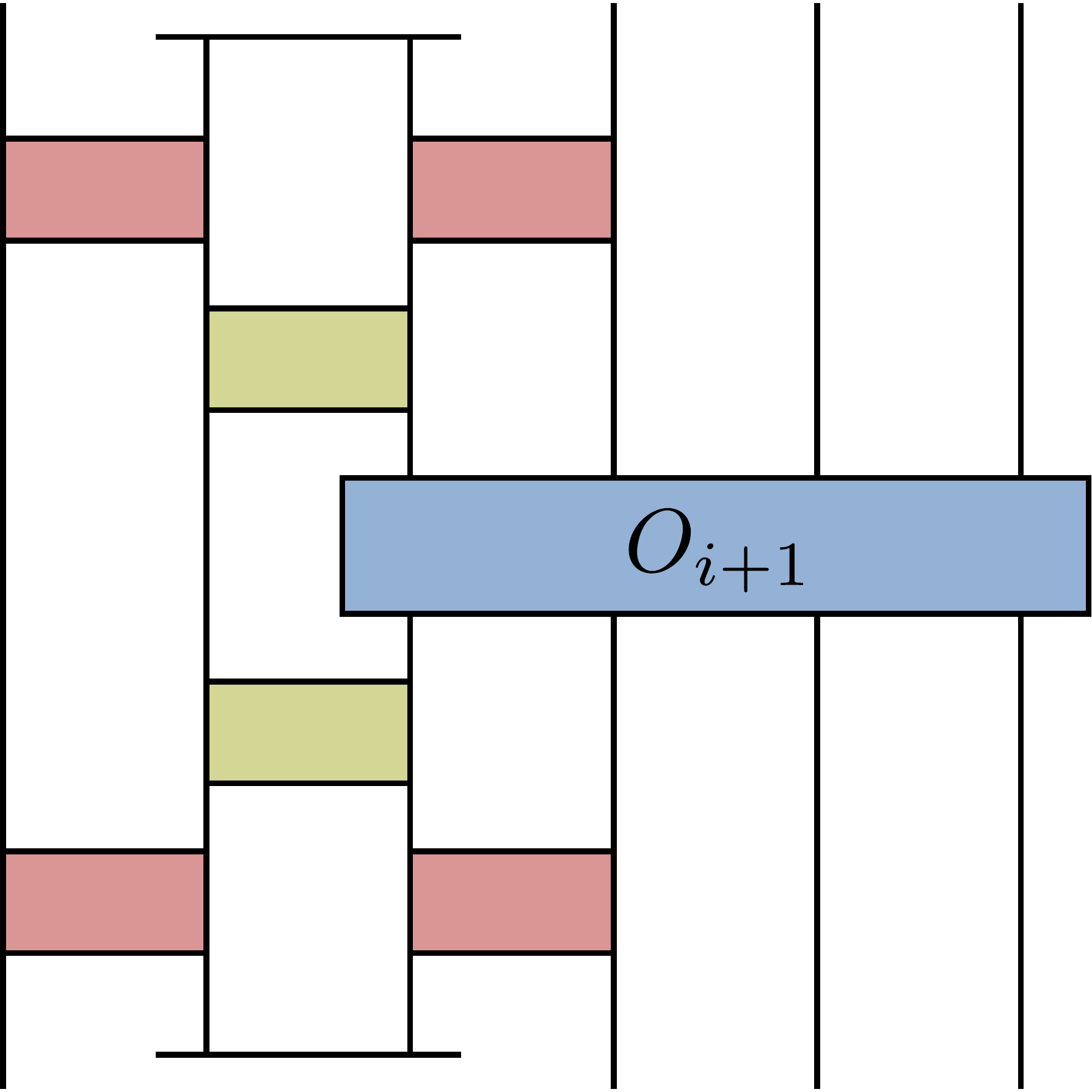} \hspace{0.05cm} 
    (s)\includegraphics[scale=0.15]{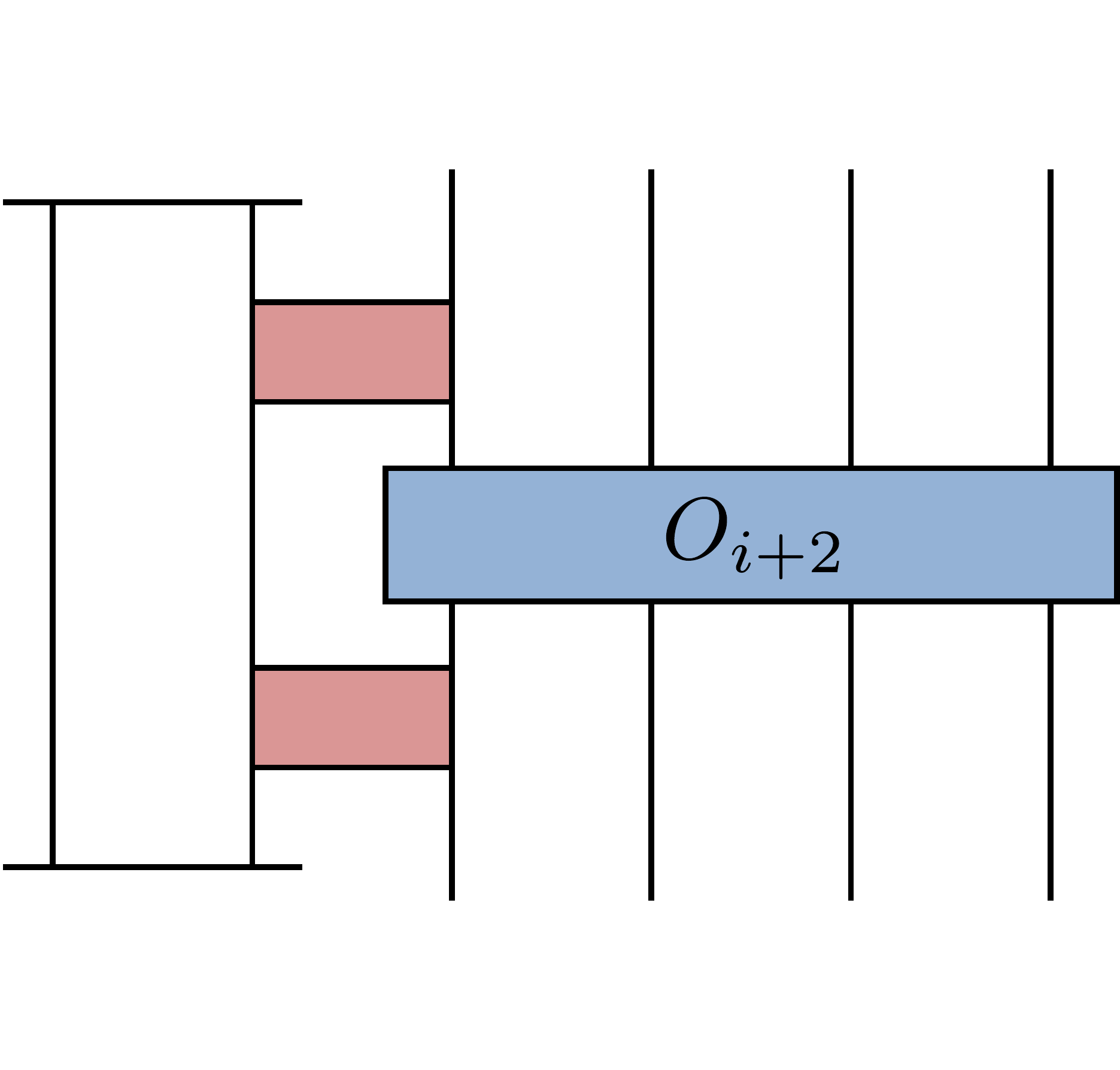} \hspace{0.05cm}
    (t)\includegraphics[scale=0.15]{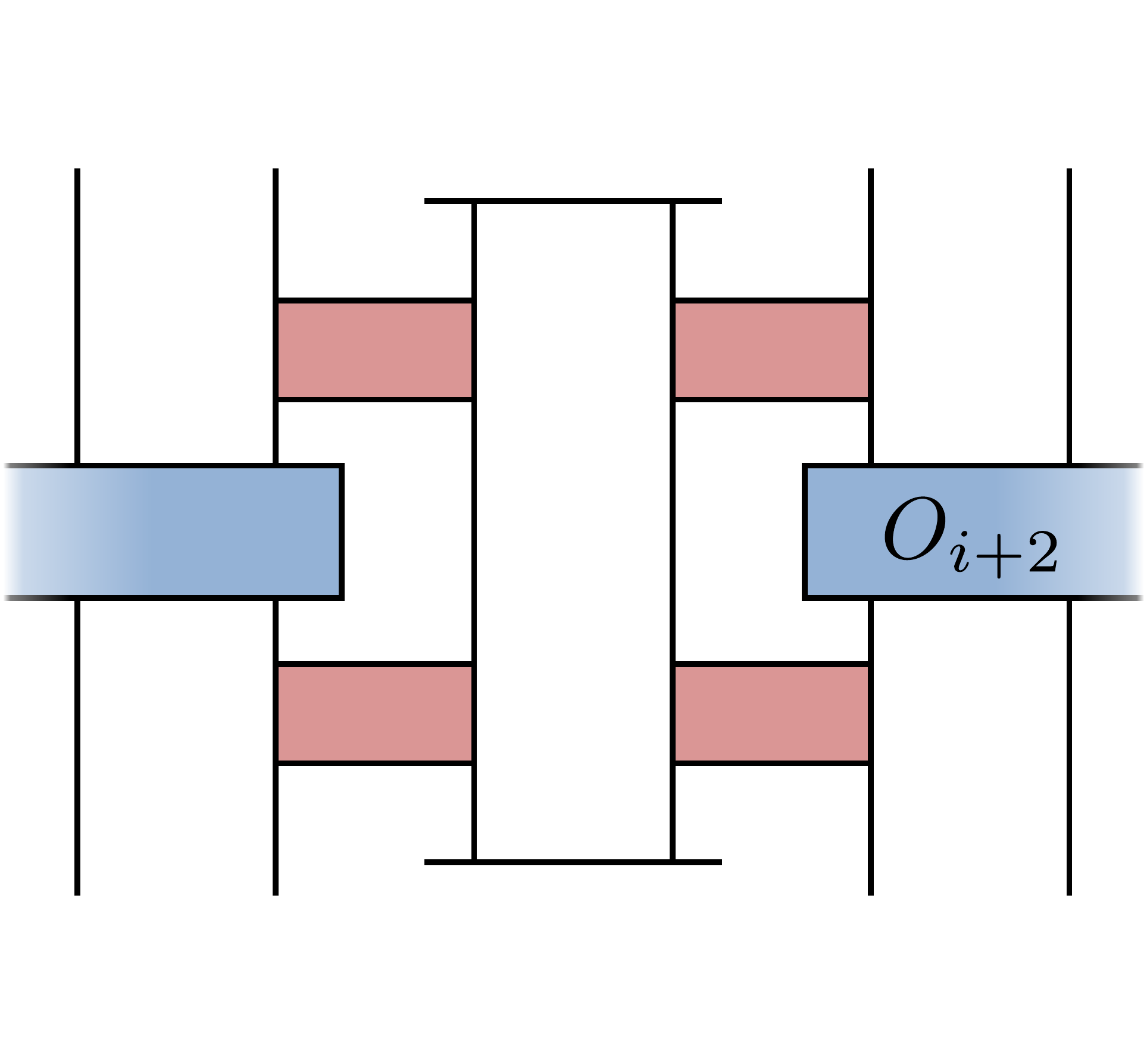}  
    \caption{
        Tensor network diagrams showing the raising of a pair of two-site operators for 
        (a) $i-1$ and $i+1$ ($\mathcal{O}(\chi^{8})$), 
        (b) $i-2$ and $i$ ($\mathcal{O}(\chi^{8})$), 
        (c) $i-2$ and $i+1$ ($\mathcal{O}(\chi^{8})$), 
        (d) $i-1$ and $i+2$ ($\mathcal{O}(\chi^{8})$), 
        (e) $i$ and $i+2$ ($\mathcal{O}(\chi^{8})$), and
        (f) $i+2$ and $i-2$ ($\mathcal{O}(\chi^{8})$).
        Raising of a three-site operator to the next level of coarse-graining for 
        (g) $i-3$ ($\mathcal{O}(\chi^{8})$), 
        (h) $i-2$ ($\mathcal{O}(\chi^{9})$), 
        (i) $i-1$ ($\mathcal{O}(\chi^{8})$), 
        (j) $i$ ($\mathcal{O}(\chi^{8})$), 
        (k) $i+1$ ($\mathcal{O}(\chi^{9})$), and
        (l) $i+2$ ($\mathcal{O}(\chi^{8})$).
        Raising of a four-site operator for 
        (m) $i-4$ ($\mathcal{O}(\chi^{10})$), 
        (n) $i-3$ ($\mathcal{O}(\chi^{11})$), 
        (o) $i-2$ ($\mathcal{O}(\chi^{10})$), 
        (p) $i-1$ ($\mathcal{O}(\chi^{10})$), 
        (q) $i$ ($\mathcal{O}(\chi^{10})$), 
        (r) $i+1$ ($\mathcal{O}(\chi^{11})$), 
        (s) $i+2$ ($\mathcal{O}(\chi^{10})$), and the PBC case
        (t) $i-2$ or $i+2$ when the system is six sites in size ($\mathcal{O}(\chi^{10})$).
    \label{fig:corr_corr}
    \label{fig:corr_make3_1}
    \label{fig:corr_make3_2}
    \label{fig:corr_make3_3}
    \label{fig:corr_make3_4}
    \label{fig:corr_make4}
    \label{fig:corr_prop3_1}
    \label{fig:corr_prop3_2}
    \label{fig:corr_prop3_3}
    \label{fig:corr_prop3_4}
    \label{fig:corr_prop3_5}
    \label{fig:corr_prop3_6}
    \label{fig:corr_prop4_1}
    \label{fig:corr_prop4_2}
    \label{fig:corr_prop4_3}
    \label{fig:corr_prop4_4}
    \label{fig:corr_prop4_5}
    \label{fig:corr_prop4_6}
    \label{fig:corr_prop4_7}
    \label{fig:corr_prop4_PBC6}
    }
\end{figure*}
\begin{figure*}
    (a)\includegraphics[scale=0.15]{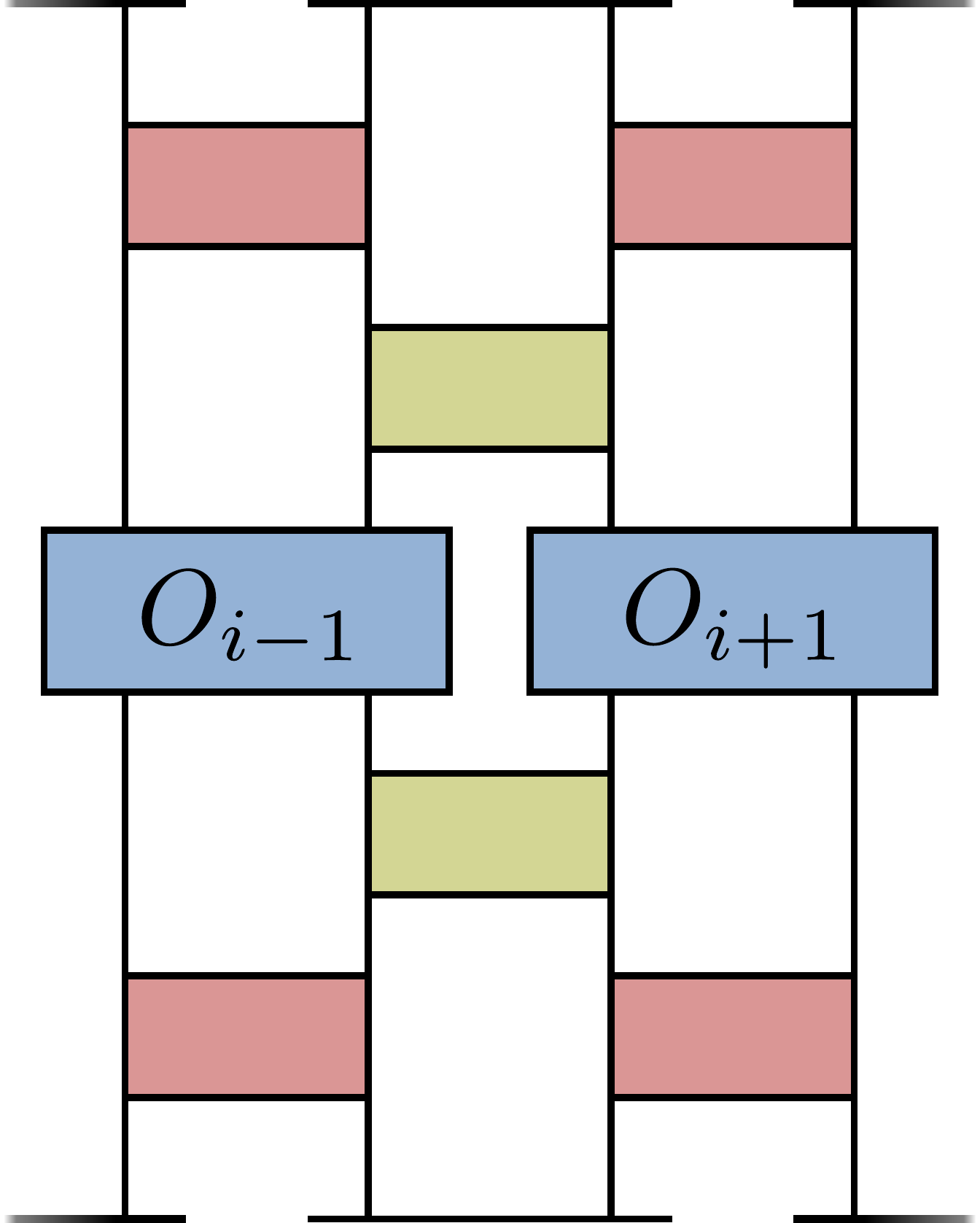} \hspace{0.05cm} 
    (b)\includegraphics[scale=0.15]{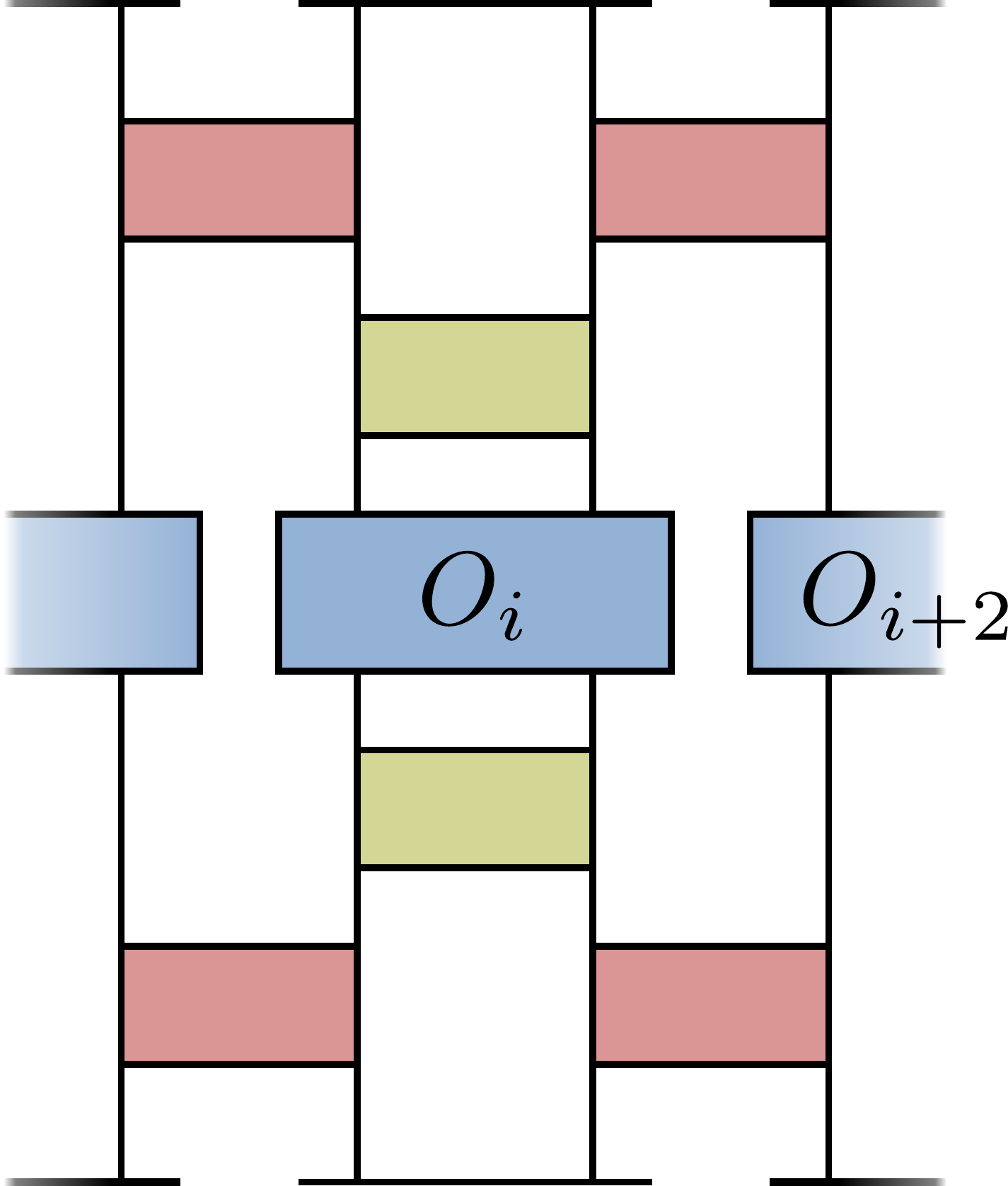} \hspace{0.05cm}
    (c)\includegraphics[scale=0.15]{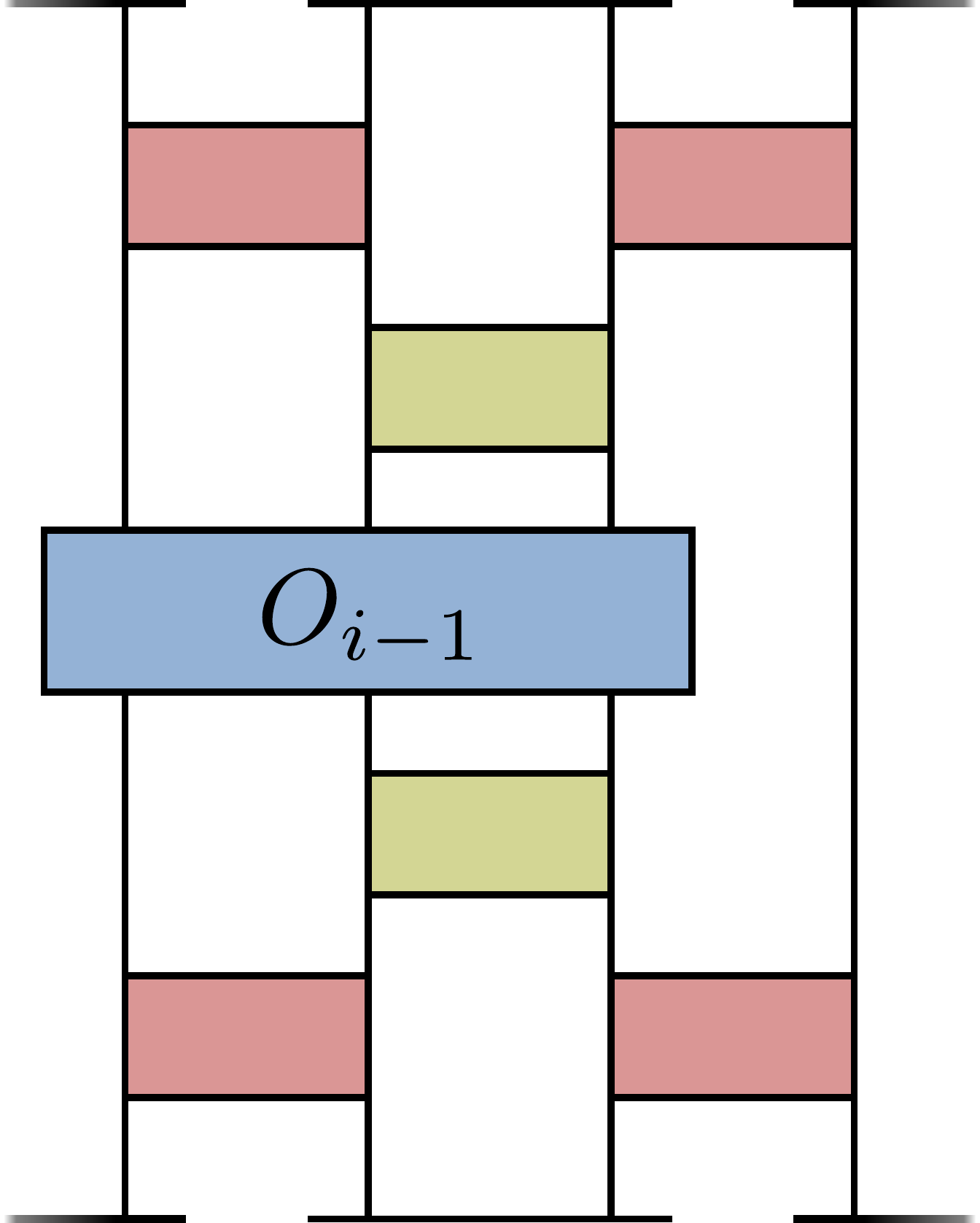} \hspace{0.05cm}
    (d)\includegraphics[scale=0.15]{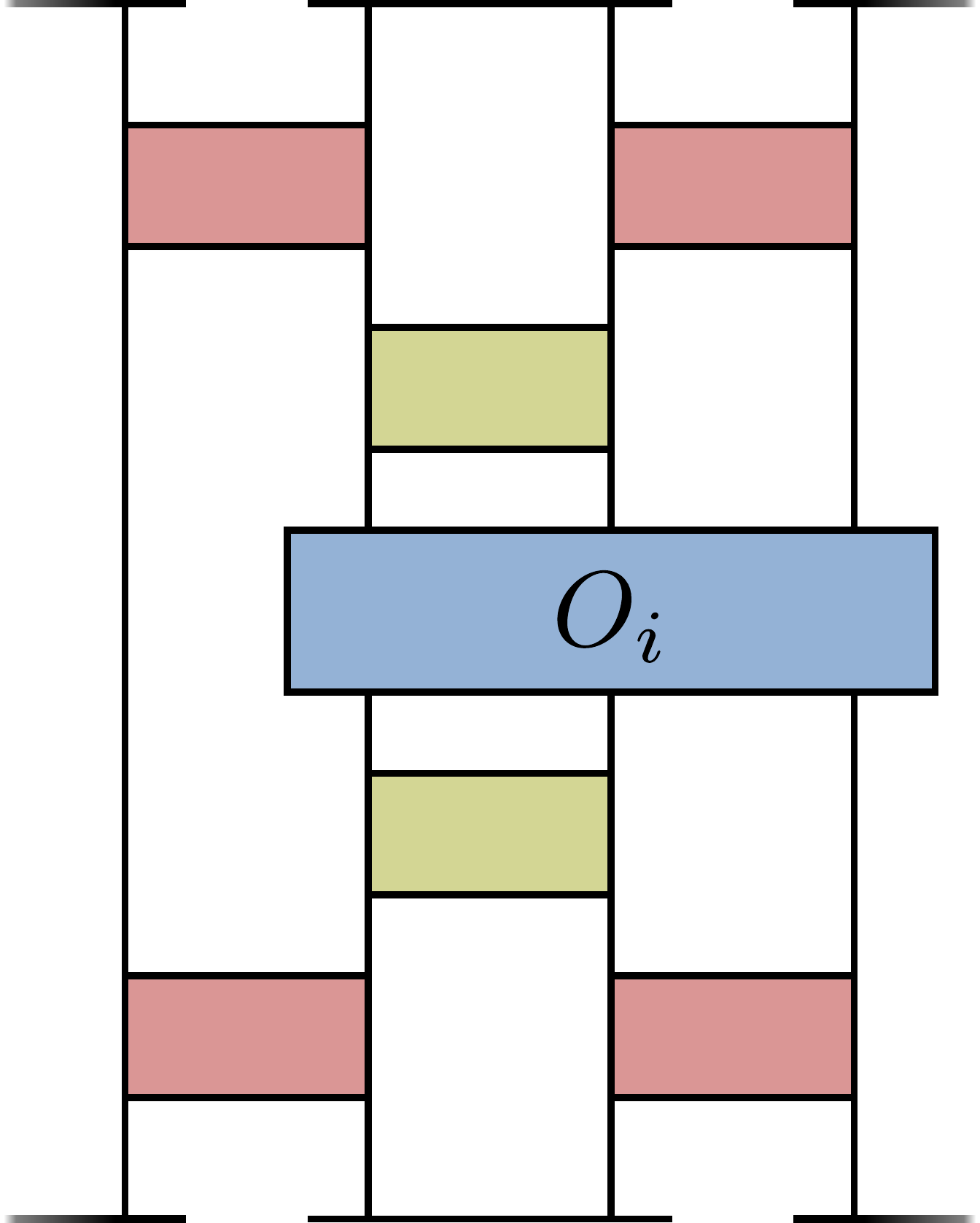} \hspace{0.05cm}
    (e)\includegraphics[scale=0.15]{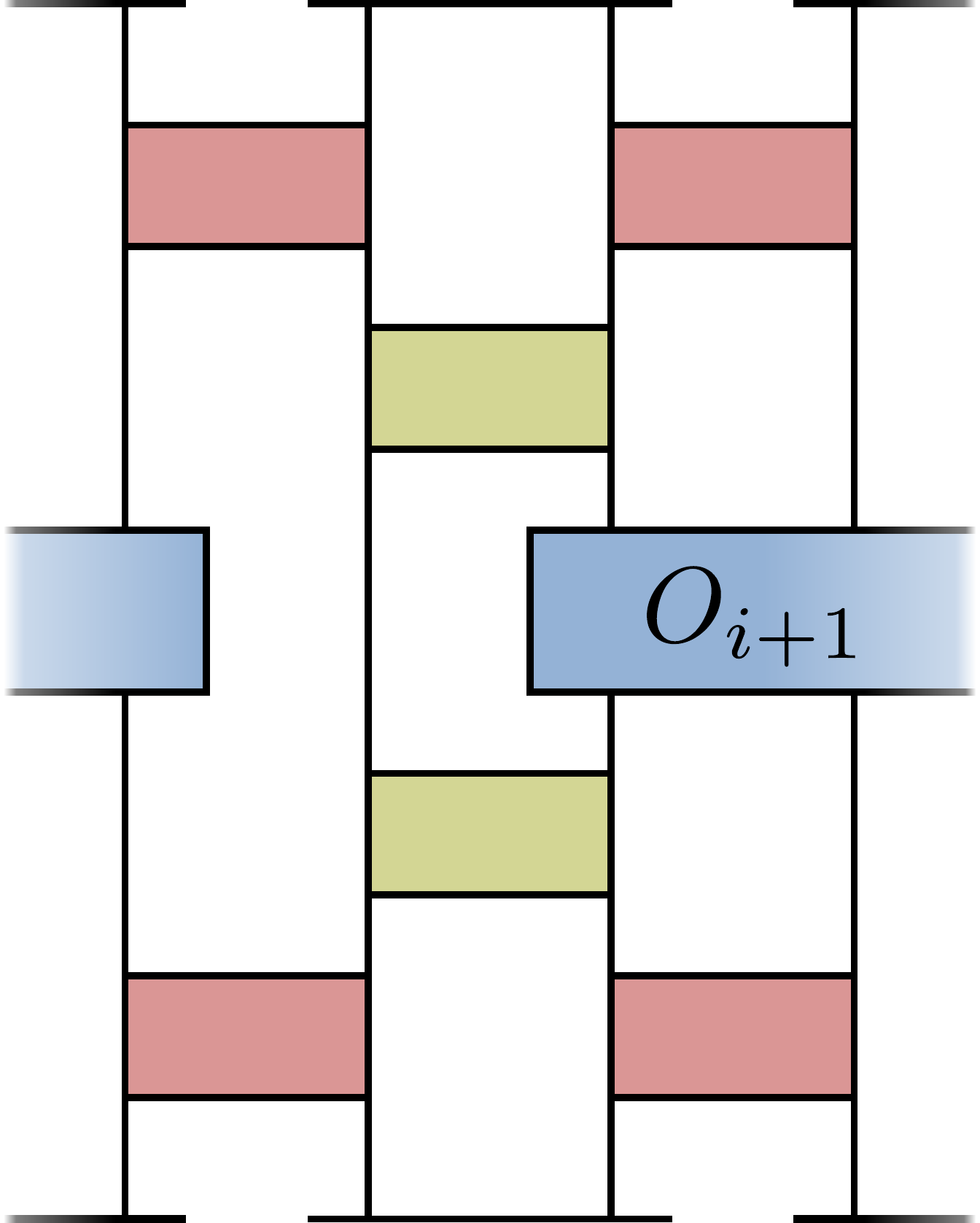} \hspace{0.05cm} 
    (f)\includegraphics[scale=0.15]{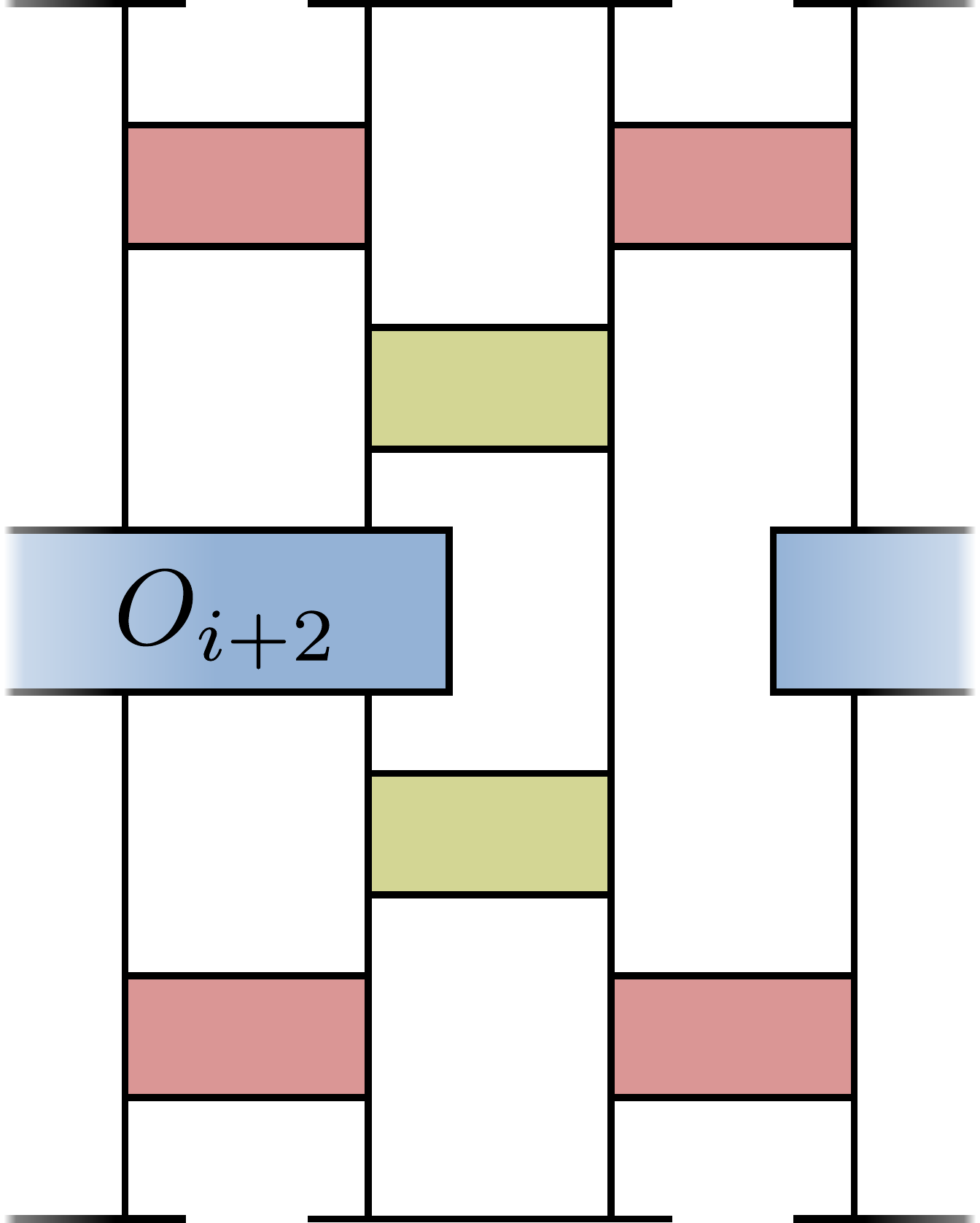} \vspace{0.4cm} \\
    (g)\includegraphics[scale=0.15]{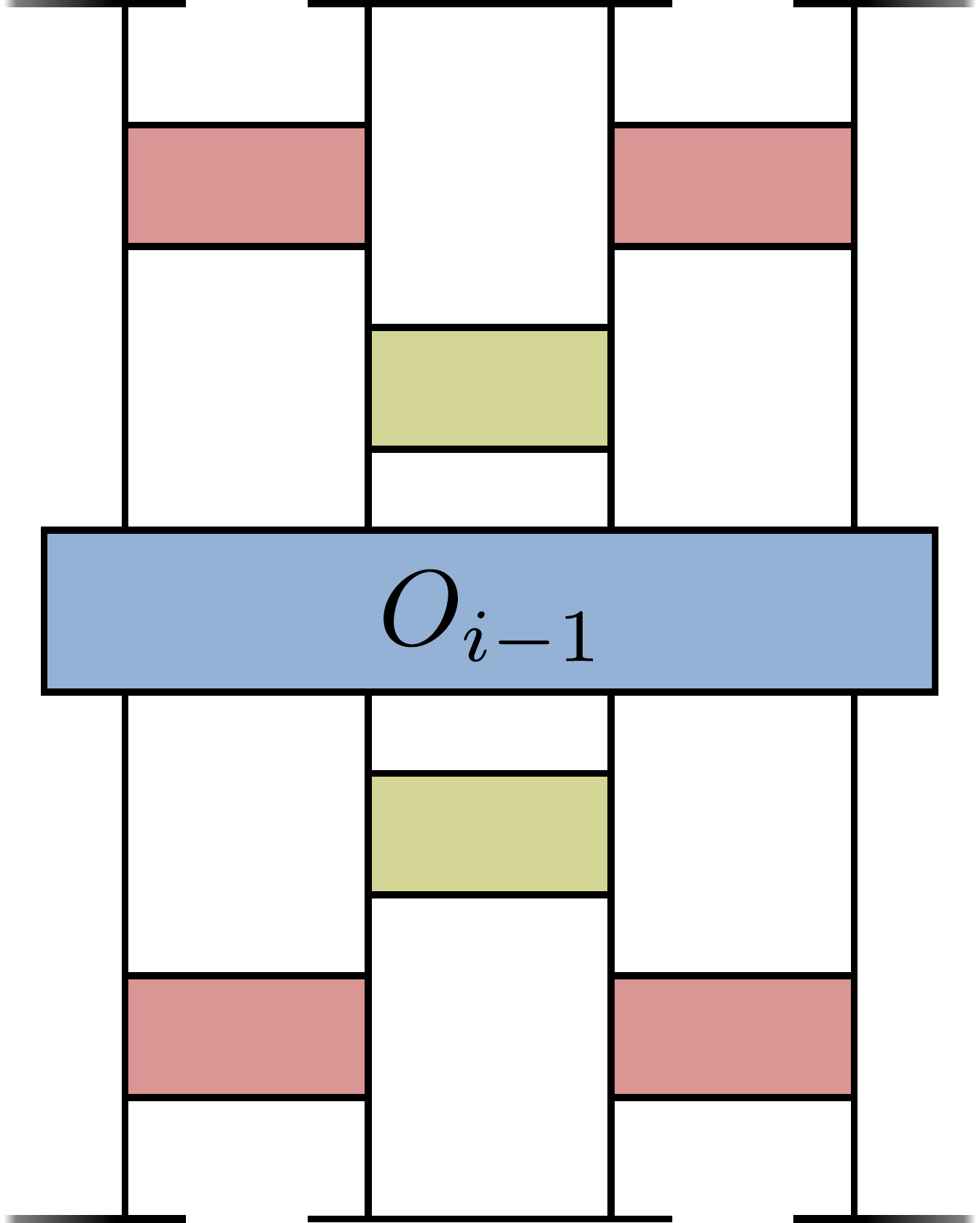} \hspace{0.05cm}
    (h)\includegraphics[scale=0.15]{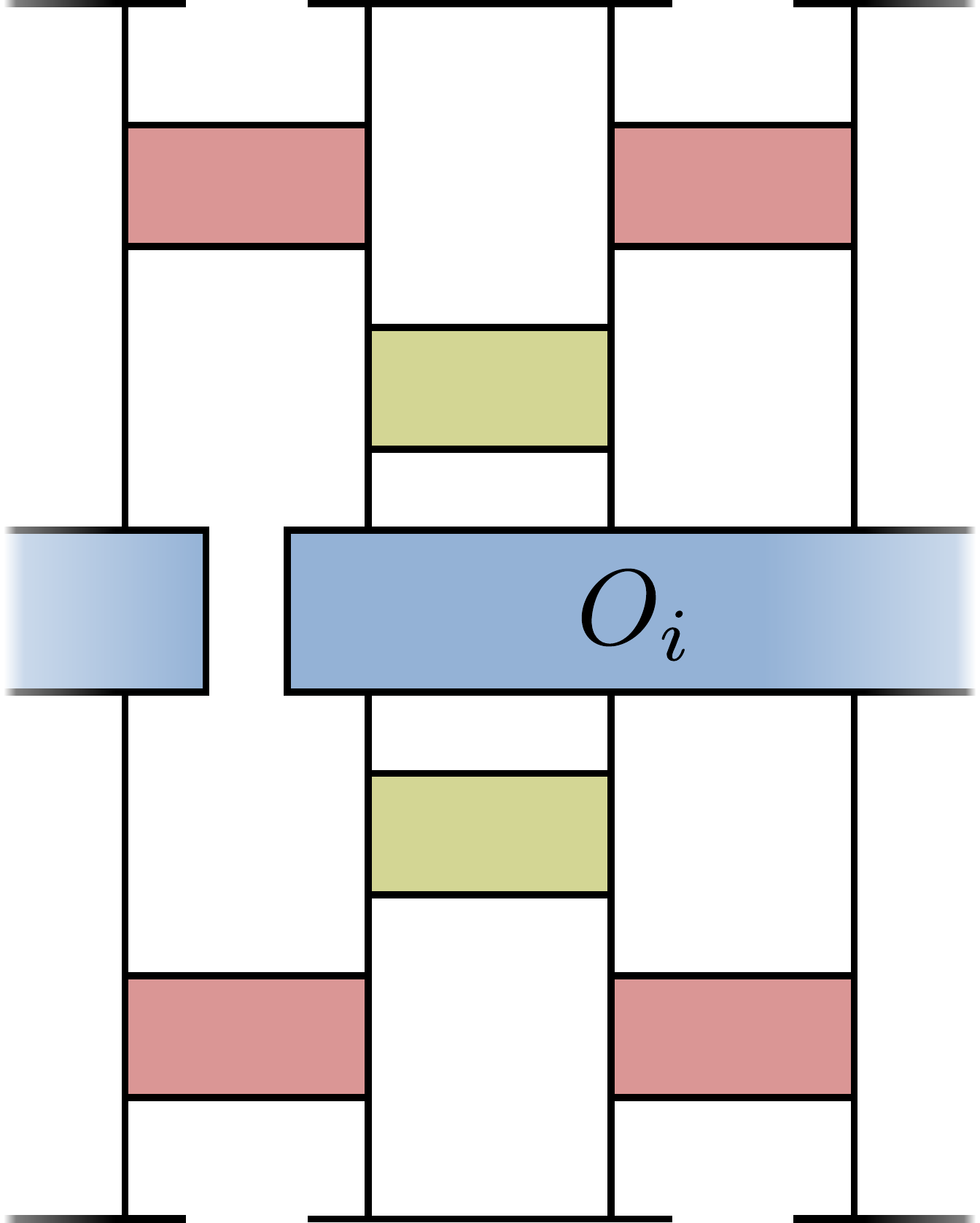} \hspace{0.05cm}  
    (i)\includegraphics[scale=0.15]{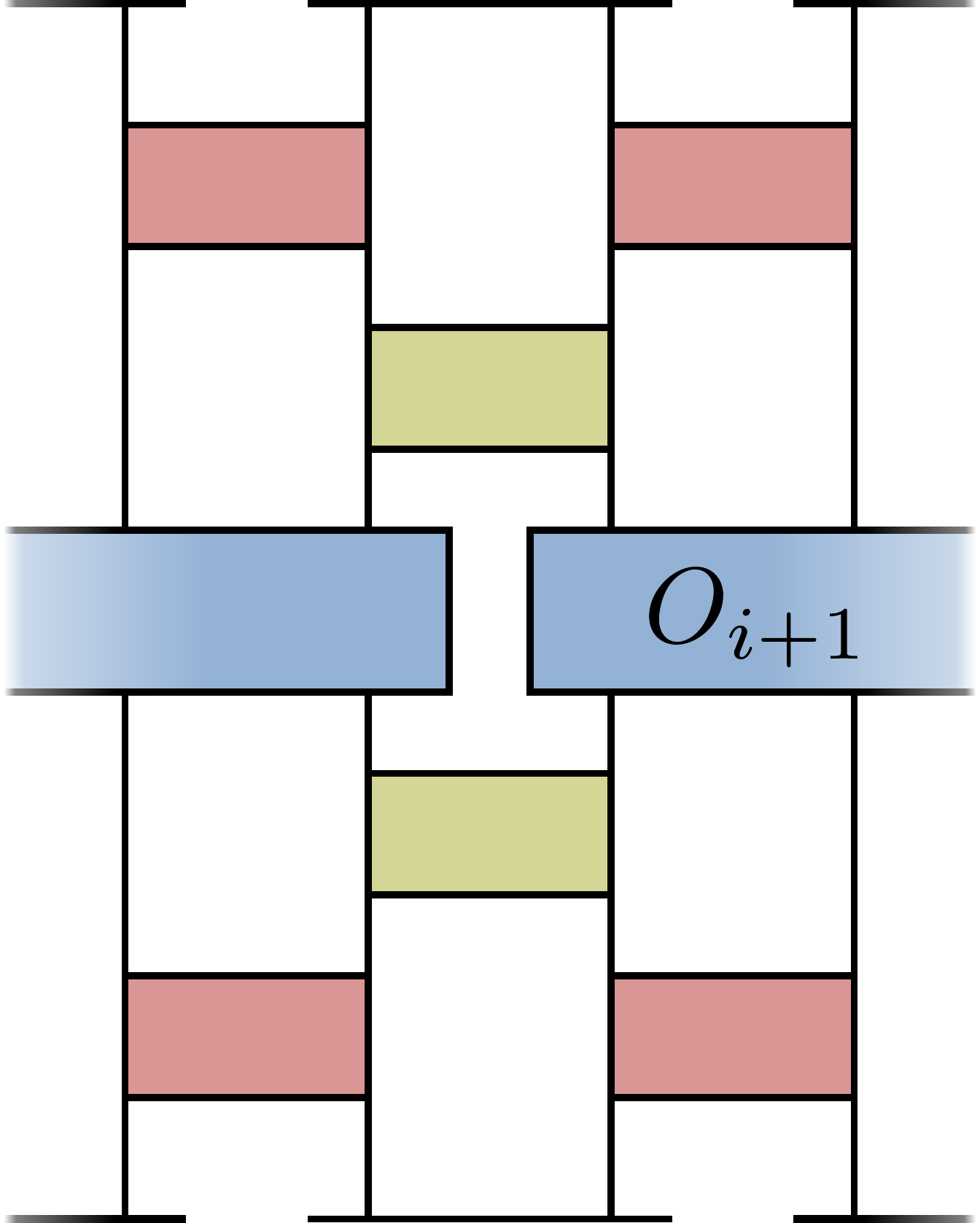} \hspace{0.05cm}
    (j)\includegraphics[scale=0.15]{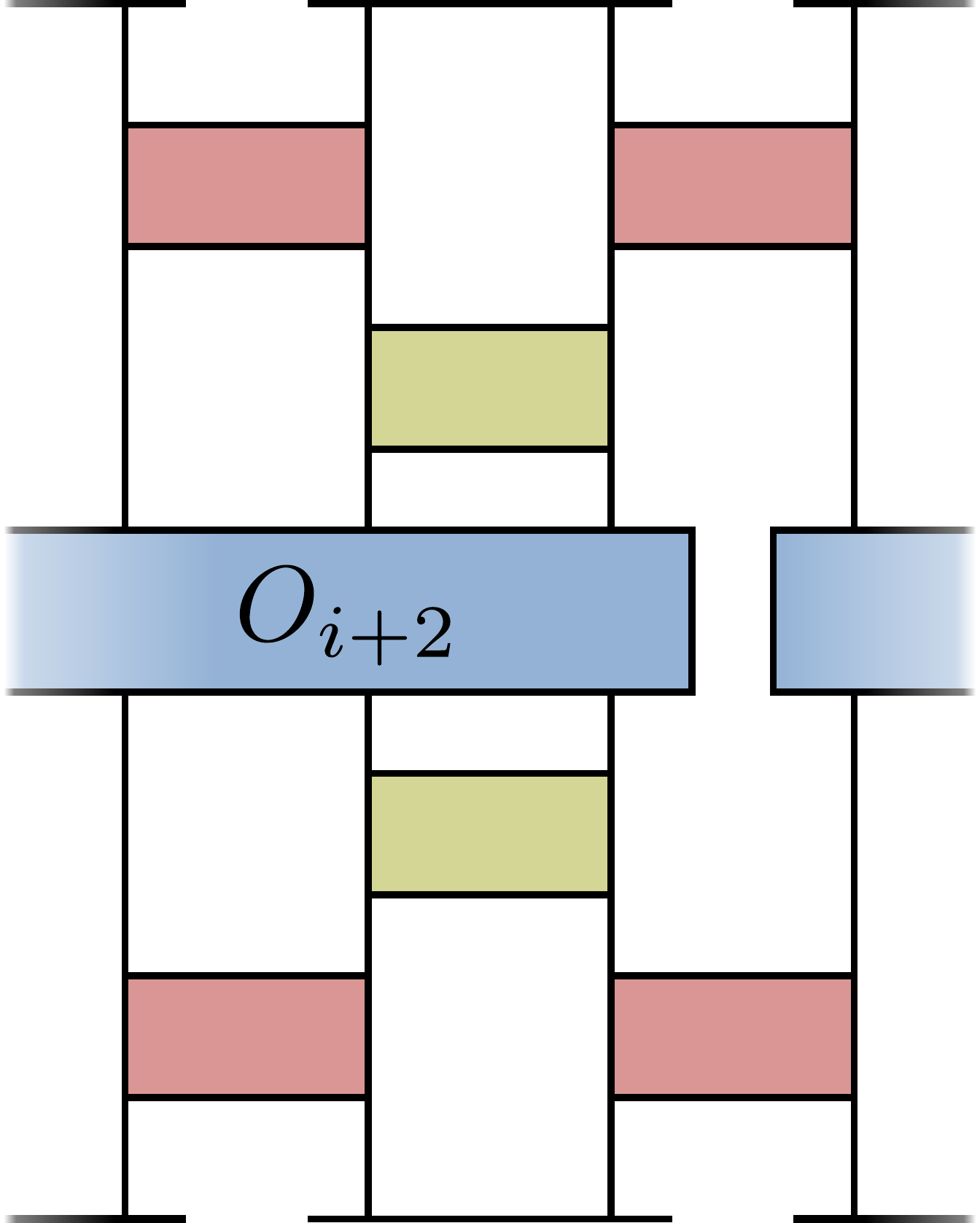} 
    \caption{
        Tensor network diagrams showing the contraction of the correlation functions at the top of the network for: A pair of two-site operators at sites
        (a) $i-1$ and $i+1$ ($\mathcal{O}(\chi^{6})$), and 
        (b) $i$ and $i+2$ ($\mathcal{O}(\chi^{6})$).
        A three-site operator at site
        (c) $i-1$ ($\mathcal{O}(\chi^{7})$),
        (d) $i$ ($\mathcal{O}(\chi^{7})$), 
        (e) $i+1$ ($\mathcal{O}(\chi^{7})$), and
        (f) $i+2$ ($\mathcal{O}(\chi^{7})$).
        A four-site operator at
        (g) $i-1$ ($\mathcal{O}(\chi^{8})$),
        (h) $i$ ($\mathcal{O}(\chi^{8})$), 
        (i) $i+1$ ($\mathcal{O}(\chi^{8})$), and
        (j) $i+2$ ($\mathcal{O}(\chi^{8})$).
    \label{fig:corr_corr_PBC1_top}
    \label{fig:corr_corr_PBC2_top}
    \label{fig:corr_prop3_PBC1_top}
    \label{fig:corr_prop3_PBC2_top}
    \label{fig:corr_prop3_PBC3_top}
    \label{fig:corr_prop3_PBC4_top}
    \label{fig:corr_prop4_PBC1_top}
    \label{fig:corr_prop4_PBC2_top}
    \label{fig:corr_prop4_PBC3_top}
    \label{fig:corr_prop4_PBC4_top}
    }
\end{figure*}

\subsection{\label{sec:ee}Entanglement entropy}
\begin{figure}
    \includegraphics[width=0.7\columnwidth]{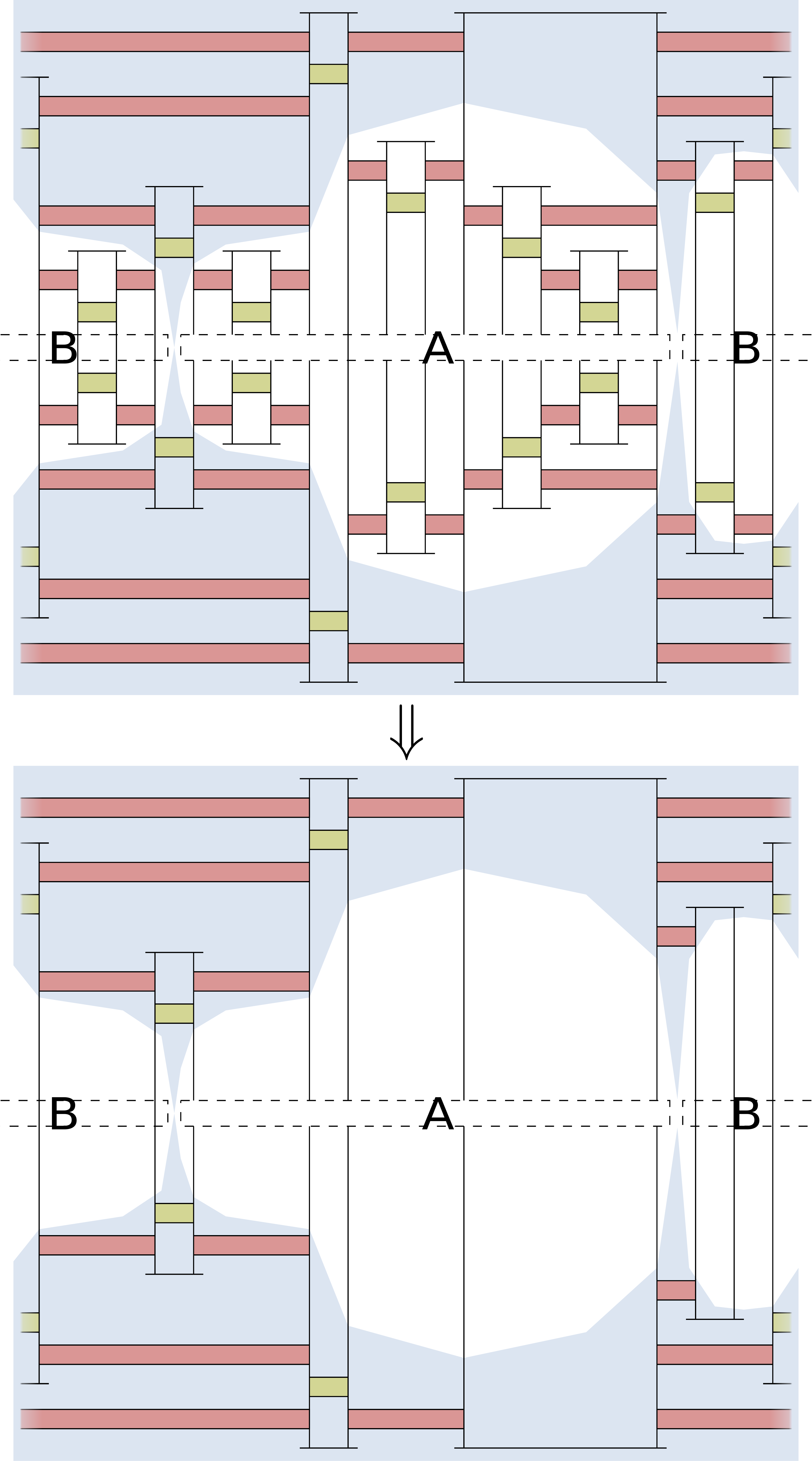}
    \caption{
        Tensor network diagram showing the calculation of entanglement entropy via a reduced density matrix.
        In a similar manner to the correlation function in Fig.\ \ref{fig:dMERA_corr_full}, the tensors that are completely within either subsystem to not need to be contracted as they do not contribute to the entanglement entropy.
    \label{fig:dMERA_dm_full}
    }
\end{figure}
The entanglement entropy $S_{A}$ of a region $A$ given by
\begin{equation}
    S_{A} = -\text{Tr} \rho_{A} \text{log}_{2} \rho_{A},
    \label{eq:entropy}
\end{equation}
where $\rho_{A}$ is the reduced density matrix, obtained by tracing over region $B$.
The contraction of $\rho_{A}$ in general is inefficient as the computational complexity grows exponentially with system size.
The construction of the dMERA wavefunction makes the problem significantly easier as we only need to contract the tensors that join blocks $A$ and $B$ \cite{GolR14,TagEV09}, as illustrated by Fig.\ \ref{fig:dMERA_dm_full}. 
This makes it possible to calculate $S_{A}$ when $\chi$ and $L$ are relatively small.
We note that in all cases the entanglement is still captured by the tensor network and it is just the calculation of the entropy that is problematic.

\section{\label{sec:XX_update}Energy minimization of the XX model}
As discussed in sec.\ \ref{sec:XX_results}, dMERA for the $\chi = 2$ XX model has a simplified form where the $u$ tensors have no variables and the $w_{L,R}$ have one angle each.
The number of parameters in this tensor network is then $L - 3$.
This number comes from the fact that there are $L/2 - 1$ coarse-graining blocks as the top one is trivial, each containing two $w$ tensors.
The final pair of angles are also correlated; the difference between the two being the only free parameter.
The update process can be performed by simply differentiating to find the minimum energy.
For example, we require the minimum energy
\begin{equation}
    E_{\text{min}} = \min_{\theta} \text{tr} \left[ \Upsilon w(\theta) \right].
\end{equation}
Writing the environment $\Upsilon$, and $w$ as matrices as in Eq.\ (\ref{eqn:XX_w_matrix}), the energy can be written as
\begin{align}
    E &= \text{tr} \left[ \Upsilon w(\theta) \right] \nonumber \\
      &= \Upsilon_{1,1} + \Upsilon_{4,4} + \left( \Upsilon_{2,2} + \Upsilon_{3,3} \right) \cos{\theta} \nonumber \\
      &\qquad + \left( \Upsilon_{2,3} - \Upsilon_{3,2} \right) \sin{\theta},
\end{align}
where $\Upsilon_{i,j}$ are elements of the environment.
Differentiation shows that the stationary points are at
\begin{equation}
    \tan{\theta} = \frac{ \Upsilon_{2,3} - \Upsilon_{3,2} }{ \Upsilon_{2,2} + \Upsilon_{3,3}},
\end{equation}
and the second derivative is used to find if it is a maximum or minimum.
The environment is then rebuilt with the new $w$ and the process is repeated until convergence.

%


\begin{thebibliography}{45}%
\makeatletter
\providecommand \@ifxundefined [1]{%
 \@ifx{#1\undefined}
}%
\providecommand \@ifnum [1]{%
 \ifnum #1\expandafter \@firstoftwo
 \else \expandafter \@secondoftwo
 \fi
}%
\providecommand \@ifx [1]{%
 \ifx #1\expandafter \@firstoftwo
 \else \expandafter \@secondoftwo
 \fi
}%
\providecommand \natexlab [1]{#1}%
\providecommand \enquote  [1]{``#1''}%
\providecommand \bibnamefont  [1]{#1}%
\providecommand \bibfnamefont [1]{#1}%
\providecommand \citenamefont [1]{#1}%
\providecommand \href@noop [0]{\@secondoftwo}%
\providecommand \href [0]{\begingroup \@sanitize@url \@href}%
\providecommand \@href[1]{\@@startlink{#1}\@@href}%
\providecommand \@@href[1]{\endgroup#1\@@endlink}%
\providecommand \@sanitize@url [0]{\catcode `\\12\catcode `\$12\catcode
  `\&12\catcode `\#12\catcode `\^12\catcode `\_12\catcode `\%12\relax}%
\providecommand \@@startlink[1]{}%
\providecommand \@@endlink[0]{}%
\providecommand \url  [0]{\begingroup\@sanitize@url \@url }%
\providecommand \@url [1]{\endgroup\@href {#1}{\urlprefix }}%
\providecommand \urlprefix  [0]{URL }%
\providecommand \Eprint [0]{\href }%
\providecommand \doibase [0]{http://dx.doi.org/}%
\providecommand \selectlanguage [0]{\@gobble}%
\providecommand \bibinfo  [0]{\@secondoftwo}%
\providecommand \bibfield  [0]{\@secondoftwo}%
\providecommand \translation [1]{[#1]}%
\providecommand \BibitemOpen [0]{}%
\providecommand \bibitemStop [0]{}%
\providecommand \bibitemNoStop [0]{.\EOS\space}%
\providecommand \EOS [0]{\spacefactor3000\relax}%
\providecommand \BibitemShut  [1]{\csname bibitem#1\endcsname}%
\let\auto@bib@innerbib\@empty
\bibitem [{\citenamefont {Anderson}(1958)}]{And58}%
  \BibitemOpen
  \bibfield  {author} {\bibinfo {author} {\bibfnamefont {P.~W.}\ \bibnamefont
  {Anderson}},\ }\href@noop {} {\bibfield  {journal} {\bibinfo  {journal}
  {Phys. Rev.}\ }\textbf {\bibinfo {volume} {109}},\ \bibinfo {pages} {1492}
  (\bibinfo {year} {1958})}\BibitemShut {NoStop}%
\bibitem [{\citenamefont {Basko}\ \emph {et~al.}(2006)\citenamefont {Basko},
  \citenamefont {Aleiner},\ and\ \citenamefont {Altshuler}}]{BasAA06}%
  \BibitemOpen
  \bibfield  {author} {\bibinfo {author} {\bibfnamefont {D.}~\bibnamefont
  {Basko}}, \bibinfo {author} {\bibfnamefont {I.}~\bibnamefont {Aleiner}}, \
  and\ \bibinfo {author} {\bibfnamefont {B.}~\bibnamefont {Altshuler}},\ }\href
  {\doibase http://dx.doi.org/10.1016/j.aop.2005.11.014} {\bibfield  {journal}
  {\bibinfo  {journal} {Ann. Phys.}\ }\textbf {\bibinfo {volume} {321}},\
  \bibinfo {pages} {1126 } (\bibinfo {year} {2006})}\BibitemShut {NoStop}%
\bibitem [{\citenamefont {Nandkishore}\ and\ \citenamefont
  {Huse}(2015)}]{NanH15}%
  \BibitemOpen
  \bibfield  {author} {\bibinfo {author} {\bibfnamefont {R.}~\bibnamefont
  {Nandkishore}}\ and\ \bibinfo {author} {\bibfnamefont {D.~A.}\ \bibnamefont
  {Huse}},\ }\href {\doibase 10.1146/annurev-conmatphys-031214-014726}
  {\bibfield  {journal} {\bibinfo  {journal} {Annu. Rev. Condens. Matter
  Phys.}\ }\textbf {\bibinfo {volume} {6}},\ \bibinfo {pages} {15} (\bibinfo
  {year} {2015})}\BibitemShut {NoStop}%
\bibitem [{\citenamefont {Altman}\ and\ \citenamefont {Vosk}(2015)}]{AltV15}%
  \BibitemOpen
  \bibfield  {author} {\bibinfo {author} {\bibfnamefont {E.}~\bibnamefont
  {Altman}}\ and\ \bibinfo {author} {\bibfnamefont {R.}~\bibnamefont {Vosk}},\
  }\href {\doibase 10.1146/annurev-conmatphys-031214-014701} {\bibfield
  {journal} {\bibinfo  {journal} {Annu. Rev. Condens. Matter Phys.}\ }\textbf
  {\bibinfo {volume} {6}},\ \bibinfo {pages} {383} (\bibinfo {year}
  {2015})}\BibitemShut {NoStop}%
\bibitem [{\citenamefont {Abanin}\ and\ \citenamefont
  {Papi\'{c}}(2017)}]{AbaP17}%
  \BibitemOpen
  \bibfield  {author} {\bibinfo {author} {\bibfnamefont {D.~A.}\ \bibnamefont
  {Abanin}}\ and\ \bibinfo {author} {\bibfnamefont {Z.}~\bibnamefont
  {Papi\'{c}}},\ }\href {\doibase 10.1002/andp.201700169} {\bibfield  {journal}
  {\bibinfo  {journal} {Ann. Phys. (Berlin)}\ }\textbf {\bibinfo {volume}
  {529}},\ \bibinfo {pages} {1700169} (\bibinfo {year} {2017})}\BibitemShut
  {NoStop}%
\bibitem [{\citenamefont {White}(1992)}]{Whi92}%
  \BibitemOpen
  \bibfield  {author} {\bibinfo {author} {\bibfnamefont {S.~R.}\ \bibnamefont
  {White}},\ }\href {\doibase 10.1103/PhysRevLett.69.2863} {\bibfield
  {journal} {\bibinfo  {journal} {Phys. Rev. Lett.}\ }\textbf {\bibinfo
  {volume} {69}},\ \bibinfo {pages} {2863} (\bibinfo {year}
  {1992})}\BibitemShut {NoStop}%
\bibitem [{\citenamefont {{{\"O}stlund}}\ and\ \citenamefont
  {Rommer}(1995)}]{OstR95}%
  \BibitemOpen
  \bibfield  {author} {\bibinfo {author} {\bibfnamefont {S.}~\bibnamefont
  {{{\"O}stlund}}}\ and\ \bibinfo {author} {\bibfnamefont {S.}~\bibnamefont
  {Rommer}},\ }\href@noop {} {\bibfield  {journal} {\bibinfo  {journal} {Phys.
  Rev. Lett.}\ }\textbf {\bibinfo {volume} {75}},\ \bibinfo {pages} {3537}
  (\bibinfo {year} {1995})}\BibitemShut {NoStop}%
\bibitem [{\citenamefont {Schollw\"ock}(2005)}]{Sch05}%
  \BibitemOpen
  \bibfield  {author} {\bibinfo {author} {\bibfnamefont {U.}~\bibnamefont
  {Schollw\"ock}},\ }\href {\doibase 10.1103/RevModPhys.77.259} {\bibfield
  {journal} {\bibinfo  {journal} {Rev. Mod. Phys.}\ }\textbf {\bibinfo {volume}
  {77}},\ \bibinfo {pages} {259} (\bibinfo {year} {2005})}\BibitemShut
  {NoStop}%
\bibitem [{\citenamefont {Schollw{\"{o}}ck}(2011)}]{Sch11}%
  \BibitemOpen
  \bibfield  {author} {\bibinfo {author} {\bibfnamefont {U.}~\bibnamefont
  {Schollw{\"{o}}ck}},\ }\href {\doibase 10.1016/j.aop.2010.09.012} {\bibfield
  {journal} {\bibinfo  {journal} {Ann. Phys.}\ }\textbf {\bibinfo {volume}
  {326}},\ \bibinfo {pages} {96} (\bibinfo {year} {2011})}\BibitemShut
  {NoStop}%
\bibitem [{\citenamefont {Juozapavi\ifmmode~\check{c}\else \v{c}\fi{}ius}\
  \emph {et~al.}(1997)\citenamefont {Juozapavi\ifmmode~\check{c}\else
  \v{c}\fi{}ius}, \citenamefont {Caprara},\ and\ \citenamefont
  {Rosengren}}]{JuoCR97}%
  \BibitemOpen
  \bibfield  {author} {\bibinfo {author} {\bibfnamefont {A.}~\bibnamefont
  {Juozapavi\ifmmode~\check{c}\else \v{c}\fi{}ius}}, \bibinfo {author}
  {\bibfnamefont {S.}~\bibnamefont {Caprara}}, \ and\ \bibinfo {author}
  {\bibfnamefont {A.}~\bibnamefont {Rosengren}},\ }\href {\doibase
  10.1103/PhysRevB.56.11097} {\bibfield  {journal} {\bibinfo  {journal} {Phys.
  Rev. B}\ }\textbf {\bibinfo {volume} {56}},\ \bibinfo {pages} {11097}
  (\bibinfo {year} {1997})}\BibitemShut {NoStop}%
\bibitem [{\citenamefont {Juozapavi\ifmmode~\check{c}\else \v{c}\fi{}ius}\
  \emph {et~al.}(1999)\citenamefont {Juozapavi\ifmmode~\check{c}\else
  \v{c}\fi{}ius}, \citenamefont {Urba}, \citenamefont {Caprara},\ and\
  \citenamefont {Rosengren}}]{JuoUCR99}%
  \BibitemOpen
  \bibfield  {author} {\bibinfo {author} {\bibfnamefont {A.}~\bibnamefont
  {Juozapavi\ifmmode~\check{c}\else \v{c}\fi{}ius}}, \bibinfo {author}
  {\bibfnamefont {L.}~\bibnamefont {Urba}}, \bibinfo {author} {\bibfnamefont
  {S.}~\bibnamefont {Caprara}}, \ and\ \bibinfo {author} {\bibfnamefont
  {A.}~\bibnamefont {Rosengren}},\ }\href {\doibase 10.1103/PhysRevB.60.14771}
  {\bibfield  {journal} {\bibinfo  {journal} {Phys. Rev. B}\ }\textbf {\bibinfo
  {volume} {60}},\ \bibinfo {pages} {14771} (\bibinfo {year}
  {1999})}\BibitemShut {NoStop}%
\bibitem [{\citenamefont {Goldsborough}\ and\ \citenamefont
  {R\"omer}(2014)}]{GolR14}%
  \BibitemOpen
  \bibfield  {author} {\bibinfo {author} {\bibfnamefont {A.~M.}\ \bibnamefont
  {Goldsborough}}\ and\ \bibinfo {author} {\bibfnamefont {R.~A.}\ \bibnamefont
  {R\"omer}},\ }\href {\doibase 10.1103/PhysRevB.89.214203} {\bibfield
  {journal} {\bibinfo  {journal} {Phys. Rev. B}\ }\textbf {\bibinfo {volume}
  {89}},\ \bibinfo {pages} {214203} (\bibinfo {year} {2014})}\BibitemShut
  {NoStop}%
\bibitem [{\citenamefont {Ruggiero}\ \emph {et~al.}(2016)\citenamefont
  {Ruggiero}, \citenamefont {Alba},\ and\ \citenamefont {Calabrese}}]{RugAC16}%
  \BibitemOpen
  \bibfield  {author} {\bibinfo {author} {\bibfnamefont {P.}~\bibnamefont
  {Ruggiero}}, \bibinfo {author} {\bibfnamefont {V.}~\bibnamefont {Alba}}, \
  and\ \bibinfo {author} {\bibfnamefont {P.}~\bibnamefont {Calabrese}},\ }\href
  {\doibase 10.1103/PhysRevB.94.035152} {\bibfield  {journal} {\bibinfo
  {journal} {Phys. Rev. B}\ }\textbf {\bibinfo {volume} {94}},\ \bibinfo
  {pages} {035152} (\bibinfo {year} {2016})}\BibitemShut {NoStop}%
\bibitem [{\citenamefont {Ma}\ \emph {et~al.}(1979)\citenamefont {Ma},
  \citenamefont {Dasgupta},\ and\ \citenamefont {Hu}}]{MaDH79}%
  \BibitemOpen
  \bibfield  {author} {\bibinfo {author} {\bibfnamefont {S.-k.}\ \bibnamefont
  {Ma}}, \bibinfo {author} {\bibfnamefont {C.}~\bibnamefont {Dasgupta}}, \ and\
  \bibinfo {author} {\bibfnamefont {C.-k.}\ \bibnamefont {Hu}},\ }\href
  {http://link.aps.org/doi/10.1103/PhysRevLett.43.1434} {\bibfield  {journal}
  {\bibinfo  {journal} {Phys. Rev. Lett.}\ }\textbf {\bibinfo {volume} {43}},\
  \bibinfo {pages} {1434} (\bibinfo {year} {1979})}\BibitemShut {NoStop}%
\bibitem [{\citenamefont {Dasgupta}\ and\ \citenamefont {Ma}(1980)}]{DasM80}%
  \BibitemOpen
  \bibfield  {author} {\bibinfo {author} {\bibfnamefont {C.}~\bibnamefont
  {Dasgupta}}\ and\ \bibinfo {author} {\bibfnamefont {S.-k.}\ \bibnamefont
  {Ma}},\ }\href {\doibase 10.1103/PhysRevB.22.1305} {\bibfield  {journal}
  {\bibinfo  {journal} {Phys. Rev. B}\ }\textbf {\bibinfo {volume} {22}},\
  \bibinfo {pages} {1305} (\bibinfo {year} {1980})}\BibitemShut {NoStop}%
\bibitem [{\citenamefont {Lin}\ \emph {et~al.}(2017)\citenamefont {Lin},
  \citenamefont {Kao}, \citenamefont {Chen},\ and\ \citenamefont
  {Lin}}]{LinKCL17}%
  \BibitemOpen
  \bibfield  {author} {\bibinfo {author} {\bibfnamefont {Y.-P.}\ \bibnamefont
  {Lin}}, \bibinfo {author} {\bibfnamefont {Y.-J.}\ \bibnamefont {Kao}},
  \bibinfo {author} {\bibfnamefont {P.}~\bibnamefont {Chen}}, \ and\ \bibinfo
  {author} {\bibfnamefont {Y.-C.}\ \bibnamefont {Lin}},\ }\href {\doibase
  10.1103/PhysRevB.96.064427} {\bibfield  {journal} {\bibinfo  {journal} {Phys.
  Rev. B}\ }\textbf {\bibinfo {volume} {96}},\ \bibinfo {pages} {064427}
  (\bibinfo {year} {2017})}\BibitemShut {NoStop}%
\bibitem [{\citenamefont {Vidal}(2007)}]{Vid07}%
  \BibitemOpen
  \bibfield  {author} {\bibinfo {author} {\bibfnamefont {G.}~\bibnamefont
  {Vidal}},\ }\href {http://link.aps.org/doi/10.1103/PhysRevLett.99.220405}
  {\bibfield  {journal} {\bibinfo  {journal} {Phys. Rev. Lett.}\ }\textbf
  {\bibinfo {volume} {99}},\ \bibinfo {pages} {220405} (\bibinfo {year}
  {2007})}\BibitemShut {NoStop}%
\bibitem [{\citenamefont {Vidal}(2008{\natexlab{a}})}]{Vid08_2}%
  \BibitemOpen
  \bibfield  {author} {\bibinfo {author} {\bibfnamefont {G.}~\bibnamefont
  {Vidal}},\ }\href {\doibase 10.1103/PhysRevLett.101.110501} {\bibfield
  {journal} {\bibinfo  {journal} {Phys. Rev. Lett.}\ }\textbf {\bibinfo
  {volume} {101}},\ \bibinfo {pages} {110501} (\bibinfo {year}
  {2008}{\natexlab{a}})}\BibitemShut {NoStop}%
\bibitem [{\citenamefont {Evenbly}\ and\ \citenamefont {Vidal}(2009)}]{EveV09}%
  \BibitemOpen
  \bibfield  {author} {\bibinfo {author} {\bibfnamefont {G.}~\bibnamefont
  {Evenbly}}\ and\ \bibinfo {author} {\bibfnamefont {G.}~\bibnamefont
  {Vidal}},\ }\href {\doibase 10.1103/PhysRevB.79.144108} {\bibfield  {journal}
  {\bibinfo  {journal} {Phys. Rev. B}\ }\textbf {\bibinfo {volume} {79}},\
  \bibinfo {pages} {144108} (\bibinfo {year} {2009})}\BibitemShut {NoStop}%
\bibitem [{\citenamefont {Pfeifer}\ \emph {et~al.}(2009)\citenamefont
  {Pfeifer}, \citenamefont {Evenbly},\ and\ \citenamefont {Vidal}}]{PfeEV09}%
  \BibitemOpen
  \bibfield  {author} {\bibinfo {author} {\bibfnamefont {R.~N.~C.}\
  \bibnamefont {Pfeifer}}, \bibinfo {author} {\bibfnamefont {G.}~\bibnamefont
  {Evenbly}}, \ and\ \bibinfo {author} {\bibfnamefont {G.}~\bibnamefont
  {Vidal}},\ }\href {\doibase 10.1103/PhysRevA.79.040301} {\bibfield  {journal}
  {\bibinfo  {journal} {Phys. Rev. A}\ }\textbf {\bibinfo {volume} {79}},\
  \bibinfo {pages} {040301} (\bibinfo {year} {2009})}\BibitemShut {NoStop}%
\bibitem [{\citenamefont {Evenbly}\ \emph {et~al.}(2010)\citenamefont
  {Evenbly}, \citenamefont {Pfeifer}, \citenamefont {Pic\'o}, \citenamefont
  {Iblisdir}, \citenamefont {Tagliacozzo}, \citenamefont {McCulloch},\ and\
  \citenamefont {Vidal}}]{EvePPI10}%
  \BibitemOpen
  \bibfield  {author} {\bibinfo {author} {\bibfnamefont {G.}~\bibnamefont
  {Evenbly}}, \bibinfo {author} {\bibfnamefont {R.~N.~C.}\ \bibnamefont
  {Pfeifer}}, \bibinfo {author} {\bibfnamefont {V.}~\bibnamefont {Pic\'o}},
  \bibinfo {author} {\bibfnamefont {S.}~\bibnamefont {Iblisdir}}, \bibinfo
  {author} {\bibfnamefont {L.}~\bibnamefont {Tagliacozzo}}, \bibinfo {author}
  {\bibfnamefont {I.~P.}\ \bibnamefont {McCulloch}}, \ and\ \bibinfo {author}
  {\bibfnamefont {G.}~\bibnamefont {Vidal}},\ }\href {\doibase
  10.1103/PhysRevB.82.161107} {\bibfield  {journal} {\bibinfo  {journal} {Phys.
  Rev. B}\ }\textbf {\bibinfo {volume} {82}},\ \bibinfo {pages} {161107}
  (\bibinfo {year} {2010})}\BibitemShut {NoStop}%
\bibitem [{\citenamefont {Evenbly}\ and\ \citenamefont {Vidal}(2014)}]{EveV14}%
  \BibitemOpen
  \bibfield  {author} {\bibinfo {author} {\bibfnamefont {G.}~\bibnamefont
  {Evenbly}}\ and\ \bibinfo {author} {\bibfnamefont {G.}~\bibnamefont
  {Vidal}},\ }\href {\doibase 10.1007/s10955-014-0983-1} {\bibfield  {journal}
  {\bibinfo  {journal} {J. Stat. Phys.}\ }\textbf {\bibinfo {volume} {157}},\
  \bibinfo {pages} {931} (\bibinfo {year} {2014})}\BibitemShut {NoStop}%
\bibitem [{\citenamefont {Fisher}(1992)}]{Fis92}%
  \BibitemOpen
  \bibfield  {author} {\bibinfo {author} {\bibfnamefont {D.~S.}\ \bibnamefont
  {Fisher}},\ }\href {\doibase 10.1103/PhysRevLett.69.534} {\bibfield
  {journal} {\bibinfo  {journal} {Phys. Rev. Lett.}\ }\textbf {\bibinfo
  {volume} {69}},\ \bibinfo {pages} {534} (\bibinfo {year} {1992})}\BibitemShut
  {NoStop}%
\bibitem [{\citenamefont {Fisher}(1995)}]{Fis95}%
  \BibitemOpen
  \bibfield  {author} {\bibinfo {author} {\bibfnamefont {D.~S.}\ \bibnamefont
  {Fisher}},\ }\href {\doibase 10.1103/PhysRevB.51.6411} {\bibfield  {journal}
  {\bibinfo  {journal} {Phys. Rev. B}\ }\textbf {\bibinfo {volume} {51}},\
  \bibinfo {pages} {6411} (\bibinfo {year} {1995})}\BibitemShut {NoStop}%
\bibitem [{\citenamefont {Westerberg}\ \emph {et~al.}(1995)\citenamefont
  {Westerberg}, \citenamefont {Furusaki}, \citenamefont {Sigrist},\ and\
  \citenamefont {Lee}}]{WesFSL95}%
  \BibitemOpen
  \bibfield  {author} {\bibinfo {author} {\bibfnamefont {E.}~\bibnamefont
  {Westerberg}}, \bibinfo {author} {\bibfnamefont {A.}~\bibnamefont
  {Furusaki}}, \bibinfo {author} {\bibfnamefont {M.}~\bibnamefont {Sigrist}}, \
  and\ \bibinfo {author} {\bibfnamefont {P.~A.}\ \bibnamefont {Lee}},\
  }\href@noop {} {\bibfield  {journal} {\bibinfo  {journal} {Phys. Rev. Lett.}\
  }\textbf {\bibinfo {volume} {75}},\ \bibinfo {pages} {4302} (\bibinfo {year}
  {1995})}\BibitemShut {NoStop}%
\bibitem [{\citenamefont {Altman}\ \emph {et~al.}(2004)\citenamefont {Altman},
  \citenamefont {Kafri}, \citenamefont {Polkovnikov},\ and\ \citenamefont
  {Refael}}]{AltKPR04}%
  \BibitemOpen
  \bibfield  {author} {\bibinfo {author} {\bibfnamefont {E.}~\bibnamefont
  {Altman}}, \bibinfo {author} {\bibfnamefont {Y.}~\bibnamefont {Kafri}},
  \bibinfo {author} {\bibfnamefont {A.}~\bibnamefont {Polkovnikov}}, \ and\
  \bibinfo {author} {\bibfnamefont {G.}~\bibnamefont {Refael}},\ }\href
  {\doibase 10.1103/PhysRevLett.93.150402} {\bibfield  {journal} {\bibinfo
  {journal} {Phys. Rev. Lett.}\ }\textbf {\bibinfo {volume} {93}},\ \bibinfo
  {pages} {150402} (\bibinfo {year} {2004})}\BibitemShut {NoStop}%
\bibitem [{\citenamefont {Igl\'{o}i}\ and\ \citenamefont
  {Monthus}(2005)}]{IglM05}%
  \BibitemOpen
  \bibfield  {author} {\bibinfo {author} {\bibfnamefont {F.}~\bibnamefont
  {Igl\'{o}i}}\ and\ \bibinfo {author} {\bibfnamefont {C.}~\bibnamefont
  {Monthus}},\ }\href@noop {} {\bibfield  {journal} {\bibinfo  {journal} {Phys.
  Rep.}\ }\textbf {\bibinfo {volume} {412}},\ \bibinfo {pages} {277} (\bibinfo
  {year} {2005})}\BibitemShut {NoStop}%
\bibitem [{\citenamefont {Refael}\ and\ \citenamefont {Moore}(2004)}]{RefM04}%
  \BibitemOpen
  \bibfield  {author} {\bibinfo {author} {\bibfnamefont {G.}~\bibnamefont
  {Refael}}\ and\ \bibinfo {author} {\bibfnamefont {J.~E.}\ \bibnamefont
  {Moore}},\ }\href {\doibase 10.1103/PhysRevLett.93.260602} {\bibfield
  {journal} {\bibinfo  {journal} {Phys. Rev. Lett.}\ }\textbf {\bibinfo
  {volume} {93}},\ \bibinfo {pages} {260602} (\bibinfo {year}
  {2004})}\BibitemShut {NoStop}%
\bibitem [{\citenamefont {Refael}\ and\ \citenamefont {Moore}(2007)}]{RefM07}%
  \BibitemOpen
  \bibfield  {author} {\bibinfo {author} {\bibfnamefont {G.}~\bibnamefont
  {Refael}}\ and\ \bibinfo {author} {\bibfnamefont {J.~E.}\ \bibnamefont
  {Moore}},\ }\href {\doibase 10.1103/PhysRevB.76.024419} {\bibfield  {journal}
  {\bibinfo  {journal} {Phys. Rev. B}\ }\textbf {\bibinfo {volume} {76}},\
  \bibinfo {pages} {024419} (\bibinfo {year} {2007})}\BibitemShut {NoStop}%
\bibitem [{\citenamefont {Hikihara}\ \emph {et~al.}(1999)\citenamefont
  {Hikihara}, \citenamefont {Furusaki},\ and\ \citenamefont
  {Sigrist}}]{HikFS99}%
  \BibitemOpen
  \bibfield  {author} {\bibinfo {author} {\bibfnamefont {T.}~\bibnamefont
  {Hikihara}}, \bibinfo {author} {\bibfnamefont {A.}~\bibnamefont {Furusaki}},
  \ and\ \bibinfo {author} {\bibfnamefont {M.}~\bibnamefont {Sigrist}},\ }\href
  {\doibase 10.1103/PhysRevB.60.12116} {\bibfield  {journal} {\bibinfo
  {journal} {Phys. Rev. B}\ }\textbf {\bibinfo {volume} {60}},\ \bibinfo
  {pages} {12116} (\bibinfo {year} {1999})}\BibitemShut {NoStop}%
\bibitem [{\citenamefont {Wilson}(1975)}]{Wil75}%
  \BibitemOpen
  \bibfield  {author} {\bibinfo {author} {\bibfnamefont {K.~G.}\ \bibnamefont
  {Wilson}},\ }\href {http://link.aps.org/doi/10.1103/RevModPhys.47.773}
  {\bibfield  {journal} {\bibinfo  {journal} {Rev. Mod. Phys}\ }\textbf
  {\bibinfo {volume} {47}},\ \bibinfo {pages} {773} (\bibinfo {year}
  {1975})}\BibitemShut {NoStop}%
\bibitem [{\citenamefont {Evenbly}\ and\ \citenamefont {Vidal}(2013)}]{EveV13}%
  \BibitemOpen
  \bibfield  {author} {\bibinfo {author} {\bibfnamefont {G.}~\bibnamefont
  {Evenbly}}\ and\ \bibinfo {author} {\bibfnamefont {G.}~\bibnamefont
  {Vidal}},\ }in\ \href@noop {} {\emph {\bibinfo {booktitle} {Strongly
  Correlated Systems: Numerical Methods}}},\ \bibinfo {series and number}
  {Springer Series in Solid-State Sciences},\ \bibinfo {editor} {edited by\
  \bibinfo {editor} {\bibfnamefont {A.}~\bibnamefont {Avella}}\ and\ \bibinfo
  {editor} {\bibfnamefont {F.}~\bibnamefont {Mancini}}}\ (\bibinfo  {publisher}
  {Springer Berlin Heidelberg},\ \bibinfo {year} {2013})\ Chap.~\bibinfo
  {chapter} {4}\BibitemShut {NoStop}%
\bibitem [{\citenamefont {Evenbly}\ and\ \citenamefont {Vidal}(2011)}]{EveV11}%
  \BibitemOpen
  \bibfield  {author} {\bibinfo {author} {\bibfnamefont {G.}~\bibnamefont
  {Evenbly}}\ and\ \bibinfo {author} {\bibfnamefont {G.}~\bibnamefont
  {Vidal}},\ }\href {\doibase 10.1007/s10955-011-0237-4} {\bibfield  {journal}
  {\bibinfo  {journal} {J. Stat. Phys.}\ }\textbf {\bibinfo {volume} {145}},\
  \bibinfo {pages} {891} (\bibinfo {year} {2011})}\BibitemShut {NoStop}%
\bibitem [{\citenamefont {Laflorencie}(2005)}]{Laf05}%
  \BibitemOpen
  \bibfield  {author} {\bibinfo {author} {\bibfnamefont {N.}~\bibnamefont
  {Laflorencie}},\ }\href {\doibase 10.1103/PhysRevB.72.140408} {\bibfield
  {journal} {\bibinfo  {journal} {Phys. Rev. B}\ }\textbf {\bibinfo {volume}
  {72}},\ \bibinfo {pages} {140408} (\bibinfo {year} {2005})}\BibitemShut
  {NoStop}%
\bibitem [{Ite()}]{Itensor023}%
  \BibitemOpen
  \href {http://itensor.org/} {\enquote {\bibinfo {title} {{IT}ensor
  library},}\ }\bibinfo {note} {Version: 0.2.3, http://itensor.org}\BibitemShut
  {NoStop}%
\bibitem [{\citenamefont {Henelius}\ and\ \citenamefont
  {Girvin}(1998)}]{HenG98}%
  \BibitemOpen
  \bibfield  {author} {\bibinfo {author} {\bibfnamefont {P.}~\bibnamefont
  {Henelius}}\ and\ \bibinfo {author} {\bibfnamefont {S.~M.}\ \bibnamefont
  {Girvin}},\ }\href {\doibase 10.1103/PhysRevB.57.11457} {\bibfield  {journal}
  {\bibinfo  {journal} {Phys. Rev. B}\ }\textbf {\bibinfo {volume} {57}},\
  \bibinfo {pages} {11457} (\bibinfo {year} {1998})}\BibitemShut {NoStop}%
\bibitem [{\citenamefont {Fisher}(1994)}]{Fis94}%
  \BibitemOpen
  \bibfield  {author} {\bibinfo {author} {\bibfnamefont {D.~S.}\ \bibnamefont
  {Fisher}},\ }\href {\doibase 10.1103/PhysRevB.50.3799} {\bibfield  {journal}
  {\bibinfo  {journal} {Phys. Rev. B}\ }\textbf {\bibinfo {volume} {50}},\
  \bibinfo {pages} {3799} (\bibinfo {year} {1994})}\BibitemShut {NoStop}%
\bibitem [{\citenamefont {Vitagliano}\ \emph {et~al.}(2010)\citenamefont
  {Vitagliano}, \citenamefont {Riera},\ and\ \citenamefont
  {Latorre}}]{VitRL10}%
  \BibitemOpen
  \bibfield  {author} {\bibinfo {author} {\bibfnamefont {G.}~\bibnamefont
  {Vitagliano}}, \bibinfo {author} {\bibfnamefont {A.}~\bibnamefont {Riera}}, \
  and\ \bibinfo {author} {\bibfnamefont {J.~I.}\ \bibnamefont {Latorre}},\
  }\href {http://stacks.iop.org/1367-2630/12/i=11/a=113049} {\bibfield
  {journal} {\bibinfo  {journal} {New J. Phys.}\ }\textbf {\bibinfo {volume}
  {12}},\ \bibinfo {pages} {113049} (\bibinfo {year} {2010})}\BibitemShut
  {NoStop}%
\bibitem [{\citenamefont {Ram\'{i}rez}\ \emph {et~al.}(2014)\citenamefont
  {Ram\'{i}rez}, \citenamefont {Rodr\'{i}guez-Laguna},\ and\ \citenamefont
  {Sierra}}]{RamRS14}%
  \BibitemOpen
  \bibfield  {author} {\bibinfo {author} {\bibfnamefont {G.}~\bibnamefont
  {Ram\'{i}rez}}, \bibinfo {author} {\bibfnamefont {J.}~\bibnamefont
  {Rodr\'{i}guez-Laguna}}, \ and\ \bibinfo {author} {\bibfnamefont
  {G.}~\bibnamefont {Sierra}},\ }\href
  {http://stacks.iop.org/1742-5468/2014/i=10/a=P10004} {\bibfield  {journal}
  {\bibinfo  {journal} {J. Stat. Mech. Theor. Exp.}\ }\textbf {\bibinfo
  {volume} {2014}},\ \bibinfo {pages} {P10004} (\bibinfo {year}
  {2014})}\BibitemShut {NoStop}%
\bibitem [{\citenamefont {Quito}\ \emph {et~al.}(2015)\citenamefont {Quito},
  \citenamefont {Hoyos},\ and\ \citenamefont {Miranda}}]{QuiHM15}%
  \BibitemOpen
  \bibfield  {author} {\bibinfo {author} {\bibfnamefont {V.~L.}\ \bibnamefont
  {Quito}}, \bibinfo {author} {\bibfnamefont {J.~A.}\ \bibnamefont {Hoyos}}, \
  and\ \bibinfo {author} {\bibfnamefont {E.}~\bibnamefont {Miranda}},\ }\href
  {http://link.aps.org/doi/10.1103/PhysRevLett.115.167201} {\bibfield
  {journal} {\bibinfo  {journal} {Phys. Rev. Lett.}\ }\textbf {\bibinfo
  {volume} {115}},\ \bibinfo {pages} {167201} (\bibinfo {year}
  {2015})}\BibitemShut {NoStop}%
\bibitem [{\citenamefont {Hyatt}\ \emph {et~al.}(2017)\citenamefont {Hyatt},
  \citenamefont {Garrison},\ and\ \citenamefont {Bauer}}]{HyaGB17}%
  \BibitemOpen
  \bibfield  {author} {\bibinfo {author} {\bibfnamefont {K.}~\bibnamefont
  {Hyatt}}, \bibinfo {author} {\bibfnamefont {J.~R.}\ \bibnamefont {Garrison}},
  \ and\ \bibinfo {author} {\bibfnamefont {B.}~\bibnamefont {Bauer}},\ }\href
  {https://arxiv.org/abs/1704.01974} {\bibfield  {journal} {\bibinfo  {journal}
  {arXiv:1704.01974 [cond-mat.str-el]}\ } (\bibinfo {year} {2017})}\BibitemShut
  {NoStop}%
\bibitem [{\citenamefont {Pfeifer}\ \emph
  {et~al.}(2014{\natexlab{a}})\citenamefont {Pfeifer}, \citenamefont
  {Haegeman},\ and\ \citenamefont {Verstraete}}]{PfeHV14}%
  \BibitemOpen
  \bibfield  {author} {\bibinfo {author} {\bibfnamefont {R.~N.~C.}\
  \bibnamefont {Pfeifer}}, \bibinfo {author} {\bibfnamefont {J.}~\bibnamefont
  {Haegeman}}, \ and\ \bibinfo {author} {\bibfnamefont {F.}~\bibnamefont
  {Verstraete}},\ }\href {\doibase 10.1103/PhysRevE.90.033315} {\bibfield
  {journal} {\bibinfo  {journal} {Phys. Rev. E}\ }\textbf {\bibinfo {volume}
  {90}},\ \bibinfo {pages} {033315} (\bibinfo {year}
  {2014}{\natexlab{a}})}\BibitemShut {NoStop}%
\bibitem [{\citenamefont {Pfeifer}\ \emph
  {et~al.}(2014{\natexlab{b}})\citenamefont {Pfeifer}, \citenamefont {Evenbly},
  \citenamefont {Singh},\ and\ \citenamefont {Vidal}}]{PfeESV14}%
  \BibitemOpen
  \bibfield  {author} {\bibinfo {author} {\bibfnamefont {R.~N.~C.}\
  \bibnamefont {Pfeifer}}, \bibinfo {author} {\bibfnamefont {G.}~\bibnamefont
  {Evenbly}}, \bibinfo {author} {\bibfnamefont {S.}~\bibnamefont {Singh}}, \
  and\ \bibinfo {author} {\bibfnamefont {G.}~\bibnamefont {Vidal}},\ }\href
  {https://arxiv.org/abs/1402.0939} {\bibfield  {journal} {\bibinfo  {journal}
  {arXiv:1402.0939 [physics.comp-ph]}\ } (\bibinfo {year}
  {2014}{\natexlab{b}})}\BibitemShut {NoStop}%
\bibitem [{\citenamefont {Vidal}(2008{\natexlab{b}})}]{Vid08}%
  \BibitemOpen
  \bibfield  {author} {\bibinfo {author} {\bibfnamefont {G.}~\bibnamefont
  {Vidal}},\ }\href@noop {} {\bibfield  {journal} {\bibinfo  {journal}
  {{arXiv:0707.1454v2}}\ } (\bibinfo {year} {2008}{\natexlab{b}})}\BibitemShut
  {NoStop}%
\bibitem [{\citenamefont {Tagliacozzo}\ \emph {et~al.}(2009)\citenamefont
  {Tagliacozzo}, \citenamefont {Evenbly},\ and\ \citenamefont
  {Vidal}}]{TagEV09}%
  \BibitemOpen
  \bibfield  {author} {\bibinfo {author} {\bibfnamefont {L.}~\bibnamefont
  {Tagliacozzo}}, \bibinfo {author} {\bibfnamefont {G.}~\bibnamefont
  {Evenbly}}, \ and\ \bibinfo {author} {\bibfnamefont {G.}~\bibnamefont
  {Vidal}},\ }\href {\doibase 10.1103/PhysRevB.80.235127} {\bibfield  {journal}
  {\bibinfo  {journal} {Phys. Rev. B}\ }\textbf {\bibinfo {volume} {80}},\
  \bibinfo {pages} {235127} (\bibinfo {year} {2009})}\BibitemShut {NoStop}%
\end{thebibliography}
\end{document}